\documentclass[a4paper,UKenglish,cleveref, autoref, thm-restate]{lipics-v2019}

\usepackage[]{todonotes}
\usepackage[inline]{enumitem}
\usepackage{bbm}
\usepackage{wrapfig}
\usepackage{subcaption}
\usepackage{stmaryrd}
\usepackage{listings}
\usepackage{MnSymbol}
\allowdisplaybreaks

\newcommand{\noflow}{\not\rcurvearrowright}

\newcommand{\step}{p}

\newcommand{\InVars}{X}
\newcommand{\OutVars}{Y}
\newcommand{\AllVars}{Z}

\newcommand{\InfFlow}{F}
\newcommand{\Assumption}{\mathcal{A}}
\newcommand{\Guarantee}{\mathcal{G}}
\newcommand{\Prop}{\mathcal{P}}
\newcommand{\DProp}[1]{\Prop^{#1}}

\newcommand{\ID}[1]{\text{Id}_{#1}}

\newcommand{\PAs}{\hat{\Assumption}}

\newcommand{\IFRef}{\preceq}
\newcommand{\IFCompt}{\sim}
\newcommand{\IFImpl}{\models}
\newcommand{\IFEnv}{\models}

\newcommand{\IFCompos}{\otimes}
\newcommand{\IFSharedRef}{\sqcap}

\newcommand{\Compl}[1]{\overline{#1}}
\newcommand{\AComp}{\Compl{\Assumption}}
\newcommand{\GComp}{\Compl{\Guarantee}}
\newcommand{\PropComp}{\Compl{\Prop}}

\newcommand{\InfFlowC}{f}

\newcommand{\Flows}{\mathcal{M}}
\newcommand{\Env}{\mathcal{E}}

\newcommand{\CFCompos}{\otimes}

\newcommand{\Simul}{H}

\newcommand{\IFFul}{\mathbb{F}}
\newcommand{\States}{Q}
\newcommand{\std}{q}
\newcommand{\InitState}{\hat{\std}}
\newcommand{\Transition}{\delta}
\newcommand{\AssumptionFul}{\mathbb{A}}
\newcommand{\GuaranteeFul}{\mathbb{G}}
\newcommand{\PropFul}{\mathbb{P}}

\newcommand{\IFCFul}{\mathbbm{f}}
\newcommand{\FlowsFul}{\mathbb{M}}
\newcommand{\EnvFul}{\mathbb{E}}

\newcommand{\OutStep}{\Transition^{\OutVars}}

\newcommand{\InStep}{\Transition^{\InVars}}

\newcommand{\alphabet}{\AllVars}
\newcommand{\setTraces}{T}
\newcommand{\allSetTraces}{\mathbb{T}}
\newcommand{\system}{S}
\newcommand{\setsetTraces}{\textbf{T}}

\newcommand{\strongS}[1]{\llbracket #1 \rrbracket_s}
\newcommand{\awareS}[1]{\llbracket #1 \rrbracket_a}
\newcommand{\unstructS}[1]{\llbracket #1 \rrbracket_u}

\newcommand{\tAnd}{\text{ and }}
\newcommand{\tOr}{\text{ or }}

\newcommand{\tIf}{\text{ if }}
\newcommand{\tthen}{\text{ then }}
\newcommand{\tThen}{\text{, then }}
\newcommand{\tIff}{\text{ iff }}
\newcommand{\tSt}{\text{ s.t.\ }}

\newcommand{\key}{\textsf{key}}
\newcommand{\can}{\textsf{can}}
\newcommand{\ecu}{\textsf{ecu}}
\newcommand{\imm}{\textsf{imm}}
\newcommand{\deb}{\textsf{deb}}

\newcommand{\ifkey}{\InfFlow_{\emph{key}}}
\newcommand{\ifcan}{\InfFlow_{\emph{can}}}
\newcommand{\ifecu}{\InfFlow_{\emph{ecu}}}
\newcommand{\ifimm}{\InfFlow_{\emph{imm}}}
\newcommand{\ifteam}{\InfFlow_{\emph{team}}}

\newcommand{\ifnotwell}{G}
\newcommand{\ifnotwellnotimp}{h}
\newcommand{\ifcnotwell}{g}
\newcommand{\ifenotwell}{g_{\Env}}

\title{Information-Flow Interfaces} 

\titlerunning{Information-Flow Interfaces} 

\author{Ezio Bartocci}{Technische Universit\"at Wien, Vienna, Austria \and \url{http://www.eziobartocci.com} }{ezio.bartocci@tuwien.ac.at}{https://orcid.org/0000-0002-8004-6601}{This research was supported by the Austrian FWF project W1255-N23}

\author{Thomas Ferr\`ere}{Imagination Technologies, Kings Langley, UK }{thomas.ferrere@imgtec.com}{https://orcid.org/0000-0001-5199-3143}{}

\author{Thomas A. Henzinger}{IST Austria, Klosterneuburg, Austria \and \url{http://pub.ist.ac.at/~tah/} }{tah@ist.ac.at}{https://orcid.org/0000-0002-2985-7724}{}

\author{Dejan Nickovic}{AIT Austrian Institute of Technology, Vienna, Austria }{dejan.nickovic@ait.ac.at}{https://orcid.org/0000-0001-5468-0396}{}

\author{Ana Oliveira da Costa}{Technische Universit\"at Wien, Vienna, Austria }{ana.costa@tuwien.ac.at}{}{This research was supported by the Austrian FWF project W1255-N23}

\authorrunning{E. Bartocci et al.} 

\Copyright{E. Bartocci and T. Ferr\`ere and T. A. Henzinger and D. Nickovic and A. Oliveira da Costa} 



\relatedversion{} 

\nolinenumbers 

\EventEditors{}
\EventNoEds{2}
\EventYear{2020}
\EventLogo{}
\SeriesVolume{}
\ArticleNo{}

\begin{document}

\maketitle

\begin{abstract}
Contract-based design is a promising methodology for taming the complexity of developing 
sophisticated systems. A formal contract distinguishes between \emph{assumptions}, which are constraints 
that the designer of a component puts on the environments in which the component can be used safely, 
and \emph{guarantees}, which are promises that the designer asks from the team that implements the component. 
A theory of formal contracts can be formalized as an \emph{interface theory}, which supports the composition 
and refinement of both assumptions and guarantees.

Although there is a rich landscape of contract-based design methods that address 
functional and extra-functional  properties,
we present the first interface theory that is designed for ensuring system-wide 
security properties, thus paving the way for a science of safety and security co-engineering.
Our framework provides a refinement relation and a composition operation that  
support both incremental design and independent 
implementability. We develop our theory for both {\em stateless} and {\em stateful} interfaces.
We illustrate the applicability of our framework with an example inspired from the 
automotive domain.
Finally, we provide three plausible trace semantics to stateful
information-flow interfaces and we show that only two correspond 
to temporal logics for specifying hyperproperties, while the third 
defines a new class of hyperproperties that lies between the other
two classes.

\end{abstract}


\section{Introduction}
\label{sec:introduction}

The rise of pervasive information and communication technologies seen
in cyber-physical systems, internet of things, and blockchain services
has been accompanied by a tremendous growth in the size and complexity
of systems~\cite{RatasichKGGSB19}.  Subtle dependencies involving
multiple architectural layers and unforeseen environmental
interactions can expose these systems to cyber-attacks.  This problem
is further exacerbated by the heterogeneous nature of their
constituent components, which are often developed independently by
different teams or providers.  In such a scenario, defining and
enforcing security requirements across components at an early stage of
the design process becomes a necessity.  This engineering approach is
called \emph{security-by-design}.  Although in recent years there has
been impressive progress in the verification of security properties
for individual system components, the science of compositional
security design~\cite{Mantel02} is still in its infancy.

Security policies are usually enforced by restricting the flow of
information in a system~\cite{schneider2000}.  \emph{Information-flow policies}
define which information a user or a software/hardware component is
allowed to observe by interacting with another component.  They cover
both confidentiality and integrity constraints.  Confidentiality
limits the availability of information, for example, by forbidding
confidential data to leak to public variables.  Integrity, dually,
concerns the modification of information, by prohibiting untrusted
agents to interfere with trusted data.  The goal of information-flow
control is to ensure that a system as a whole satisfies the desired
policies.  This task is challenging because it needs to account also
for implicit flows and side-channels from which an attacker can
infer information by combining multiple observations.  From a
formal-language perspective, such security vulnerabilities are not
characterized by properties of a single system execution, but rather
by properties of sets of execution traces, which are called
\emph{hyperproperties}~\cite{ClarksonS10}.  Much recent work on formal
methods for security has focused on model checking various classes of
hyperproperties \cite{ClarksonFKMRS14,CoenenFST19,FinkbeinerHT18}.

In this paper, we present a \emph{contract-based
design}~\cite{BenvenisteCNPRR18} approach for information-flow
properties.  Contract-based design provides a formal framework for
building complex systems from individual components, mixing both
top-down and bottom-up steps.  A top-down step decomposes and refines
system-wide requirements; a bottom-up step assembles a system by
combining available components.  A formal contract distinguishes
between \emph{assumptions}, which are constraints that the designer
of a component puts on the environments in which the component can
be used safely, and \emph{guarantees}, which are promises that the
designer asks from the team that implements the component.  A theory
of formal contracts can be formalized as an \emph{interface theory},
which supports the composition and refinement of both assumptions and
guarantees \cite{de2001interface,dealfaro2005,tripakis2011theory}.
While there is a rich landscape of interface theories for functional
and extra-functional properties
\cite{ChakrabartiAHS03,AlfaroHS02,DavidLLNW10}, we present the first
interface theory that is designed for ensuring system-wide security
properties, thus paving the way for a science of safety and security
co-engineering.

Our theory is based on information-flow assumptions as well as
information-flow guarantees.  As an interface theory, our theory
supports both {\em incremental design} and {\em independent
implementability}~\cite{dealfaro2005}.  Incremental design
allows the composition of different system parts, each coming with
their own assumptions and guarantees, without requiring additional
knowledge of the overall design context.  Independent
implementability enables the separate refinement of different system
parts by different teams that, without gaining additional information
about each other's design choices, can still be certain that their
designs, once combined, preserve the specified system-wide
requirements.  While in previous interface theories, the environment
of a component is held responsible for meeting assumptions, and
the implementation of the component for the guarantees, there are
cases of information-flow violations for which blame cannot be
assigned uniquely to the implementation or the environment.  In
information-flow interfaces we therefore introduce, besides
assumptions and guarantees, a new, third type of constraint---called
\emph{properties}---whose enforcement is the shared responsibility
of the implementation and the environment.

We develop our framework for both {\em stateless} and {\em stateful}
interfaces.  Stateless information-flow interfaces are built from
primitive information-flow constraints---assumptions, guarantees, and
properties---of the form ``the value of a variable $y$ is always
independent of the value of another variable~$x$.''  Stateful
information-flow interfaces add a temporal dimension, e.g.,
``the value of $y$ is independent of $x$ \emph{until} the value
of $z$ is independent of~$x$.''  The temporal dimension is introduced
through a natural notion of state and state transition for interfaces,
not through logical operators.  Besides proving that our calculus of
information-flow interfaces satisfies the principles of incremental
design and independent implementability, we show that among three
plausible ways of giving a trace semantics to stateful
information-flow interfaces, only two correspond to temporal logics
for specifying hyperproperties, namely HyperLTL~\cite{ClarksonFKMRS14} 
and linear time epistemic logic~\cite{fagin2003reasoning,bozzelli2015unifying}.  The third, and arguably most
natural, trace interpretation of stateful information-flow interfaces
defines a new class of hyperproperties that lies between the other
two classes.

%

\section{Motivating Example}
\label{sec:motivation}

We illustrate the main concepts of our information flow interface theory using an {\em electronic vehicle immobilizer} (EVI) example. 
An EVI is a security device handling a transponder key and used by car manufacturers to prevent hot wiring a car, thus prevent car theft.
If the transponder authentication fails, then the engine control unit blocks the car's ignition. 
The communication between the immobilizer and other components in the vehicle takes place through the controller area network (CAN), a serial communication 
technology that is commonly used in automobile architectures to connect electronic control units (ECUs). This 
communication protocol does not include native support of security-related features. 
Thus, it is the responsibility of the components that use the bus to enforce confidentiality and integrity policies.

\begin{wrapfigure}{r}{0.45\textwidth}
\centering
\scalebox{0.5}{ 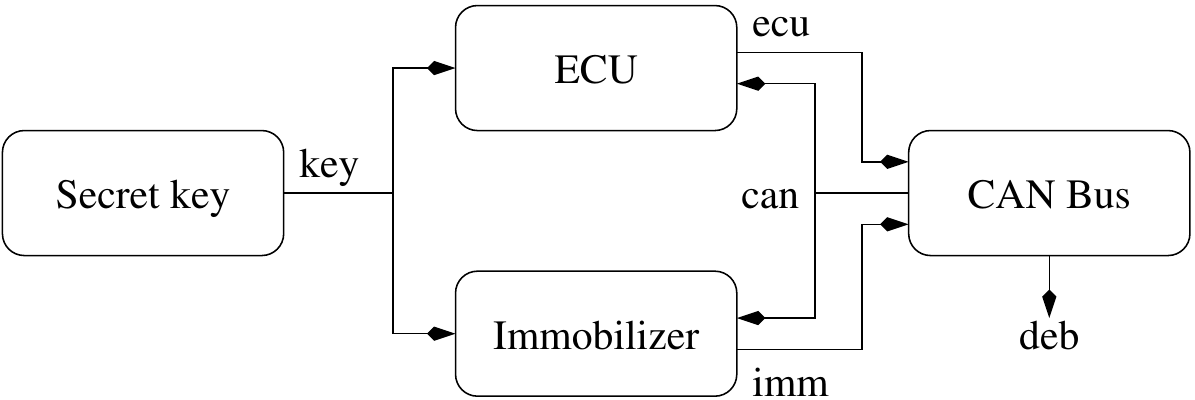 }
\caption{High-level view on immobilizer feature requirements.}
\label{fig:imm}
\end{wrapfigure}
In this example, we model the communication of an automotive engine control with an immobilizer through the CAN.  
The authentication follows a challenge-response protocol.
The communication session starts with the engine's ECU sending a freshly generated random number encrypted with a secret key known to both of the devices. The immobilizer replies to this challenge with an appropriate response encrypted with the same key. 
Figure~\ref{fig:imm} depicts a high-level overview of this model, which must enforce the following security property: the secret key shall never leak to the environment via the CAN bus.
In the remainder of this section, we illustrate how we can use an interface theory for information flow policies to implement the above specification in 
(1) a top-down, and (2) a bottom-up design fashion.

%
\noindent \textbf{Top-down design:}  We first demonstrate \emph{stepwise refinement of a global specification}: different engineering teams can independently implement 
subsystems, without violating the overall specification. 
We start by illustrating this process for the stateless case in Fig.~\ref{fig:design}~(a).
We start with the interface $\InfFlow$ that represents the overall (closed) system when it is in operation mode.
It specifies the global property that information from \key\ is not allowed to flow to \can\ or \deb. 
We assume that the secret key and the CAN bus are standard components provided by third-party suppliers. Our goal is to design the remaining sub-system consisting 
of the immobilizer and the ECU. This gives us a natural decomposition of $\InfFlow$ into three interfaces: (1) $\ifkey$ specifying the secret key, (2) $\ifcan$ specifying 
the CAN bus, and (3) $\ifteam$ specifying the sub-system that we want to further develop. We note that the property from $\InfFlow$ becomes an assumption in 
$\ifteam$.

The proposed decomposition does not work because $\ifteam$ \emph{is not compatible} with $\ifcan$---their composition would 
violate the assumption from $\ifteam$ by enabling the secret key to flow to the CAN via the ECU or the immobilizer. 
These two interfaces can be made compatible by strengthening guarantees of $\ifteam$ and 
forbidding the key to flow to the ECU and the immobilizer, resulting in the interface $\ifteam'$.
We further decompose $\ifteam'$ into two interfaces: (1) $\ifecu$ specifying the 
ECU component, and (2) $\ifimm$ specifying the immobilizer component. 
We note that the composition of $\ifkey$, $\ifcan$, $\ifecu$ and $\ifimm$ 
refines the original system-level specification $\InfFlow$.
Finally, we implement the four interfaces, derived from the overall specification, in components $f_{\emph{key}}$, $f_{\emph{ecu}}$, $f_{\emph{imm}}$ 
and $f_{\emph{can}}$. We note that the implementation of $f_{\emph{ecu}}$ and $f_{\emph{imm}}$ could be done 
independently, by two separate teams. The ECU (immobilizer) component guarantees that 
the secret key does not flow to its output port and works correctly in any environment that forbids other means of the secret key flowing 
to the CAN bus.

\begin{figure}
\centering
\scalebox{0.44}{ 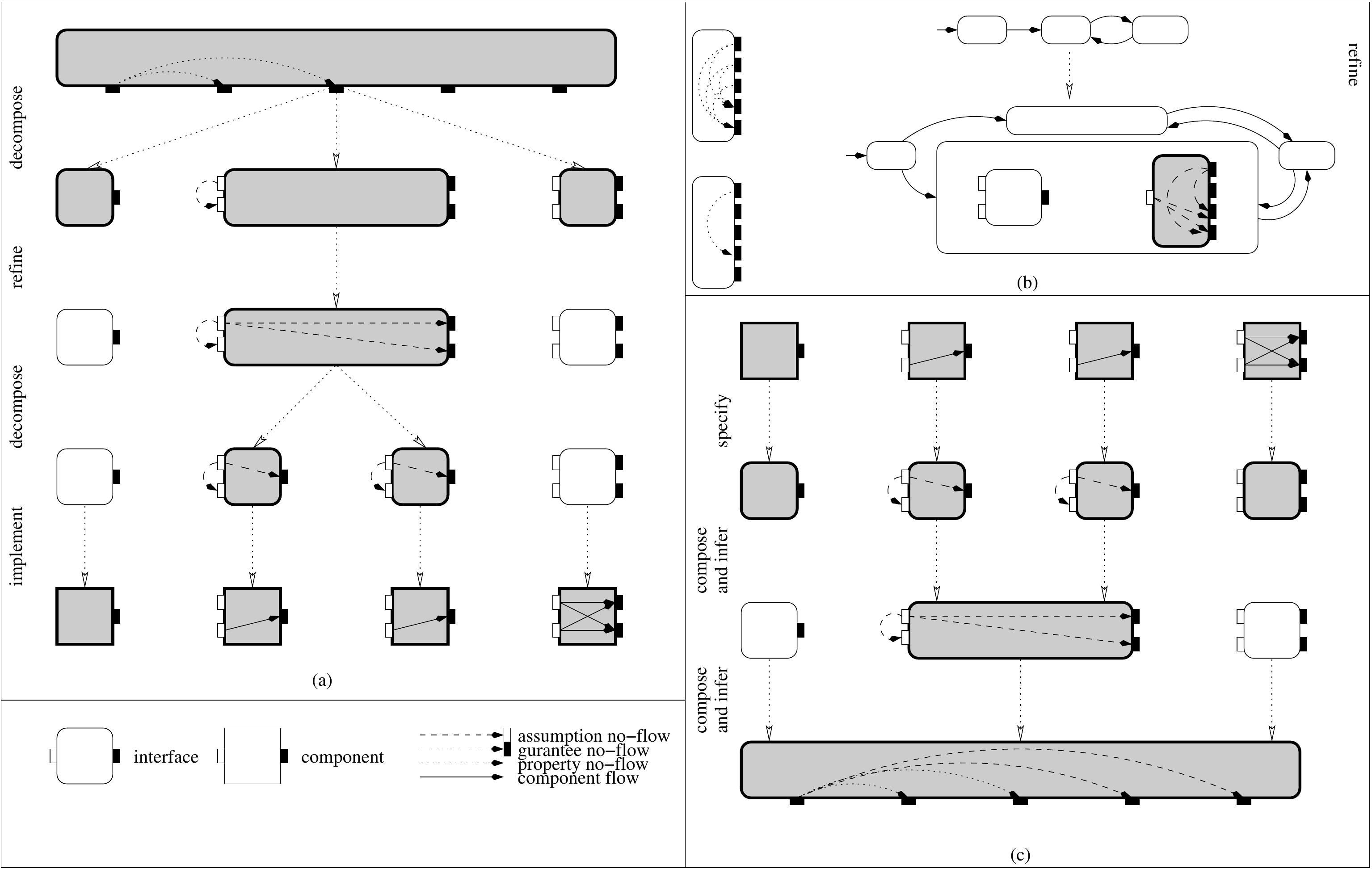 }
\vspace{-2ex}
\caption{EVI system: top-down design for (a) stateless and (b) stateful interface, and (c) bottom-up design.}
\vspace{-4.5ex}
\label{fig:design}
\end{figure}

The system being designed can be in one of the following modes: \emph{initialization}, \emph{operational} or \emph{debug}.
This is illustrated in the stateful interface \(\IFFul\) in Fig.~\ref{fig:design}~(b).
\emph{Stateful interfaces} are finite state machines with every state being labeled with a stateless interface. 
During initialization no information flows to CAN or the debug port, this state ends when an immobilizer communicates with the car.
Then, the operation mode is decorated 
with the same $\InfFlow$ interface from the stateless example while the initialization is decorated with the interface
\(\InfFlow_{\emph{init}}\) in Fig.~\ref{fig:design}~(b). Debug mode specified in
interface \(\InfFlow_{\emph{debug}}\) allows all information to flow to the debug port.

In the stateful interface \(\IFFul'\) we illustrate a refinement of the initial specification in  \(\IFFul\).
We consider the case that the team needs to accommodate two different architectures for the operational mode, for example due to backward compatibility constraints.
Then, in addition to the decomposition of \(\InfFlow\) showcased in Fig.~\ref{fig:design}~(a), it is specified an alternative  in which
the immobilizer is a third-party part. Note that \(\IFCompos\) denotes interface composition.


\noindent \textbf{Bottom-up design:} The bottom-up design, illustrated in  Fig.~\ref{fig:design}~(c), is to a large extent symmetric to the top-down approach. 
We start with the available secret key, ECU, immobilizer and CAN components, specified by interfaces $\ifkey$, $\ifecu$, $\ifimm$ and $\ifcan$. 
The main design step consists in composing a set of interfaces and \emph{inferring new global properties of the composition}. 
These properties are flows that cannot be created given the current set of assumptions and guarantees.
In the example we infer two global properties, that the secret key can never flow either to the 
CAN or to the debug port, because those ports can only be accessed through \(\ecu\) or \(\imm\) ports to which \(\key\) cannot flow.


In summary, our framework defines relations on both stateless and stateful interfaces that allow to check if:
\begin{itemize}
\item a given interface refines (or abstracts) the current specification;
\item two interfaces are compatible for composition, i.e.\ they respect  each other assumptions;
\item a specification is consistent, i.e.\ if the interface assumptions and guarantees are sufficient to ensure its property;
\item information-flows in a component define an implementation of a given interface.
\end{itemize}
Moreover, we define a composition operator which infers new global specification from the local information-flow restrictions specified by the interfaces being composed.
We can then combine this operator with the refinement relation to check if a system decomposition meets this system's specification.

%

\section{Stateless Interfaces}
\label{sec:stateless}

In this section, we introduce a stateless interface and component algebra for secure information flow. 
This algebra is based on properties of graphs linking input and output variables, and enables us to reason about security requirements in a hierarchical manner. 
We say that information flows from one variable to another when the value of the first influences the second.
We are interested in the \emph{structural} properties of information flow within a system and 
define relations abstracting flows, \emph{flow relations}, as being both reflexive and transitively closed.
Proofs are in the appendix~\ref{sec:app:stateless}.

\emph{Information-flow components} abstract implementations of a system by a \emph{flow relation}.
While {\em information-flow interface} specifies forbidden flows in an open system by defining
three kinds of constraints: assumptions, guarantees, and properties.
The \emph{assumption} characterizes flows that we assume are not part of the environment 
while the \emph{guarantee} describes all flows the system forbids and that are local to it. 
The \emph{property} qualifies the forbidden flows at the interaction between the system and its environment.
Hence, it represents a requirement on the closed system that needs to be enforced by guarantees on the open system and assumptions on its environment.
The distinction between the guarantee and the property is akin to the distinction between open systems and closed systems.
The property represents a requirement on the closed system that is enforced by a guarantee on the open system and an assumption on its environment.

\begin{definition}
\label{def:component:interface}
Let \(\InVars\) and \(\OutVars\) are disjoint sets of \emph{input}
and \emph{output} variables, respectively, with $\AllVars = \InVars \cup \OutVars$ the set of all variables.
A \emph{flow relation} in \(A \times B\)
is reflexive in \(A \cap B\) and transitive in \(A \cup B\).
A \emph{ stateless information-flow component} is a tuple \((\InVars,\OutVars,\Flows)\), where
\(\Flows \subseteq  (\InVars \cup \OutVars) \times \OutVars\) is a flow relation, called \emph{flows}.
A \emph{stateless information-flow interface} is a tuple 
\((\InVars,\OutVars,\Assumption,\Guarantee, \Prop)\), where:
\(\Assumption \subseteq \AllVars \times \InVars\) is a relation, called \emph{assumption}; 
\(\Guarantee \subseteq \AllVars \times \OutVars\) is a  relation, called \emph{guarantee};
and 
\(\Prop \subseteq \AllVars \times \OutVars\) is a relation, called \emph{property}.
\end{definition}

In what follows,  \({\InfFlow = (\InVars, \OutVars, \Assumption, \Guarantee, \Prop)}\)
and \(\InfFlow' = (\InVars, \OutVars, \Assumption', \Guarantee', \Prop')\) refer to arbitrary interfaces, and \(\InfFlowC=(\InVars,\OutVars,\Flows)\) and \(\InfFlowC_{\Env} = (\OutVars, \InVars, \Env)\) to arbitrary components.


Given an interface we are interested in components that describe either one of its implementations or one of the environments it admits.
The implementation of an interface shares its input/output signature, 
while the implementation of its environment reverses it.  
We define \emph{admissible environments} and \emph{implementations} of a given interface as components that do 
not include flows forbidden by either the interface assumptions or guarantees, respectively.

\begin{definition}
\label{def:i:environment}
A component \(\InfFlowC_{\Env}\) of the form \(\InfFlowC_{\Env} = (\OutVars, \InVars, \Env)\) is called
an \emph{environment} of \(\InfFlow\).
An environment is \emph{admissible for} \(\InfFlow\), 
denoted by \(\InfFlowC_{\Env} \IFEnv \InfFlow\), iff \(\Env \subseteq \AComp\).
A component \(\InfFlowC\) of the form \(\InfFlowC = (\InVars, \OutVars, \Flows)\) \emph{implements} the interface
\(\InfFlow\), denoted by \(\InfFlowC \IFImpl \InfFlow\), iff
\(\Flows \subseteq \GComp\).
\end{definition}

Note that the sets of implementations and admissible environments are mutually exclusive, hence we can use the notation \(\models\) for both. 

An information-flow interface is \emph{well-formed} when it has at least one implementation and one admissible environment.
Therefore, all of its relations must be irreflexive. 
A well-formed interface ensures, additionally, that its property holds under any combination of its admissible environments and implementations.

\begin{example}
\label{ex:stateless:interface}
\begin{figure}
\centering
\scalebox{0.50}{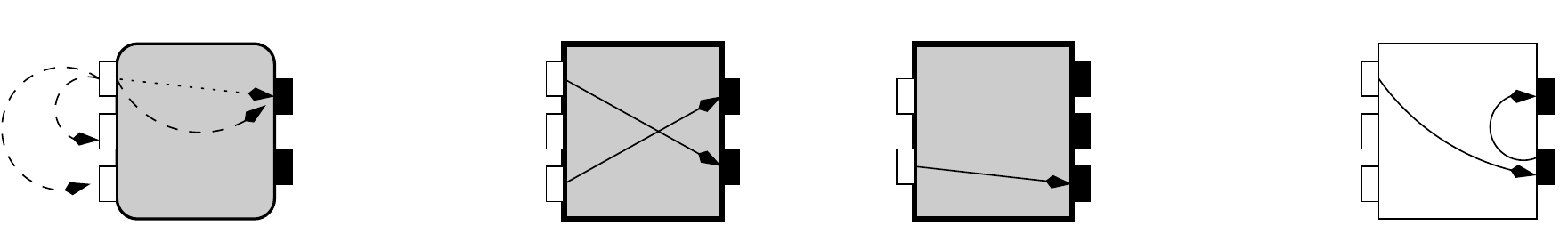 }
\caption{Example of an ill-formed interface \(\ifnotwell\) with an implementation \(\ifcnotwell\), admissible environment \(\ifenotwell\), and a component \(\ifnotwellnotimp\) that does not implement it.	
}
\label{fig:ex:notwelformed}
\end{figure}

In Figure \ref{fig:ex:notwelformed}, interface \(\ifnotwell\) requires as its only \emph{property} and \emph{guarantee}
that no flow is created from the 
input variable \(\key\) to the output variable \(\can\). 
Additionally, the interface \emph{assumes} that the environment does not create information flows from the input variable \(\key\) to both \(\imm\) and \(\ecu\).
In the same figure, we depict components  \(\ifcnotwell\)  and  \(\ifenotwell\), 
which only have flows allowed by \(\ifnotwell\)'s guarantee and assumptions, respectively.
Hence, \(\ifcnotwell\) \emph{implements} interface \(\ifnotwell\), and \(\ifenotwell\) is an \emph{admissible environment} of \(\ifnotwell\).

A flow from \(\key\) to \(\can\) can be created by combining flows from both \(\ifcnotwell\) and \(\ifenotwell\). 
In particular, the information can flow from \(\key\) to 
\(\deb\), then to \(\ecu\) and 
finally to \(\can\). This violates the \(\ifnotwell\) property.
Its assumptions and guarantees
are too weak to ensure that the property is satisfied under any combination 
of its admissible environments and implementations.
Thus,
\(\ifnotwell\) is \emph{not} a well-formed interface.
For the same reason, \(\ifnotwell\) is \emph{not} a well-formed interface.
The aforementioned path can be used to show that \(\ifnotwell\) does not satisfy 
the condition in Definition~\ref{def:i:well-formed}.
\end{example}

We note that the complement of a guarantee is not transitively closed and hence does not define a flow 
relation (the same holds for the complement of an assumption). This \emph{prevents some 
interfaces to have a maximal implementation}, i.e. a component that implements all allowed flows.
Figure~\ref{fig:ex:notwelformed} shows an interface \(\ifnotwell\) and a component \(\ifnotwellnotimp\) that does not implement it.
Both flows from \(\ifnotwellnotimp\)  (full arrows inside the box) are allowed to be implemented on their own, but not together. 
Note that their transitive closure includes the only flow forbidden by the interface---a flow from \(\key\)
to \(\can\). 
Thus, to evaluate possible flow paths between implementations and environments
we need to consider alternated paths between the complement of the guarantees 
and the complement of the assumptions.

To check that an interface is well-formed we need to verify that the flows forbidden by its property can not be created by one of its implementations 
interacting with one of its admissible environments. 
As explained above, we cannot use the transitive closure of all flows allowed by assumptions, \emph{\(\AComp\)}, and guarantees, \emph{\(\GComp\)}
to define all flows created by such interaction. We consider instead all flows created by a path of arbitrary size that alternate between pairs of \(\AComp\) and \(\GComp\).

\begin{definition}
\label{def:i:well-formed}
A \emph{no-flow relation} in \(A \times B\) is irreflexive in \(A \cap B\).
An interface \((\InVars, \OutVars, \Assumption, \Guarantee, \Prop)\) is \emph{well-formed} iff 
\(\Assumption\), \(\Guarantee\) and \(\Prop\) are \emph{no-flow relations}; and
\(((\ID{\AllVars} \cup \AComp) \circ (\GComp \circ \AComp)^* \circ \GComp) \cap \Prop = \emptyset\).
\end{definition}

We show that our definitions capture the intended semantics
for implementations and environments.
In particular, we prove that the interaction between any of implementation 
and admissible environment of a given well-formed interface cannot create flows forbidden by its property.

\begin{proposition}
\label{thm:well-formed}
For all well-formed interfaces \(\InfFlow\), and for all components \(\InfFlowC = (\InVars, \OutVars, \Flows)\)
and \(\InfFlowC_{\Env} = (\OutVars, \InVars, \Env)\):
if \(\InfFlowC \IFImpl \InfFlow\) and \(\InfFlowC_{\Env}  \IFEnv \InfFlow\),
then \((\Flows \cup \Env)^* \cap \Prop = \emptyset\).
\end{proposition}

%


We introduce below the auxiliary notion of derived properties, which are guarantees that hold under any admissible environment.
They are derived by checking for all pairs \((z,y)\) of no-flows in the guarantees that there is no alternated path 
using flows allowed by assumptions and guarantees that create a flow from \(z\) to \(y\).

\begin{definition}
\label{def:derived_p}
We denote by \(\DProp{\Assumption,\Guarantee}\) the set of derived properties from assumptions \(\Assumption\)
and guarantees \(\Guarantee\), defined as follows $\DProp{\Assumption,\Guarantee} = \Guarantee \setminus ( (\ID{\AllVars} \cup \AComp) \circ (\GComp \circ \AComp)^* \circ \GComp)$.
\end{definition}

\subsection{Composition}
\label{subsec:stateless:compos}

We now present how to {\em compose} components and interfaces. We introduce 
a \emph{compatibility} predicate that checks whether the composition of two interfaces is a well-formed interface.
We prove that these two notions support the \emph{incremental design} of systems.
We prove that our composition is both commutative and associative.
The composition of two components is obtained as the reflexive and transitive closure of union of individual component flows.

\begin{definition}
\label{def:vars:between_i}
We define the different type of variables between interfaces 
\(\InfFlow\) and \(\InfFlow'\), as~follows:
\(\OutVars_{\InfFlow, \InfFlow'} = \OutVars \cup \OutVars'\);
 \(\InVars_{\InfFlow, \InfFlow'} = (\InVars \cup \InVars') \setminus \OutVars_{\InfFlow, \InfFlow'}\);
and \(\AllVars_{\InfFlow, \InfFlow'} = \OutVars_{\InfFlow, \InfFlow'} \cup \InVars_{\InfFlow, \InfFlow'}\).
The same definition applies to components \(\InfFlowC\) and \(\InfFlowC'\).
\end{definition}

\begin{definition}
\label{def:c:composition}
The \emph{composition} of two  components \(\InfFlowC\) and \(\InfFlowC'\)
is the component defined as follows:
\(\InfFlowC \CFCompos \InfFlowC' = (
\InVars_{\InfFlowC, \InfFlowC'}, 
\OutVars_{\InfFlowC, \InfFlowC'}, 
(\Flows \cup \Flows')^*). \)
\end{definition}

We present interface composition by defining separately $\Assumption$, $\Guarantee$ and $\Prop$ of the composed interface.

%
The composition of two interfaces should not restrict the set of implementations for each interface in isolation.
We introduce below the notion of \emph{composite flows}, as the set of 
all flows that can be in the composition of any implementation of the interfaces being composed.
We compute them by considering all possible alternated paths between both interfaces.
Then, given that guarantees specify which flows cannot be implemented, 
we define \emph{composite guarantees} as the composite flows' complement.

\begin{definition}
\label{def:compos:flows}
\(\InfFlow\) and \(\InfFlow'\) \emph{composite flows} are defined as follows:
\vspace{-0.2cm}
\begin{center}
\(\GComp_{\InfFlow, \InfFlow'} = (\ID{\AllVars'} \cup \GComp_{\InfFlow}) \circ (\GComp_{\InfFlow'} \circ \GComp_{\InfFlow})^* \circ (\ID{\OutVars} \cup \GComp_{\InfFlow'}).\)
\end{center}
\end{definition}

\begin{definition}
\label{def:compos:g}
\(\InfFlow\) and \(\InfFlow'\) \emph{composite guarantees} are defined as
\(\Guarantee_{\InfFlow, \InfFlow'}\! = \!  (\AllVars_{\InfFlow, \InfFlow'}\! \times\! \OutVars_{\InfFlow, \InfFlow'})\! \setminus\! \GComp_{\InfFlow, \InfFlow'}\).
We denote them by \(\Guarantee_{\InfFlow \IFCompos \InfFlow'}\) and \(\Guarantee_{\InfFlow, \InfFlow'}\), interchangeably.
\end{definition}

%
The assumptions of the composition of multiple interfaces 
is the weakest condition on the environment that 
allows the interfaces being composed to work together. Additionally,
it must support incremental design. So, 
the admissibility of an environment must be independent of the order in which the interfaces are composed.
Naturally, all the assumptions of each interface must be considered during composition.
However, not all of them can be kept as assumptions of the composite.

The composition of two interfaces may change the classification of some input variables of the interfaces being composed to output variables of their composite (c.f.\ Definition \ref{def:vars:between_i}). This happens to variables that are common to both interfaces being composed, which we refer to as \emph{shared variables}.
Interface assumptions that point to shared variables cannot be assumptions of the composite interface, because they no longer point to an input variable.

Our solution is to compute \emph{propagated assumptions} between two interfaces.
Given an assumption pointing to a shared variable, we propagate it to all variables that
can reach the shared one. 
We define it below followed by an illustrative example.

\begin{definition}
\label{def:prop:assumptions}
The set of \emph{assumptions propagated from \(\InfFlow\) to \(\InfFlow'\)} is defined as follows:
\vspace{-0.2cm}
\begin{center}
\(\PAs_{\InfFlow \rightarrow \InfFlow'} =  
\{(z,z') \ |\ \exists s \in X\cap Y' \tSt (z,s) \in \Assumption \tAnd (z',s) \in \GComp_{\InfFlow,\InfFlow'}\}.\)
\end{center}
The set with all propagated assumptions of \(\InfFlow\) and \(\InfFlow'\) is defined as
\(\PAs_{\InfFlow, \InfFlow'} = \PAs_{\InfFlow \rightarrow \InfFlow'} \cup \PAs_{\InfFlow' \rightarrow \InfFlow}\).
\end{definition}

The composite assumptions include all pairs that point to input variables and are
in the union of the assumptions of the interfaces being composed with their propagated assumptions.

\begin{definition}
\label{def:compos:a}
The \emph{composite assumptions} of \(\InfFlow\) and \(\InfFlow'\) is defined as follows:
\begin{center}
\(\Assumption_{\InfFlow,\InfFlow'} = (\Assumption \cup \Assumption' \cup \PAs_{\InfFlow, \InfFlow'}) \cap (\AllVars_{\InfFlow, \InfFlow'}\! \times\! \InVars_{\InfFlow, \InfFlow'})\).
\end{center}
\end{definition}

\begin{example}\hfil

\begin{minipage}{0.43\textwidth}
\scalebox{0.60}{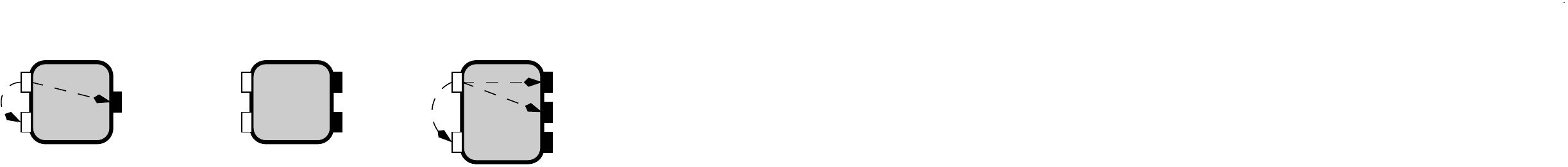 }
\captionof{figure}{Assumptions for composion of \(\InfFlow_{\emph{imm}}\) with \(\InfFlow_{\emph{can}}\).}
\label{fig:ex:compos:imm_can}
\end{minipage}
\hspace{0.1cm}
\begin{minipage}{0.51\textwidth}
In Figure \ref{fig:ex:compos:imm_can} we can see interface
\(\InfFlow_{\emph{imm}}\) which has only one assumption: \(\key\) should not flow to \(\can\).
As \(\can\) is a shared variable of the interfaces being composed, we need to compute the propagated assumptions that it generates.
\end{minipage}
\vspace{0.1cm}

In \(\InfFlow_{\emph{can}}\) both \(\ecu\) and \(\imm\) can flow to \(\can\).
So, they are allowed flows in the composite interface, as well.
Note that, by Definition \ref{def:compos:flows}, 
\(\{(\ecu,\can),(\imm,\can), (\deb, \can)\} \in  \GComp_{\InfFlow_{\emph{imm}}, \InfFlow_{\emph{can}}}\).
Hence, by \(\Assumption_{\InfFlow_{\emph{imm}}} = \{(\key,\can)\} \), we have  \(\PAs_{\InfFlow_{\emph{imm}} \rightarrow \InfFlow_{\emph{can}}} =\{(\key,\ecu), (\key,\imm), (\key,\deb)\}\).
From the derived assumptions only \((\key,\ecu)\) points to an input variable.
Besides, as explained before, the only assumption of \(\InfFlow_{\emph{imm}}\) cannot be kept after composition.
Hence \(\Assumption_{\InfFlow_{\emph{imm}}, \InfFlow_{\emph{can}}} = \{(\key,\ecu)\}\).
\end{example}

%
The properties of the composition needs to preserve the individual properties of each interface being composed.
In addition, it adds all derived properties from the assumptions and guarantees of the composite.
This allows to infer global properties from local specification.

\begin{definition}
\label{def:compos:p}
We denote by \(\Prop_{\InfFlow,\InfFlow'}\) the \emph{composite properties} of \(\InfFlow\) and \(\InfFlow'\), defined as follows:
\(\Prop_{\InfFlow,\InfFlow'} = \Prop \cup \Prop' \cup \DProp{\Assumption_{\InfFlow,\InfFlow'}, \Guarantee_{\InfFlow,\InfFlow'}}\).

\end{definition}

Interfaces composition is defined using the concepts of composite assumptions, guarantees and properties defined before.

\begin{definition}
\label{def:i:composition}
The \emph{composition} of two  interfaces \(\InfFlow\) and \(\InfFlow'\)
is the interface defined as follows:
\(\InfFlow \IFCompos \InfFlow' = (
\InVars_{\InfFlow, \InfFlow'}, 
\OutVars_{\InfFlow, \InfFlow'}, 
\Assumption_{\InfFlow, \InfFlow'}, 
\Guarantee_{\InfFlow, \InfFlow'}, 
\Prop_{\InfFlow, \InfFlow'}). \)
\end{definition}

We allow composition for any two arbitrary interfaces.
However, not all compositions result in well-formed interfaces.
For this reason, we define the notions of two interfaces being \emph{composable} and \emph{compatible}.
Composability imposes the syntactic restriction that both interface's output variables are disjoint.
Compatibility captures the semantic requirement that whenever an interface \(\InfFlow\) provides
inputs to an other interface \(\InfFlow'\), then  \(\InfFlow'\) needs to include guarantees that imply the assumptions of  \(\InfFlow\).

\begin{definition}
\label{def:composable}
\(\InfFlow\) and \(\InfFlow'\) are \emph{composable} iff 
 \(Y \cap Y' = \emptyset\).
\end{definition}

\begin{definition}
\label{def:compatible:g}
Two interfaces \(\InfFlow\) and \(\InfFlow'\) are \emph{compatible}, denoted 
\(\InfFlow \IFCompt \InfFlow'\) iff they are composable and
\(((\Assumption \cup \Assumption') \cap (\AllVars_{\InfFlow, \InfFlow'} \times \OutVars_{\InfFlow, \InfFlow'}))
\subseteq \Guarantee_{\InfFlow,\InfFlow'}.\)
\end{definition}

\begin{example}\hfil

\begin{minipage}{0.3\textwidth}
\scalebox{0.60}{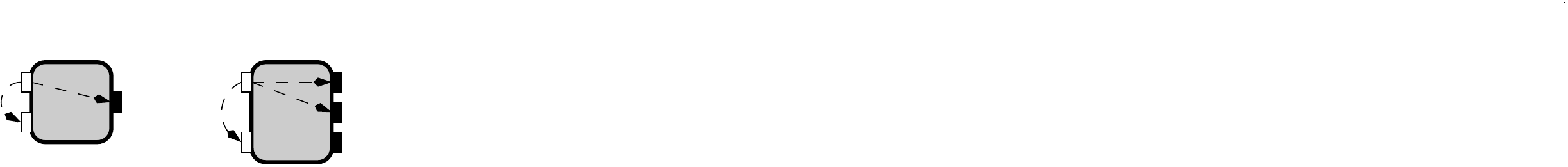 }
\captionof{figure}{Compatibility.}
\label{fig:ex:compos:comp}
\end{minipage}
\hspace{0.1cm}
\begin{minipage}{0.63\textwidth}
In Figure~\ref{fig:ex:compos:comp} we have interface
\(\InfFlow_{\emph{ecu}}\) which has only one guarantee: \(\key\) does not flow to \(\ecu\).
This interface is compatible with the composition of \(\InfFlow_{\emph{can}}\) with \(\InfFlow_{\emph{imm}}\), 
because the only assumption of \(\InfFlow_{\emph{can}} \IFCompos \InfFlow_{\emph{imm}}\) is a guarantee of \(\InfFlow_{\emph{ecu}}\).
\end{minipage}
\end{example}

We prove below that our compatibility relation is sufficient to guarantee that 
the result of composing two interfaces is a well-formed interface.

\begin{theorem}[Composition preserves well-formedness]
\label{thm:comp:wellformed}Commutativity
Let \(\InfFlow\) and \(\InfFlow'\) be well-formed interfaces.
If \(\InfFlow \IFCompt \InfFlow'\), then \(\InfFlow \IFCompos \InfFlow'\) is a well-formed interface.
\end{theorem}
%
%

The composition operator is both commutative and associative.
In particular, it supports incremental design of systems.

\begin{proposition}
\label{prop:commut}
Let $\InfFlow$ and~$\InfFlow'$ be interfaces:
\(\InfFlow \IFCompt \InfFlow' \tIff \InfFlow' \IFCompt \InfFlow\), and
\(\InfFlow \IFCompos \InfFlow' = \InfFlow' \IFCompos \InfFlow\).
\end{proposition}

\begin{theorem}[Incremental design]
\label{thm:incremental}
Let \(\InfFlow\),  \(\InfFlow'\) and  \(\InfFlow''\) be interfaces. 
If  \(\InfFlow \IFCompt \InfFlow'\) and 
\((\InfFlow \IFCompos \InfFlow') \IFCompt \InfFlow''\), then \(\InfFlow' \IFCompt \InfFlow'' \) and 
\(\InfFlow  \IFCompt (\InfFlow' \IFCompos \InfFlow'')\).
\end{theorem}
\begin{proof}[Proof sketch]
We proved first that composite assumptions are associative. 
We assume that \(\InfFlow \IFCompt \InfFlow'\) and 
\((\InfFlow \IFCompos \InfFlow') \IFCompt \InfFlow''\).
The most interesting case is when 
\((z,s)\) is an assumption of \(\InfFlow\) and we need to prove  \((z,s) \in \Guarantee_{\InfFlow, \InfFlow' \IFCompos \InfFlow''}\).
We prove this by assuming towards a contradiction that \((z,s) \in \GComp_{\InfFlow, \InfFlow' \IFCompos \InfFlow''}\). 
We illustrate it in Figure \ref{fig:thm:inc:design}. 

\vspace{-0.2cm}
\begin{minipage}{0.34\textwidth}
\hspace{-2cm}
\scalebox{0.60}{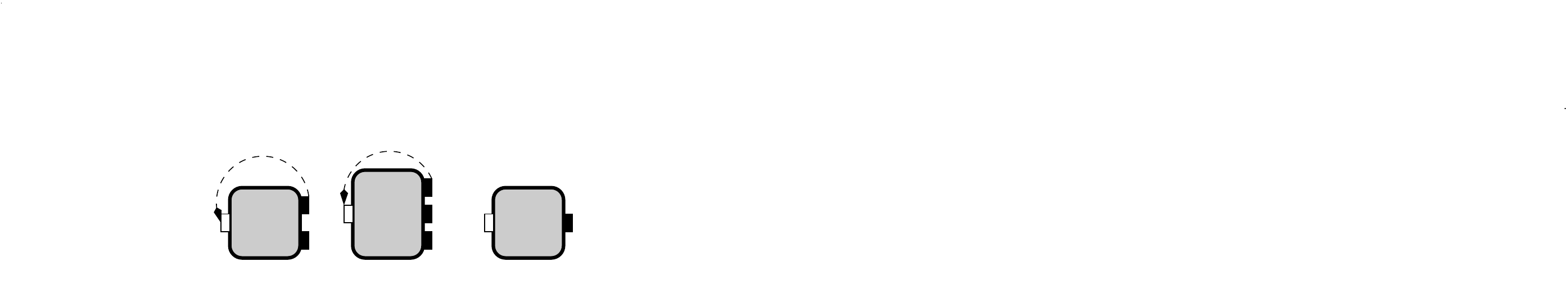 }
\captionof{figure}{Incremental design.}
\label{fig:thm:inc:design}
\end{minipage}
\begin{minipage}{0.62\textwidth}
By composite flows being associative, \((z,s) \in \GComp_{\InfFlow \IFCompos \InfFlow', \InfFlow''}\). 
By \((z,s)\) being  an assumption of \(\InfFlow\) and \((s',s)  \in \GComp_{\InfFlow \IFCompos \InfFlow'}\), then 
\((z,s')\! \in\! \PAs_{\InfFlow \rightarrow \InfFlow'}\)
and, so \({(z,s')\! \in\! \Assumption_{\InfFlow \IFCompos \InfFlow'}}\).
Moreover, \({(z,s')\! \in\! \GComp_{\InfFlow \IFCompos \InfFlow', \InfFlow''}}\), because
\(z\) can flow to \(s'\) when \(\InfFlow \IFCompos \InfFlow'\) is composed with 
\(\InfFlow''\).
This contradicts our initial assumption that \((\InfFlow \IFCompos \InfFlow') \IFCompt \InfFlow''\).
\qedhere
\end{minipage}
\end{proof}

\begin{corollary}
\label{thm:assoc}
If \(\InfFlow \IFCompt \InfFlow'\) and \(\InfFlow \IFCompos \InfFlow' \IFCompt  \InfFlow'\), then
\((\InfFlow \IFCompos \InfFlow') \IFCompos  \InfFlow'' = \InfFlow \IFCompos (\InfFlow' \IFCompos \InfFlow'').\)
\end{corollary}

Finally, we show that flows resulting from the composition of any components that 
implement two given interfaces are allowed by the composition of these interfaces.

\begin{proposition}
\label{prop:impl}
Let \(\InfFlow\) and \(\InfFlow'\) be interfaces,
and \(\InfFlowC = (\InVars, \OutVars, \Flows)\) and
\(\InfFlowC' = (\InVars', \OutVars', \Flows')\) be components.
If \(\InfFlowC \IFImpl \InfFlow\) and \(\InfFlowC' \IFImpl \InfFlow'\), then 
\(\InfFlowC \IFCompos \InfFlowC' \IFImpl \InfFlow \IFCompos \InfFlow'\).
\end{proposition}

\subsection{Refinement and Independent Implementability}
\label{subsec:stateless:ref}

We now define a refinement relation between interfaces. Intuitively, an interface $\InfFlow'$ refines $\InfFlow$ iff 
$\InfFlow'$ admits more environments than $\InfFlow$, while constraining its implementations.

\begin{definition}
\label{def:i:refines}
Interface \(\InfFlow'\) \emph{refines} \(\InfFlow\), written \(\InfFlow' \IFRef \InfFlow\), 
when \({\Assumption' \subseteq \Assumption}\), \(\Guarantee  \subseteq \Guarantee'\) and  \(\Prop \subseteq \Prop'\).
\end{definition}


\begin{proposition}\label{thm:refinement}
Let \(\InfFlow\) and \(\InfFlow'\) be interfaces s.t.\ \(\InfFlow' \IFRef \InfFlow\).
Let \(\InfFlowC = (\InVars, \OutVars, \Flows)\) and
\(\InfFlowC_{\Env} = (\OutVars, \InVars, \Env)\) be components.
\begin{enumerate*}[label=(\alph*)]
\item If \(\InfFlowC \IFImpl \InfFlow'\), then \(\InfFlowC \IFImpl \InfFlow\); and
\item if \(\InfFlowC_{\Env} \IFEnv \InfFlow\), then \(\InfFlowC_{\Env} \IFEnv \InfFlow'\).
\end{enumerate*}
\end{proposition}


Derived properties are helpful to simplify the concept of wellformedness of an interface. 
In fact, an interface is well-formed iff all no-flows in its property are in the set of derived properties from its assumptions and guarantees.

\begin{proposition}
\label{prop:well-defined_and_implements}
An interface \(\InfFlow = (\InVars, \OutVars, \Assumption, \Guarantee, \Prop)\)
is well-formed iff
\(\Prop \subseteq \DProp{\Assumption,\Guarantee}\).
\end{proposition}

\begin{theorem}[Independent implementability]
\label{thm:independent_imp}
For all well-formed interfaces \(\InfFlow_1'\), \(\InfFlow_1\) and \(\InfFlow_2\), 
if \(\InfFlow_1' \IFRef \InfFlow_1\) 
and \(\InfFlow_1 \IFCompt \InfFlow_2\), 
then \(\InfFlow_1' \IFCompt \InfFlow_2\) and \(\InfFlow_1' \IFCompos \InfFlow_2 \IFRef \InfFlow_1 \IFCompos \InfFlow_2\).
\end{theorem}
\begin{proof}[Proof sketch](Full proof in appendix \ref{sec:app:stateless})
The challenging part is to prove that
the refined composite contains all properties of the abstracted one,
i.e.\ \(\Prop_{\InfFlow_1 \IFCompos \InfFlow_2} \subseteq \Prop_{\InfFlow_1' \IFCompos \InfFlow_2}\).
We prove by induction on \(n \in \mathbb{N}\) that:
if \((z,y) \notin (\ID{\AllVars_{\InfFlow_1,\InfFlow_2}} \cup \AComp_{\InfFlow_1,\InfFlow_2}) 
\circ (\GComp_{\InfFlow_1,\InfFlow_2} \circ \AComp_{\InfFlow_1,\InfFlow_2})^n \circ \GComp_{\InfFlow_1,\InfFlow_2}\),
then \((z,y) \notin (\ID{\AllVars_{\InfFlow_1',\InfFlow_2}} \cup \AComp_{\InfFlow_1',\InfFlow_2}) 
\circ (\GComp_{\InfFlow_1',\InfFlow_2} \circ \AComp_{\InfFlow_1',\InfFlow_2})^n \circ \GComp_{\InfFlow_1',\InfFlow_2}.\)
We remark that for all \((z,s) \in \AComp_{\InfFlow_1, \InfFlow_2}\) s.t.\
there exists \((s,y) \in \Guarantee_{\InfFlow_1, \InfFlow_2}\),
then, by \(\InfFlow_1' \IFRef \InfFlow_1\),
it follows that for all \((z,s) \in \AComp_{\InfFlow_1', \InfFlow_2}\)
there exists \((s,y) \in \Guarantee_{\InfFlow_1', \InfFlow_2}\).
Hence if \((z,y) \notin \AComp_{\InfFlow_1, \InfFlow_2} \circ \GComp_{\InfFlow_1, \InfFlow_2}\), then \((z,y) \notin \AComp_{\InfFlow_1', \InfFlow_2} \circ \GComp_{\InfFlow_1', \InfFlow_2}\) as well.
\end{proof}

\subsection{Shared Refinement}

The \emph{shared refinement between two interfaces} is their most general refinement. It allows to describe components that are meant to be used in different environments by providing separate descriptions for each case.
Implementation of the shared refined interface must be able to work with both environments while guaranteeing the same properties.

In order to cope with both environments it may be necessary to allow the environment to create flows that were assumed to not be there by one of the interfaces.
This may be problematic, because a property may need that no-flow to be satisfied.
To overcome this, we introduce \emph{propagated guarantees} below.
The idea is that for each no-flow assumption \((z,x)\) that cannot be in the shared refined interface and for which it is relevant for a no-flow property \((x,y)\) to hold,
we add a no-flow guarantee \((x,y)\) that prevents a flow from the sink of that assumption to the sink of the property.

\begin{definition}
\label{def:prop:guarantees}
Let \(\InfFlow\) and \(\InfFlow'\) be two interfaces.
The \emph{propagated guarantees} \(\InfFlow\) and \(\InfFlow'\) is defined as:
\(\hat{\Guarantee}_{\InfFlow \rightarrow \InfFlow'} = 
\{(x,y) \ |\  (z,x) \in \Assumption,\ (z,x) \notin \Assumption'  \tAnd (z,y) \in \Prop \}.\) The set of all propagated assumptions is defined as 
\(\hat{\Guarantee}_{\InfFlow \rightarrow \InfFlow'} = \hat{\Guarantee}_{\InfFlow \rightarrow \InfFlow'} \cup \hat{\Guarantee}_{\InfFlow' \rightarrow \InfFlow}\).
\end{definition}

\begin{definition}
\label{def:shared:ref}
Let \(\InfFlow\) and \(\InfFlow'\) be two interfaces such that 
\(\InVars = \InVars'\) and \(\OutVars = \OutVars'\).
The \emph{shared refinement of} \(\InfFlow\) and \(\InfFlow'\) is defined as:
\(\InfFlow \IFSharedRef \InfFlow' = (\InVars, \OutVars, \Assumption \cap \Assumption', \Guarantee \cup \Guarantee' \cup \hat{\Guarantee}_{\InfFlow, \InfFlow'}, \Prop \cup \Prop')\).
\end{definition}

In the following, we prove that if interfaces are well-formed then their shared refinement is well-formed, as well. Additionally, we show that the shared refinement is the most abstract well-formed interface that refine both of them. 

\begin{theorem}
\label{thm:sharedref}
Let \(\InfFlow\), \(\InfFlow'\) and \(\InfFlow'\) be well-formed interfaces.
\(\InfFlow \IFSharedRef \InfFlow'\) is a well-formed interface; and
if \(\InfFlow'' \IFRef \InfFlow\)
and \(\InfFlow'' \IFRef \InfFlow'\), then 
\(\InfFlow'' \IFRef \InfFlow \IFSharedRef \InfFlow'\).
\end{theorem}
\begin{proof}[Proof sketch](Full proof in appendix \ref{sec:app:stateless})
The interesting step is to prove that the shared refinement is well-formed.
We assume towards a contradiction that it does not hold. Then, 
we prove the following statement about alternated paths, for all \(n \in \mathbb{N}\):
if 
\((z,y) \in ((\ID{\AllVars} \cup \AComp_{\InfFlow \IFSharedRef \InfFlow'}) \circ (\GComp_{\InfFlow \IFSharedRef \InfFlow'} \circ \AComp_{\InfFlow \IFSharedRef \InfFlow'})^n \circ \GComp_{\InfFlow \IFSharedRef \InfFlow'})\), then 
\((z,y) \in ((\ID{\AllVars} \cup \AComp) \circ (\GComp \circ \AComp)^n \circ \GComp)\).
Note that, if \((z,y) \in \Prop\), \((z,s) \in \Assumption\) and 
\((z,s) \notin \Assumption'\) then, by
definition of propagated guarantees, \((s,y) \in \hat{\Guarantee}_{\InfFlow \rightarrow \InfFlow'}\).
Thus, by definition of shared refinement, it cannot be the case that 
\((s,y) \in \GComp_{\InfFlow \IFSharedRef \InfFlow'}\), and so 
\((z,y) \notin \AComp_{\InfFlow \IFSharedRef \InfFlow'} \circ \GComp_{\InfFlow \IFSharedRef \InfFlow'}\).
\end{proof}

\subsection{Discussion}
\label{sub:stateless:discussion}

\textbf{Properties.}
Without properties in our interfaces (i.e.\ only with assumptions and guarantees)
it is not possible to define an useful algebra to specify structural properties of no-flows that supports both incremental design and independent implementability. 
Consider the top-down design of a system where we start by defining a no-flow in a closed system that we expect to be preserved along the decomposition
of this system into smaller open systems. Then, we would need assumptions or guarantees to specify global no-flows (i.e.\ restrictions that hold under any composition of its implementations with any of its environments). Then, we could not support incremental design. Compatibility criteria for 
interfaces with global no-flows would need to be too strong with intuitive and correct designs considered incompatible. Note that, for any two variables \(x\) and \(y\) there exists an interface that can create a flow between them, as long as \(x\) is a source of some flow and \(y\) is a sink of some other flow. Hence we introduce global requirements as \emph{properties} that enforces a well-formedness criteria on assumptions and guarantees while it allows interfaces to be independently specified and implemented.

\textbf{Completing interfaces.} 
There is no unique way to extend an ill-formed interface into a well-formed one.
Given an alternated path that witnesses a violation of a property,
it is sufficient  to add one of its elements to the interface specification to remove this specific violation. So, there are as many extensions to the ill-formed interface as the elements in that path, and all of them are incomparable.

\textbf{Semantics.} 
In this work we do not adopt any specific semantic characterization of information flows, as we only work with its (abstracted) structural properties.
One possible interpretation of the pairs \((x,y)\) in \emph{no-flow} relations is as \emph{independence predicates}. 
Then,  \(x\) is independent of \(y\)
if the system outputs all combinations of all the possible values of \(x\) and \(y\). 
Formally, if \(\pi\) and \(\pi'\) are observations of different runs of the system at time \(t\), with \(\pi[t](x)\) being the value of \(x\) in \(\pi\) at that time,
then there must exist a third observation \(\pi''\) such that \(\pi[t](x) = \pi''[t](x)\  \wedge\  \pi'[t](y) = \pi''[t](y)\), i.e.\ \(\pi''\) \emph{interleaves} the value off \(x\) in \(\pi\) with \(y\) in \(\pi'\).

%
\section{Stateful Interfaces}
\label{sec:staful}

We extend our theory with stateful components and interfaces. 
These are transition systems in which each state is a stateless component or interface, respectively. Proofs are in appendix~\ref{sec:app:stateful}.

\begin{definition}
\label{def:component:full}
A \emph{stateful information-flow component} \(\IFCFul\) is a tuple 
\((\InVars,\OutVars,\States, \InitState, \Transition, \FlowsFul)\), where:
\(\InVars\) and \(\OutVars\) are disjoint sets of \emph{input}
and \emph{output} variables respectively, with $\AllVars = \InVars \cup \OutVars$ the set of all variables;
\(\States\) is a set of states and \(\InitState \in \States\)  is the initial state;
\(\Transition: \States \rightarrow 2^{\States}\) is a transition relation;
\(\FlowsFul: \States \rightarrow 2^{\AllVars \times \OutVars}\) is a state labeling.
For each state \(\std \in \States\) we require that \(\FlowsFul(\std)\) is a flow relation and we denote by \(\IFCFul(\std) = (\InVars, \OutVars, \FlowsFul(\std))\) the stateless component implied by the labeling of $q$.
\end{definition}

\begin{definition}
\label{def:interface:full}
A \emph{stateful information-flow interface} \(\IFFul\) is a tuple 
\((\InVars,\OutVars,\States, \InitState, \Transition, \AssumptionFul,\GuaranteeFul, \PropFul)\), where:
\(\InVars\) and \(\OutVars\) are disjoint sets of \emph{input}
and \emph{output} variables, respectively;
\(\States\) is a set of states and \(\InitState \in \States\) is the initial state;
\(\Transition: \States \rightarrow 2^{\States}\) is a transition relation;
\(\AssumptionFul: \States \rightarrow 2^{\AllVars \times \InVars}\) is called \emph{assumption};
\(\GuaranteeFul: \States \rightarrow 2^{\AllVars \times \OutVars}\) is called \emph{guarantee}; and
 \(\PropFul: \States \rightarrow 2^{\AllVars \times \OutVars}\) is called \emph{property}.
For each state \(\std \in \States\) we require that \(\AssumptionFul(\std)\), \(\GuaranteeFul(\std)\) and \(\PropFul(\std)\)
are no-flow relations and we denote by \(\IFFul(\std) = (\InVars, \OutVars, \AssumptionFul(\std), \GuaranteeFul(\std), \PropFul(\std))\) the stateless interface implied by the assumption, guarantee and property of $q$.
\end{definition}

In what follows, 
\(\IFFul = (\InVars,\OutVars,\States, \InitState, \Transition, \AssumptionFul,\GuaranteeFul, \PropFul)\)
and 
\(\IFFul' = (\InVars',\OutVars',\States', \InitState', \Transition', \AssumptionFul',\GuaranteeFul', \PropFul')\)
are stateful interfaces, and
\(\IFCFul = (\InVars,\OutVars,\States_{\IFCFul}, \InitState_{\IFCFul}, \Transition_{\IFCFul}, \FlowsFul)\)
and 
\(\IFCFul_{\Env} = (\OutVars, \InVars, \States_{\Env}, \InitState_{\Env}, \Transition_{\Env}, \EnvFul)\) are stateful components.

\begin{definition}
\label{def:well-formed:ful}
A \emph{stateful interface} \(\IFFul\) is well-formed iff \(\IFFul(\InitState)\) is a well-formed stateless interface, and
for all \(\std \in \States\) reachable from \(\InitState\) the stateless interface \(\IFFul(\std)\) is well-formed.
\end{definition}

\subsection{Implements}
\label{subsec:stateful:imp}

A stateful component \(\IFCFul\) implements a stateful interface \(\IFFul\) if there exists a simulation relation 
from \(\IFCFul\) to \(\IFFul\) such that the stateless components in the relation implement
the stateless interfaces they are related to.
Admissible environments are defined analogously.

\begin{definition}
\label{def:implements:full}
A component \(\IFCFul\) \emph{implements} the interface
\(\IFFul\), denoted by \(\IFCFul \IFImpl \IFFul\), iff\
there exists \(\Simul \subseteq \States_\IFCFul \times \States\) s.t.\
\((\InitState_\IFCFul, \InitState) \in \Simul\) and for all 
\((\std_\IFCFul, \std) \in \Simul\):
\begin{enumerate*}[label=(\roman*)]
\item \(\IFCFul(\std_\IFCFul) \IFImpl \IFFul(\std)\); and
 
\item 
if \(\std_\IFCFul' \in \Transition_\IFCFul(\std_\IFCFul)\), then 
there exists a state  \(\std' \in \Transition(\std)\)
s.t.\ \((\std_\IFCFul', \std') \in \Simul\).
\end{enumerate*}
\end{definition}

\begin{definition}
\label{def:environment:full}
A component \(\IFCFul_{\Env}\) 
is an \emph{admissible environment} for the interface
\(\IFFul\), denoted by \(\IFCFul_{\Env} \IFEnv \IFFul\), iff
there exists a relation \(\Simul \subseteq \States \times \States_\Env\) s.t.\
\((\InitState, \InitState_\Env) \in \Simul\) and for all 
\((\std, \std_\Env) \in \Simul\):
\begin{enumerate*}[label=(\roman*)]
\item \(\IFCFul(\std_\Env) \IFEnv \IFFul(\std)\); and
 
\item 
if \(\std' \in \Transition_\IFFul(\std)\), then 
there exists a state  \(\std_\Env' \in \Transition_\Env(\std_\Env)\)
s.t.\ \((\std',\std_\Env') \in \Simul\).
\end{enumerate*}
\end{definition}

As for stateless interfaces, we want to prove that 
interface's properties are satisfied 
after we compose any of its implementations \(\IFCFul\)
with any of its admissible environments \(\IFCFul_{\Env}\).
We prove this below for all  
sates \(\std\) that are common to any pair of relations witnessing
\(\IFCFul \IFImpl \IFFul\) and \(\IFCFul_{\Env} \IFEnv \IFFul\).

\begin{proposition}\label{prop:well_formed:full}
Let \(\IFFul\) be a well-formed interface, and 
\(\IFCFul \IFImpl \IFFul\) and \(\IFCFul_{\Env} \IFEnv \IFFul\). 
For all 
 \(\Simul \subseteq \States_{\IFCFul} \times \States\) 
and
\(\Simul_{\Env} \subseteq \States \times  \States_{\Env}\) 
that witness them, respectively, then:
\begin{enumerate*}[label=(\roman*)]
\item \((\FlowsFul(\InitState_\IFCFul) \cup \EnvFul(\InitState_\Env))^* \cap \PropFul(\InitState) = \emptyset\); and 
\item for all \(\std \in \States\) that are reachable from \(\InitState\), 
if \((\std_\IFCFul, \std) \in \Simul\) and \((\std, \std_\Env) \in \Simul_{\Env}\), 
then \((\FlowsFul(\std_\IFCFul) \cup \EnvFul(\std_\Env))^* \cap \PropFul(\std) = \emptyset.\)
\end{enumerate*}
\end{proposition}

\subsection{Composition}
\label{subsec:stateful:compos}

Composition of two components is defined as their synchronous product.
The composition of two interfaces is defined as their synchronous product, as well.
However, we only keep the states that are defined by the composition of two comptible stateless interfaces.
Recall that we assume that the environment is helpful, so it will respect state's assumptions. 

\begin{definition}
\label{def:compos:c:ful}
Let \(\IFCFul\) and \(\IFCFul'\) be two components.
Their \emph{composition} is defined as the tuple:
\(\IFCFul \CFCompos \IFCFul' = (\InVars_{\IFCFul, \IFCFul'}, \OutVars_{\IFCFul, \IFCFul'},
\States_{\IFCFul, \IFCFul'}, \InitState_{\IFCFul, \IFCFul'}, 
\Transition_{\IFCFul, \IFCFul'}, 
\FlowsFul_{\IFCFul, \IFCFul'})\),
where:
\(\States_{\IFCFul, \IFCFul'} = \States \times \States'\) with 
\(\InitState_{\IFCFul, \IFCFul'} = (\InitState, \InitState')\);
\((\std_2, \std_2') \in \Transition_{\IFCFul, \IFCFul'}(\std_1, \std_1)\) iff 
\(\std_2 \in \Transition(\std_1)\) and \(\std_2' \in \Transition'(\std_1')\);
for all \((\std, \std') \in \States_{\IFCFul, \IFCFul'}:\ 
\FlowsFul_{\IFCFul, \IFCFul'}(\std, \std') = (\FlowsFul(\std) \cup \FlowsFul'(\std'))^*\).
\end{definition}

\begin{definition}
\label{def:i:compositionful}
Let \(\IFFul\) and \(\IFFul'\) be two interfaces.
Their \emph{composition} is defined as the tuple:
\(\IFFul \IFCompos \IFFul' = (\InVars_{\IFFul, \IFFul'}, \OutVars_{\IFFul, \IFFul'},
\States_{\IFFul, \IFFul'}, \InitState_{\IFFul, \IFFul'}, 
\Transition_{\IFFul, \IFFul'}, 
\AssumptionFul_{\IFFul, \IFFul'}, 
\GuaranteeFul_{\IFFul, \IFFul'},
\PropFul_{\IFFul, \IFFul'} ),\)
where: \(\InitState_{\IFFul, \IFFul'} = (\InitState, \InitState')\) and 
\(\States_{\IFFul, \IFFul'} = \{\InitState_{\IFFul, \IFFul'}\} \cup \{(q,q') \ | \ \IFFul(\std) \IFCompt \IFFul'(\std')\}\); 
\((\std_2, \std_2') \in \Transition_{\IFFul, \IFFul'}(\std_1, \std'_1)\) iff 
\(\std_2 \in \Transition(\std_1)\) and \(\std_2' \in \Transition'(\std_1')\);
for all  \((\std, \std') \in \States_{\IFFul, \IFFul'}:\ 
\IFFul_{\IFFul, \IFFul'}(\std, \std') = \IFFul(\std) \otimes \IFFul'(\std')\).
\end{definition}

Two stateful interfaces are compatible if the stateless interfaces defined by their initial states are compatible. Note that this implies that all of the states reachable from the initial state are defined by the composition of two compatible stateless interfaces, as well.

\begin{definition}
\label{def:compatible:g}
Two interfaces \(\IFFul\) and \(\IFFul'\) are \emph{compatible}, denoted by
\(\IFFul \IFCompt \IFFul'\), iff  \(\IFFul(\InitState) \IFCompt \IFFul'(\InitState')\).
\end{definition}

We prove below propositions related to the composition of stateful interfaces that follow from results from the stateless interfaces.

\begin{proposition}
\label{thm:prop:ful}
For all interfaces $\IFFul$ and~$\IFFul'$:
\(\IFFul \IFCompt \IFFul' \tIff \IFFul' \IFCompt \IFFul\).
\end{proposition}

\begin{proposition}[Composition preserves well-formedness]
\label{prop:comp:wellformed:ful}
For all well-formed interfaces \(\IFFul_1\) and \(\IFFul_2\):
If \(\IFFul_1 \IFCompt \IFFul_2\), then \(\IFFul_1 \IFCompos \IFFul_2\) is a well-formed interface.
\end{proposition}

\begin{proposition}[Incremental design]
\label{prop:incremental:ful}
For all interfaces $\mathbb{F}$, $\mathbb{G}$ and $\mathbb{I}$, if  \(\mathbb{F} \IFCompt \mathbb{G}\) and 
\((\mathbb{F} \IFCompos \mathbb{G}) \IFCompt \mathbb{I}\), 
then \(\mathbb{G} \IFCompt \mathbb{I}\) and 
\(\mathbb{F}  \IFCompt (\mathbb{G} \IFCompos \mathbb{I})\).
\end{proposition}

\begin{proposition}
\label{prop:impl:ful}
If \(\IFCFul \IFImpl \IFFul\) and \(\mathbbm{g} \IFImpl \mathbb{G}\), 
then \(\IFCFul \IFCompos \mathbbm{g} \IFImpl \IFFul \IFCompos \mathbb{G}\).
\end{proposition}
\begin{proof}
Let \(\Simul_\IFCFul\) and \(\Simul_\mathbbm{g}\) be the witnesses for 
\(\IFCFul \IFImpl \IFFul\) and 
\(\mathbbm{g} \IFImpl \mathbb{G}\).
Then, the relation 
\({\Simul = \{((\std_\IFCFul, \std_\mathbbm{g}), (\std_\IFFul, \std_\mathbb{G}))
\ |\ \std_\IFFul \in \Simul_\IFCFul(\std_\IFCFul) \tAnd \std_\mathbb{G} \in \Simul_{\mathbbm{g}}(\std_\mathbbm{g})\}}\) witnesses \(\IFCFul \IFCompos \mathbbm{g} \IFImpl \IFFul \IFCompos \mathbb{G}\).
\end{proof}

\subsection{Refinement}
\label{subsec:stateful:ref}

Given an interface, we define transitions parameterized by no-flows on its input variables (i.e.\ with fixed assumptions) or on its output variables 
(i.e.\ with fixed guarantees and properties).

\begin{definition}
\label{def:moves:in:out}
Let \(\IFFul\) be an interface.
\emph{Input transitions} from a given state \(\std \in \States\) are defined as 
\(\InStep(q) = \{ \InStep(q, \Assumption) \ | \ \Assumption \subseteq \AllVars\times \InVars\}\) with \(\InStep(q, \Assumption) = \{q' \in \delta(q) \ |\ \AssumptionFul(q') = \Assumption \}\).
\emph{Output transitions} from a given state \(\std \in \States\) are defined as 
\(\OutStep(q) = \{\OutStep(q, \Guarantee, \Prop) \ | \ \Guarantee \subseteq \AllVars \times \OutVars \tAnd \Prop \subseteq \AllVars \times \OutVars\}\) with \(\OutStep(q, \Guarantee, \Prop) = \{q' \in \delta(q) \ |\   \GuaranteeFul(q') = \Guarantee \tAnd \PropFul(q') = \Prop\}\).
\end{definition}

Interface \(\IFFul_R\) refines \(\IFFul_A\), if all output steps of \(\IFFul_R\)
can be simulated by \(\IFFul_A\), while  all input steps of \(\IFFul_A\)
can be simulated by \(\IFFul_R\).
This corresponds to alternating alternating refinement, as 
introduced in 
\cite{alur1998alternating}.

\begin{definition}
\label{def:ref:full}
Interface \(\IFFul_R\) \emph{refines} \(\IFFul_A\), written \(\IFFul_R \IFRef \IFFul_A\), iff
there exists a relation \(\Simul \subseteq \States_R \times \States_A\) s.t.\
\((\InitState_R, \InitState_A)\in \Simul\)  and
for all \((\std_R, \std_A) \in \Simul\):
\begin{enumerate*}[label=(\roman*)]
\item \(\IFFul_R(\std_R) \IFRef \IFFul_A(\std_A)\);
\item 
for all set of states \(O \in \OutStep_R(\std_R)\),
there exists \({O' \in \OutStep_A(\std_A)}\)~s.t.
for all  set of states  \(I' \in \InStep_A(\std_A)\),
there exists \(I \in \InStep_R(\std_R)\)~s.t.
\((O\cap I) \times(O'\cap I') \subseteq \Simul\).
\end{enumerate*}
\end{definition}

\begin{example}
\begin{figure}
\scalebox{0.45}{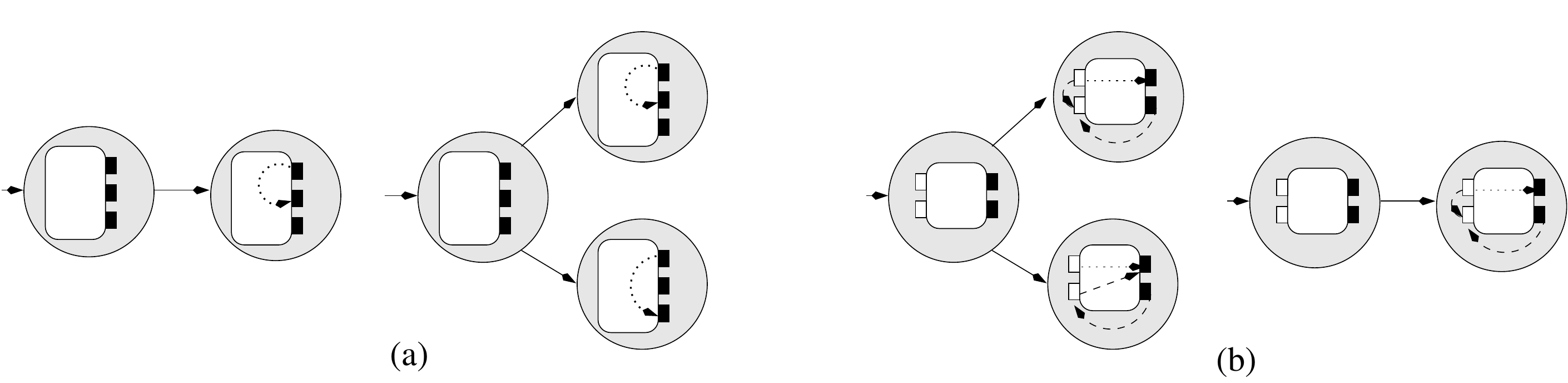 }
\caption{Refined interfaces with witness: 
(a)\(\{(\InitState_1, \InitState_1'), (\std_2, \std_2')\}\); and 
(b) \(\{(\InitState_1, \InitState_1'), (\std_2, \std_2'), (\std_3, \std_2')\}\). }
\label{fig:ref:ful}
\end{figure}
In Figure~\ref{fig:ref:ful} we depict two examples of refined stateful interfaces.

In Figure~\ref{fig:ref:ful}(a) the stateless interface in each state only uses
output ports and it only specifies properties. The initial state of both stateful interfaces 
is the same, so they clearly refine each other. 
As there are no assumptions and guarantees, then, by Definition \ref{def:ref:full},
we need to check that for all successors of the initial state in the refined interface \(\std_s\), 
there exists a successor of the initial state in the abstract interface \(\std'_s\)
such that \(\PropFul_A(\std'_s) \subseteq \PropFul_R(\std_s)\). This holds for the states \((\std_2, \std_2')\).
Hence the relation \(\{(\InitState_1, \InitState_1'), (\std_2, \std_2')\}\) witnesses the refinement. 
Note that the refined interface is obtained by removing a nondeterministic choice on the transition function.

The witness relation for the refinement depicted in Figure \ref{fig:ref:ful}(b) is
\(\{(\InitState_1, \InitState_1'), (\std_2, \std_2'), (\std_3, \std_2')\}\).
The initial states are the same, so the condition (i) in Definition \ref{def:ref:full} is trivially satisfied.
The refined interface has two distinct output transitions from the initial state \(\InitState_1\).
It can either go to state \(\std_2\) by choosing the set of guarantees and proposition with only one element \((x,y)\)
or it can transition to state \(\std_3\) by committing to the set of no-flows \(\{(x,y), (x',y)\}\) for the guarantees and 
\(\{(x,y)\}\) as property. From the initial state of the abstract interface, there exists only one input transition possible, 
to assume that \(x\) does not flow to \(x'\) and \(y'\) does not flow to \(x\).
The following holds for both states accessible from the initial state in the refined interface:
\(\AssumptionFul_R(\std_2) \subseteq \AssumptionFul_A(\std'_2)\) and  \(\AssumptionFul_R(\std_3) \subseteq \AssumptionFul_A(\std'_2)\).
The refined interface specifies an alternative transition from the initial state (represented by state \(\std_3\)) 
that allows more environments while restricting the implementation and preserving the property.
\end{example}

\begin{theorem}\label{thm:refinement:full}
Let \(\IFFul' \IFRef \IFFul\).
\begin{enumerate*}[label=(\alph*)]
\item If \(\IFCFul \IFImpl \IFFul'\), then \(\IFCFul \IFImpl \IFFul\).

\item If \(\IFCFul_{\Env} \IFEnv \IFFul\), then \(\IFCFul_{\Env} \IFEnv \IFFul'\).
\end{enumerate*}
\end{theorem}
\begin{proof}
Let \(\Simul_{\IFRef} \subseteq \States_1 \times \States_2\) be the relation witnessing
\(\Simul_{\IFRef} \subseteq \States_1 \times \States_2\).
\begin{enumerate*}[label=(\alph*)]
\item 
Let \(\Simul_{\IFImpl} \subseteq \States_\IFCFul \times \States_1\) be the relation witnessing \(\IFCFul \IFImpl \IFFul_1\). 
Then, \(\Simul =  \Simul_{\IFImpl} \circ \Simul_{\IFRef}\) witnesses \(\IFCFul \IFImpl \IFFul\).
\item Let  \(\Simul_{\IFImpl} \subseteq \States_2 \times \States_\Env\) be the relation witnessing \(\IFCFul_{\Env} \IFEnv \IFFul\). Then, \(\Simul = \Simul_{\IFRef} \circ  \Simul_{\IFImpl}\) witnesses  \(\IFCFul_{\Env} \IFEnv \IFFul'\).
\end{enumerate*}
\end{proof}

\begin{theorem}[Independent implementability]
\label{thm:independent_imp:ful}
For all well-formed interfaces \(\IFFul_1'\), \(\IFFul_1\) and \(\IFFul_2\), 
if \(\IFFul_1' \IFRef \IFFul_1\) and \(\IFFul_1 \IFCompt \IFFul_2\), 
then \(\IFFul_1' \IFCompt \IFFul_2\) and \(\IFFul_1' \IFCompos \IFFul_2 \IFRef \IFFul_1 \IFCompos \IFFul_2\).
\end{theorem}
\begin{proof}
Let
\(\Simul_\IFRef \subseteq \States_1' \times \States_1\) be the relation that witnesses 
\(\IFFul_1' \IFRef \IFFul_1\). Then, the relation 
\(\Simul = \{((\std_{\IFFul_1'}, \std_{\IFFul_2}),(\std_{\IFFul_1}, \std_{\IFFul_2})) \ |\  (\std_{\IFFul_1'},\std_{\IFFul_1})\! \in\! \Simul_\IFRef  \tAnd \IFFul_1(\std_{\IFFul_1}) \IFCompt  \IFFul_2(\std_{\IFFul_2}) \}\)witnesses \(\IFFul_1'\! \IFCompos\! \IFFul_2 \IFRef \IFFul_1\! \IFCompos\! \IFFul_2\).
\end{proof}

\subsection{Trace Semantics}
\label{sec:exp}

In this section, we explore three set semantics for a particular stateful interface.
For simplicity, we restrict our attention to boolean variables.
Trace properties \(\setTraces\) are defined as sets of infinite traces over the set of propositional variables, that is, \(\setTraces \subseteq \alphabet^\omega\). The
set \(\allSetTraces = 2^{\alphabet^\omega}\) defines the set of all trace properties.
A system's implementation is characterized by a set of its executions \(\system\) represented as a set of traces.
An interface characterizes a set of implementations,thus it defines sets of sets of traces, \(\setsetTraces \subseteq 2^{\allSetTraces}\).

\begin{wrapfigure}{l}{0.34\textwidth}
\centering
\scalebox{0.45}{ 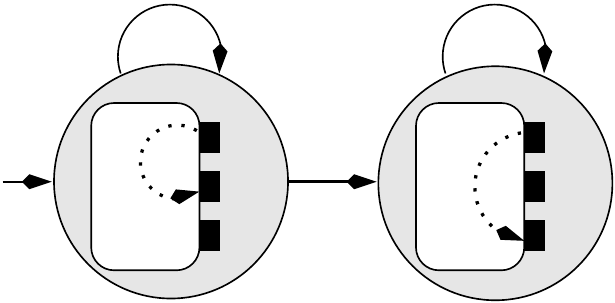 }
\caption{Stateful interface \(\IFFul_{\mathcal{U}}\).}
\label{fig:comparison:hyperltl}
\end{wrapfigure}Interface \(\IFFul_{\mathcal{U}}\), in Figure~\ref{fig:comparison:hyperltl},
specifies that \(x\) does not flow to \(y\) until \(x\) does not flow to \(z\) and it stays in that state. 
In this example, no-flows are interpreted as introduced in subsection \ref{sub:stateless:discussion}. 
We use the abbreviation \(x_1[t] \noflow_{3} y_{2}[t] \equiv x \neq y\ \wedge \ \pi_1[t](x) = \pi_3[t](x)\  \wedge\  \pi_2[t](y) = \pi_3[t](y)\),
where  \(\pi[t](x)\) denotes  the \emph{valuation of variable \(x\) at time \(t\) in \(\pi\)}.

To capture our intended interpretation for the no-flow predicate
\(\pi\) and \(\pi'\) need to be universally quantified while \(\pi''\) needs to be existentially quantified.
The quantification over time admits different interpretations.
We can, for instance, force all traces of an implementation to be synchronous concerning the states in that interface. Then, state transitions are satisfied at the same time in all traces of an implementation. This is the \emph{strong {no-flow}} interpretation over time.
Alternatively, we can require that for any arbitrary observation of a system execution all valuations witnessing a no-flow requirement w.r.t.\ other observations need to be in the same time point.
In this case, state transitions must be synchronous with respect to individual traces. 
This is the \emph{structure-aware no-flow} interpretation over time.
Finally, we may only require that the witnesses are presented in some time which may differ for each triple of traces. 
We refer to this interpretation as \emph{unstructured no-flow}.

\begin{definition}
\label{def:interface:semantics}
We define that \emph{\(x_1 \noflow_3 y_2\) until it always holds that  \(x_1 \noflow_3 z_2\)} as:
\vspace{-0.3cm}
\begin{align*}
\phi_{\mathcal{U}}(t, x_1, y_{2}, z_{2}, \pi_3) \equiv   \ 0 < t\ \wedge\ 
(\forall  0 \leq t' < t\  x_1[t'] \noflow_3 y_{2}[t']) \ \wedge 
(\forall t \leq t''\ x_1[t''] \noflow_3 z_{2}[t'']).
\end{align*}
\vspace{-0.3cm}
There are three plausible set semantics for \(\IFFul_{\mathcal{U}}\), which differ in the position of \(\exists t\):
\begin{description}
\item[Strong no-flow]
\(\strongS{\IFFul_{\mathcal{U}}} = 
\{\setTraces \ |\ \exists t\ \forall \pi_1\! \in\! \setTraces\ \forall \pi_2\! \in\! \setTraces\ \exists \pi_3\! \in\! \setTraces: \phi_{\mathcal{U}}(t, x_1, y_{2}, z_{2}, \pi_3)\}\);
\item[Structure-aware no-flow]
\(\awareS{\IFFul_{\mathcal{U}}} = 
\{\setTraces \ |\ \forall \pi_1\! \in\! \setTraces\ \exists t\  \forall \pi_2\! \in\! \setTraces\ \exists \pi_3\! \in\! \setTraces: \phi_{\mathcal{U}}(t, x_1, y_{2}, z_{2}, \pi_3)\}\);
\item[Unstructured no-flow]
\(\unstructS{\IFFul_{\mathcal{U}}} = 
\{\setTraces \ |\ \forall \pi_1\! \in\! \setTraces\ \forall \pi_2\! \in\! \setTraces\ \exists \pi_3\! \in\! \setTraces\ \exists t: \phi_{\mathcal{U}}(t, x_1, y_{2}, z_{2}, \pi_3)\}\).
\end{description}
\end{definition}

Clearly, \(\strongS{\IFFul_{\mathcal{U}}} \subseteq \awareS{\IFFul_{\mathcal{U}}} \subseteq  \unstructS{\IFFul_{\mathcal{U}}}\). The other direction does not hold. Hence 
these set semantics are not equivalent.

\begin{theorem}\label{thm:semantics:comparison}
\(\unstructS{\IFFul_{\mathcal{U}}} \not \subseteq \awareS{\IFFul_{\mathcal{U}}} \not \subseteq \strongS{\IFFul_{\mathcal{U}}}.\)
\end{theorem}
\begin{proof}
\(\setTraces_u \in \unstructS{\IFFul_{\mathcal{U}}}\) but \(\setTraces_u \notin \awareS{\IFFul_{\mathcal{U}}}\). And, 
\(\setTraces_a \in \awareS{\IFFul_{\mathcal{U}}}\) but \(\setTraces_a \notin \strongS{\IFFul_{\mathcal{U}}}\).

\begin{minipage}{0.4\textwidth}
\begin{tabular}{c|ccc|ccc|ccc|}
&\multicolumn{3}{c}{t=0}&\multicolumn{3}{c}{t=1}&\multicolumn{3}{c|}{1 < t}\\
&x & y & z & x & y & z & x & y & z\\ 
\hline 
\(\pi_1\)&0 & 0 & 0 & 1 & 1 & 0 & 0 & 0 & 0 \\
\(\pi_2\)&0 & 0 & 0 & 0 & 0 & 0 & 0 & 0 & 0 \\
\(\pi_3\)&0 & 0 & 0 & 0 & 1 & 1 & 0 & 0 & 0 \\
\multicolumn{9}{c}{}
\end{tabular} 
\label{tab:ex:expressiv:model:1}
\captionof{table}{Set of traces \(\setTraces_u\).}
\end{minipage}
\hspace{0.7cm}
\begin{minipage}{0.4\textwidth}
\begin{tabular}{c|ccc|ccc|ccc|}
&\multicolumn{3}{c}{t=0}&\multicolumn{3}{c}{t=1}&\multicolumn{3}{c|}{1 < t}\\
&x & y & z & x & y & z & x & y & z\\ 
\hline 
\(\pi_1\)&0 & 0 & 0 & 0 & 0 & 1 & 0 & 0 & 0 \\
\(\pi_2\)&0 & 0 & 0 & 1 & 1 & 0 & 0 & 0 & 0 \\
\(\pi_3\)&0 & 0 & 0 & 1 & 0 & 0 & 0 & 0 & 0 \\
\(\pi_4\)&0 & 0 & 0 & 0 & 0 & 0 & 0 & 0 & 0 \\
\end{tabular} 
\label{tab:ex:expressiv:model:2}
\captionof{table}{Set of traces \(\setTraces_a\).}
\end{minipage}
\end{proof}

Linear temporal logic~\cite{Pnueli77} (LTL) cannot express the properties introduced in Definition~\ref{def:interface:semantics}~\cite{ClarksonS10,ClarksonFKMRS14}.
LTL extended with knowledge (linear time epistemic logic~\cite{bozzelli2015unifying}) 
under the synchronous perfect recall semantics
can express the strong no-flow interpretation of \(\IFFul_{\mathcal{U}}\) \cite{bozzelli2015unifying}.
The unstructured no-flow semantics can be specified in HyperLTL~\cite{ClarksonFKMRS14}.
HyperLTL extends LTL by allowing quantification over traces, which occur at the beginning of the formula. Epistemic temporal logic and HyperLTL 
have incomparable expressive power~\cite{bozzelli2015unifying}. To the best of our knowledge, there is no temporal formalism that supports the structure-aware semantics.

\subsection{Discussion}

The composition operation on stateful information-flow interfaces can be generalized to distinguish between compatible and incompatible transitions of interfaces when they are composed.
Usually this is done by labeling transitions with letters from an alphabet, so that only transitions with the same letter can be synchronized.
While necessary for practical modeling, we omit this technical generalization to allow the reader to focus on the novelties of our formalism, which are the information-flow constraints (environment assumptions, implementation guarantees, and global properties) at each state of an interface.   

%

%

\section{Related Work}
\label{sec:related}

Interface theories belong to the broader area of contract-based 
design~\cite{BenvenisteCNPRR18,BenvenisteCFMPS07,BauerDHLLNW12,DammHJPS11,quinton2008}.  The use of contracts to design systems was popularized by 
Bertrand Meyer~\cite{meyer1992applying}, following the earlier ideas introduced by Floyd 
and Hoare \cite{Floyd1967Flowcharts,Hoare69}. 
Hoare 
logic allows to reason formally about refinement and composition for sequential 
imperative programs.  Other language-based techniques have been proved useful to verify and 
enforce information flow policies~\cite{sabelfeld2003language}.  Examples range from type 
systems~\cite{focardi2011types} to program analysis using program-dependency graphs 
(PDGs)~\cite{hammer2009flow,tooljoana2013atps}. These techniques are tailored for specific implementation languages. 
In our approach we are 
we aim at working with composition and refinement notions that are independent of the language adopted for the implementations.
Therefore, language oriented techniques are 
not directly related to our work.

Interface automata~\cite{de2001interfaceaut} (IA), is a stateful interface 
language designed to capture interfaces' input/output (I/O) temporal aspect.  
IA allows the specification of \emph{input} 
and \emph{output} actions through which a component can
interact with the environment, while  \emph{hidden} actions represent 
 the internal transitions that cannot be observed by the 
environment. This formalism has been later further enriched to 
include also extra functional requirements such 
as resource~\cite{ChakrabartiAHS03}, timing~\cite{AlfaroHS02,DavidLLNW10} and 
 security~\cite{LEE2010107} requirements.
 
The work in~\cite{LEE2010107} presents a variant 
of IA called  \emph{Interface for structure and security} (ISS) 
that enables the specification of two different kind of actions:
one type for actions related to low confidential information 
that are accessible to all users and another type related to high 
confidential information that are accessible to users with appropriate rights.  
In this setting the authors present a bisimulation based 
notion of non-interference (BNNI) that checks whether the 
system behaves in the same way when high actions
are performed or when they are considered hidden actions.
In a recent paper~\cite{5750424}, the same authors notice ISS and BNNI 
may fail to detect an information leakage and they propose 
an alternative refinement-based notion of non-interference.
Our approach is orthogonal to IA and their extensions: we do 
not characterise the type of actions of each component, 
but only their input/output ports  defining 
explicitly the information flow relations among them.

Closer to our approach is~\cite{de2001interface}, which 
defines theories for both stateless and deterministic stateful 
interfaces where the assumptions on the environment are
satisfiable predicates over input variables and the guarantees 
are predicates over output variables. These theories support 
both stepwise-refinement and independent implementation of components
and they have proven to be very successful to reason compositionally 
about trace properties~\cite{BenvenisteCNPRR18}. We are 
not aware of extensions that provide explicit support for security
policies such as information flow.  Interface theories require input and output ports to be disjoint, 
and so it cannot express properties that relate them. On the 
contrary, our information flow interfaces, to ensure compositionality, require
a third relation ({\em property}) that specifies forbidden information flows for which 
the component and the environment must share responsibility to enforce them.

Our approach took inspiration from the work on \emph{relational interfaces} (RIs)
in~\cite{tripakis2011theory}.  RIs specify the legal inputs that the environment is allowed 
to provide to the component and for each the legal input, what are the legal outputs that 
the component can generate when provided with that input. 
A contract in RIs is  expressed
 as first-order logic (FOL) formulas that are evaluated over individual traces. 
 Hence, RIs can only relate input and output values in a trace, and not across 
 multiple traces.  Our formalism can instead 
 specify information flow requirements that are related to sets of systems' executions.


Temporal logics~\cite{Pnueli77}, like LTL or CTL* are used to specify trace properties of reactive systems. 
HyperLTL and HyperCTL*~\cite{ClarksonFKMRS14} extend temporal logics by introducing quantifiers 
over path variables. They allow relating multiple executions and expressing for instance
information-flow security properties~\cite{ClarksonS10,ClarksonFKMRS14}.
Epistemic temporal logics~\cite{bozzelli2015unifying} provide the knowledge connective with an 
implicit quantification over traces.
All these extensions reason about closed systems while our approach allows compositional reasoning about open systems.

%

\section{Conclusion}
\label{sec:conclusion}

We propose a novel interface theory to specify information flow properties. 
Our framework includes both \emph{stateless} and \emph{stateful} interfaces
 and supports both incremental design and independent 
implementability.  We also provide three plausible trace semantics for the class
of stateful interfaces with boolean variables.  We show that, while two of these semantics 
correspond to temporal logics for specifying information-flow properties, for the third (\emph{structure-aware no-flow}),
we do not know of any temporal formalism supporting it.

As future work, we will explore how to extend our theory with sets of \emph{must-flows}, i.e.\ support for 
modal specifications \cite{raclet2011modal}. This will enable, for example, to specify flows that a state \(\std\) 
must implement so that the system can transition to a different state, which is useful to specify declassification of information.
Another direction is to investigate   the expressiveness of the 
\emph{structure-aware no-flow} trace semantics, characterizing further the 
class of hyperproperties \cite{ClarksonS10} that can be expressed in our formalism.

\bibliography{flow_policies}

\begin{thebibliography}{10}

\bibitem{alur1998alternating}
Rajeev Alur, Thomas~A Henzinger, Orna Kupferman, and Moshe~Y Vardi.
\newblock Alternating refinement relations.
\newblock In {\em International Conference on Concurrency Theory}, pages
  163--178. Springer, 1998.

\bibitem{BauerDHLLNW12}
Sebastian~S. Bauer, Alexandre David, Rolf Hennicker, Kim~Guldstrand Larsen,
  Axel Legay, Ulrik Nyman, and Andrzej Wasowski.
\newblock Moving from specifications to contracts in component-based design.
\newblock In {\em Proc. of {FASE} 2012: the 15th International Conference on
  Fundamental Approaches to Software Engineering}, volume 7212 of {\em Lecture
  Notes in Computer Science}, pages 43--58. Springer, 2012.
\newblock \href {http://dx.doi.org/10.1007/978-3-642-28872-2}
  {\path{doi:10.1007/978-3-642-28872-2}}.

\bibitem{BenvenisteCFMPS07}
Albert Benveniste, Beno{\^{\i}}t Caillaud, Alberto Ferrari, Leonardo Mangeruca,
  Roberto Passerone, and Christos Sofronis.
\newblock Multiple viewpoint contract-based specification and design.
\newblock In {\em Proc. of {FMCO} 2007: the 6th International Symposium on
  Formal Methods for Components and Objects}, volume 5382 of {\em Lecture Notes
  in Computer Science}, pages 200--225. Springer, 2008.
\newblock \href {http://dx.doi.org/10.1007/978-3-540-92188-2\_9}
  {\path{doi:10.1007/978-3-540-92188-2\_9}}.

\bibitem{BenvenisteCNPRR18}
Albert Benveniste, Beno{\^{\i}}t Caillaud, Dejan Nickovic, Roberto Passerone,
  Jean{-}Baptiste Raclet, Philipp Reinkemeier, Alberto~L.
  Sangiovanni{-}Vincentelli, Werner Damm, Thomas~A. Henzinger, and Kim~G.
  Larsen.
\newblock Contracts for system design.
\newblock {\em Foundations and Trends in Electronic Design Automation},
  12(2-3):124--400, 2018.
\newblock \href {http://dx.doi.org/10.1561/1000000053}
  {\path{doi:10.1561/1000000053}}.

\bibitem{bozzelli2015unifying}
Laura Bozzelli, Bastien Maubert, and Sophie Pinchinat.
\newblock Unifying hyper and epistemic temporal logics.
\newblock In {\em International Conference on Foundations of Software Science
  and Computation Structures}, pages 167--182. Springer, 2015.

\bibitem{ChakrabartiAHS03}
Arindam Chakrabarti, Luca de~Alfaro, Thomas~A. Henzinger, and Mari{\"{e}}lle
  Stoelinga.
\newblock Resource interfaces.
\newblock In {\em Proc. of {EMSOFT} 2003: the Third International Conference on
  Embedded Software}, volume 2855 of {\em LNCS}, pages 117--133. Springer,
  2003.
\newblock \href {http://dx.doi.org/10.1007/978-3-540-45212-6\_9}
  {\path{doi:10.1007/978-3-540-45212-6\_9}}.

\bibitem{ClarksonFKMRS14}
Michael~R. Clarkson, Bernd Finkbeiner, Masoud Koleini, Kristopher~K. Micinski,
  Markus~N. Rabe, and C{\'{e}}sar S{\'{a}}nchez.
\newblock Temporal logics for hyperproperties.
\newblock In {\em Proc. of {POST} 2014: the Third International Conference on
  Principles of Security and Trust}, volume 8414 of {\em Lecture Notes in
  Computer Science}, pages 265--284. Springer, 2014.
\newblock \href {http://dx.doi.org/10.1007/978-3-642-54792-8}
  {\path{doi:10.1007/978-3-642-54792-8}}.

\bibitem{ClarksonS10}
Michael~R. Clarkson and Fred~B. Schneider.
\newblock Hyperproperties.
\newblock {\em Journal of Computer Security}, 18(6):1157--1210, 2010.
\newblock \href {http://dx.doi.org/10.3233/JCS-2009-0393}
  {\path{doi:10.3233/JCS-2009-0393}}.

\bibitem{CoenenFST19}
Norine Coenen, Bernd Finkbeiner, C{\'{e}}sar S{\'{a}}nchez, and Leander
  Tentrup.
\newblock Verifying hyperliveness.
\newblock In {\em Proc. of {CAV} 2019: the 31st International Conference on
  Computer Aided Verification}, volume 11561 of {\em Lecture Notes in Computer
  Science}, pages 121--139. Springer, 2019.
\newblock \href {http://dx.doi.org/10.1007/978-3-030-25540-4}
  {\path{doi:10.1007/978-3-030-25540-4}}.

\bibitem{DammHJPS11}
Werner Damm, Hardi Hungar, Bernhard Josko, Thomas Peikenkamp, and Ingo
  Stierand.
\newblock Using contract-based component specifications for virtual integration
  testing and architecture design.
\newblock In {\em Proc. of {DATE} 2011: Design, Automation and Test in Europe},
  pages 1023--1028. {IEEE}, 2011.
\newblock \href {http://dx.doi.org/10.1109/DATE.2011.5763167}
  {\path{doi:10.1109/DATE.2011.5763167}}.

\bibitem{DavidLLNW10}
Alexandre David, Kim~G. Larsen, Axel Legay, Ulrik Nyman, and Andrzej Wasowski.
\newblock Timed {I/O} automata: a complete specification theory for real-time
  systems.
\newblock In {\em Proc. of {HSCC} 2010: the 13th {ACM} International Conference
  on Hybrid Systems: Computation and Control}, pages 91--100. {ACM}, 2010.

\bibitem{de2001interfaceaut}
Luca De~Alfaro and Thomas~A. Henzinger.
\newblock Interface automata.
\newblock In {\em Proc. of the 8th European Software Engineering Conference
  held jointly with 9th {ACM} {SIGSOFT} International Symposium on Foundations
  of Software Engineering 2001}, pages 109--120. ACM, 2001.

\bibitem{de2001interface}
Luca De~Alfaro and Thomas~A. Henzinger.
\newblock Interface theories for component-based design.
\newblock In {\em International Workshop on Embedded Software}, pages 148--165.
  Springer, 2001.

\bibitem{dealfaro2005}
Luca de~Alfaro and Thomas~A. Henzinger.
\newblock Interface-based design.
\newblock In Manfred Broy, Johannes Gr{\"u}nbauer, David Harel, and Tony Hoare,
  editors, {\em Engineering Theories of Software Intensive Systems}, pages
  83--104, Dordrecht, 2005. Springer Netherlands.

\bibitem{AlfaroHS02}
Luca de~Alfaro, Thomas~A. Henzinger, and Mari{\"{e}}lle Stoelinga.
\newblock Timed interfaces.
\newblock In {\em Proc. of {EMSOFT} 2002: the Second International Conference
  on Embedded Software}, volume 2491 of {\em LNCS}, pages 108--122. Springer,
  2002.
\newblock \href {http://dx.doi.org/10.1007/3-540-45828-X}
  {\path{doi:10.1007/3-540-45828-X}}.

\bibitem{fagin2003reasoning}
Ronald Fagin, Yoram Moses, Joseph~Y Halpern, and Moshe~Y Vardi.
\newblock {\em Reasoning about knowledge}.
\newblock MIT press, 2003.

\bibitem{FinkbeinerHT18}
Bernd Finkbeiner, Christopher Hahn, and Hazem Torfah.
\newblock Model checking quantitative hyperproperties.
\newblock In {\em Proc. of {CAV} 2018: the 30th International Conference on
  Computer Aided Verification}, volume 10981 of {\em Lecture Notes in Computer
  Science}, pages 144--163. Springer, 2018.
\newblock \href {http://dx.doi.org/10.1007/978-3-319-96145-3}
  {\path{doi:10.1007/978-3-319-96145-3}}.

\bibitem{Floyd1967Flowcharts}
Robert~W. Floyd.
\newblock Assigning meanings to programs.
\newblock {\em Proceedings of Symposium on Applied Mathematics}, 19:19--32,
  1967.

\bibitem{focardi2011types}
Riccardo Focardi and Matteo Maffei.
\newblock Types for security protocols.
\newblock {\em Formal Models and Techniques for Analyzing Security Protocols.
  Cryptology and Information Security Series}, 5:143--181, 2011.

\bibitem{tooljoana2013atps}
J{\"u}rgen Graf, Martin Hecker, and Martin Mohr.
\newblock Using {JOANA} for information flow control in {Java} programs - a
  practical guide.
\newblock In {\em Proceedings of the 6th Working Conference on Programming
  Languages (ATPS'13)}, Lecture Notes in Informatics (LNI) 215, pages 123--138.
  Springer Berlin / Heidelberg, February 2013.

\bibitem{hammer2009flow}
Christian Hammer and Gregor Snelting.
\newblock Flow-sensitive, context-sensitive, and object-sensitive information
  flow control based on program dependence graphs.
\newblock {\em International Journal of Information Security}, 8(6):399--422,
  2009.

\bibitem{Hoare69}
C.~A.~R. Hoare.
\newblock An axiomatic basis for computer programming.
\newblock {\em Commun. ACM}, 12(10):576--580, October 1969.
\newblock URL: \url{http://doi.acm.org/10.1145/363235.363259}, \href
  {http://dx.doi.org/10.1145/363235.363259} {\path{doi:10.1145/363235.363259}}.

\bibitem{5750424}
M.~{Lee} and P.~R. {D'Argenio}.
\newblock A refinement based notion of non-interference for interface automata:
  Compositionality, decidability and synthesis.
\newblock In {\em 2010 XXIX International Conference of the Chilean Computer
  Science Society}, pages 280--289, Nov 2010.
\newblock \href {http://dx.doi.org/10.1109/SCCC.2010.14}
  {\path{doi:10.1109/SCCC.2010.14}}.

\bibitem{LEE2010107}
Matias Lee and Pedro~R. D'Argenio.
\newblock Describing secure interfaces with interface automata.
\newblock In {\em Proceedings of the 7th International Workshop on Formal
  Engineering approaches to Software Components and Architectures (FESCA
  2010)}, volume 264(1) of {\em Electronic Notes in Theoretical Computer
  Science}, pages 107 -- 123, 2010.
\newblock \href {http://dx.doi.org/https://doi.org/10.1016/j.entcs.2010.07.008}
  {\path{doi:https://doi.org/10.1016/j.entcs.2010.07.008}}.

\bibitem{Mantel02}
Heiko Mantel.
\newblock On the composition of secure systems.
\newblock In {\em Proc. of 2002 {IEEE} Symposium on Security and Privacy},
  pages 88--101. {IEEE} Computer Society, 2002.
\newblock \href {http://dx.doi.org/10.1109/SECPRI.2002.1004364}
  {\path{doi:10.1109/SECPRI.2002.1004364}}.

\bibitem{meyer1992applying}
Bertrand Meyer.
\newblock Applying `design by contract'.
\newblock {\em Computer}, 25(10):40--51, 1992.

\bibitem{Pnueli77}
Amir Pnueli.
\newblock The temporal logic of programs.
\newblock In {\em Proc. of FOCS77: the 18th Annual Symposium on Foundations of
  Computer Science}, pages 46--57. {IEEE} Computer Society, 1977.
\newblock \href {http://dx.doi.org/10.1109/SFCS.1977.32}
  {\path{doi:10.1109/SFCS.1977.32}}.

\bibitem{quinton2008}
S.~Quinton and S.~Graf.
\newblock Contract-based verification of hierarchical systems of components.
\newblock In {\em 2008 Sixth IEEE International Conference on Software
  Engineering and Formal Methods}, pages 377--381. {IEEE} Computer Society, Nov
  2008.
\newblock \href {http://dx.doi.org/10.1109/SEFM.2008.28}
  {\path{doi:10.1109/SEFM.2008.28}}.

\bibitem{raclet2011modal}
Jean-Baptiste Raclet, Eric Badouel, Albert Benveniste, Beno{\^\i}t Caillaud,
  Axel Legay, and Roberto Passerone.
\newblock A modal interface theory for component-based design.
\newblock {\em Fundamenta Informaticae}, 108(1-2):119--149, 2011.

\bibitem{RatasichKGGSB19}
Denise Ratasich, Faiq Khalid, Florian Geissler, Radu Grosu, Muhammad Shafique,
  and Ezio Bartocci.
\newblock A roadmap toward the resilient internet of things for cyber-physical
  systems.
\newblock {\em {IEEE} Access}, 7:13260--13283, 2019.
\newblock \href {http://dx.doi.org/10.1109/ACCESS.2019.2891969}
  {\path{doi:10.1109/ACCESS.2019.2891969}}.

\bibitem{sabelfeld2003language}
Andrei Sabelfeld and Andrew~C Myers.
\newblock Language-based information-flow security.
\newblock {\em IEEE Journal on selected areas in communications}, 21(1):5--19,
  2003.

\bibitem{schneider2000}
Fred~B. Schneider.
\newblock Enforceable security policies.
\newblock {\em ACM Transactions on Information and System Security},
  3(1):30--50, 2000.
\newblock \href {http://dx.doi.org/10.1145/353323.353382}
  {\path{doi:10.1145/353323.353382}}.

\bibitem{tripakis2011theory}
Stavros Tripakis, Ben Lickly, Thomas~A. Henzinger, and Edward~A. Lee.
\newblock A theory of synchronous relational interfaces.
\newblock {\em ACM Transactions on Programming Languages and Systems (TOPLAS)},
  33(4):14, 2011.

\end{thebibliography}

\appendix

\section{Stateless Interfaces: Proofs and Auxiliary Results}
\label{sec:app:stateless}

\begin{lemma}\label{lemma:disjointY}
If  \(F \IFCompt G \) and \(F \IFCompos G \IFCompt I \), then
their output variables are pairwise disjoint, in particular,
\(\OutVars_{F} \cap \OutVars_{G} = \emptyset\), \(\OutVars_{F} \cap \OutVars_{I} = \emptyset\),
and \(\OutVars_{G} \cap \OutVars_{I} = \emptyset\).
\end{lemma}
\begin{proof} Trivial.
\end{proof}

\begin{lemma}
\label{lemma:mono:g}
If  \(\InfFlow_1 \IFCompt \InfFlow_2\) and \({\InfFlow_1 \IFCompos \InfFlow_2 \IFCompt \InfFlow_3}\), 
then:
\(\DProp{\Assumption_{\InfFlow_1 \IFCompos \InfFlow_2},\Guarantee_{\InfFlow_1 \IFCompos \InfFlow_2}} \subseteq 
\DProp{\Assumption_{\InfFlow_1 \IFCompos \InfFlow_2, \InfFlow_3}, \Guarantee_{\InfFlow_1 \IFCompos \InfFlow_2, \InfFlow_3}}\).
\end{lemma}
\begin{proof}
Follows from 
\(\AComp_{\InfFlow_1 \IFCompos \InfFlow_2, \InfFlow_3} \subseteq \AComp_{\InfFlow_1 \IFCompos \InfFlow_2}\)
and
\(\GComp_{\InfFlow_1 \IFCompos \InfFlow_2, \InfFlow_3} \subseteq \GComp_{\InfFlow_1 \IFCompos \InfFlow_2}\).
\end{proof}

From now on, \(\InfFlow = (\InVars, \OutVars, \Assumption, \Guarantee, \Prop)\) is an interface, and
\(\InfFlowC = (\InVars, \OutVars, \Flows)\)
and \(\InfFlowC_{\Env} = (\OutVars, \InVars, \Env)\) are components.

\subsection{Proof for Proposition \ref{thm:well-formed}}

\emph{Let \(\InfFlow\) be a well-formed interface. For all components \(\InfFlowC = (\InVars, \OutVars, \Flows)\)
and \(\InfFlowC_{\Env} = (\OutVars, \InVars, \Env)\): if 
\(\InfFlowC \IFImpl \InfFlow\) and \(\InfFlowC_{\Env}  \IFEnv \InfFlow\), then
 \((\Flows \cup \Env)^* \cap \Prop = \emptyset\).}

\begin{proof}Consider an arbitrary interface \(\InfFlow\).
Assume that:
\begin{enumerate*}[label=(a\arabic*)]
  \item \(\InfFlow \) is a well-formed interface, \label{thm:well_formed:a}
  \item \(\InfFlowC \IFImpl \InfFlow\), and \label{thm:well_formed:a1}
  \item \(\InfFlowC_{\Env}  \IFEnv \InfFlow\). \label{thm:well_formed:a2}
\end{enumerate*}
\begin{align*}
&(z,z') \in (\Flows \cup \Env)^* 
\ \ \overset{\text{denesting rule}}{\Leftrightarrow} \\
&(z,z') \in \Env^* \circ (\Flows \circ \Env^*)^* 
\overset{\text{flows are closed for *}}{\Leftrightarrow}&\\
&(z,z') \in  (\ID{\AllVars} \cup \Env) \circ (\Flows \circ \Env)^* \circ (\ID{\OutVars} \cup \Flows) 
\underset{\ref{thm:well_formed:a2}}{\overset{\ref{thm:well_formed:a1}}{\Rightarrow}}&\\
&(z,z') \in  (\ID{\AllVars} \cup \AComp) \circ (\GComp \circ \AComp)^* \circ (\ID{\OutVars} \cup \GComp) 
\Rightarrow &\\
& (z,z') \in  (\ID{\AllVars} \cup \AComp) \circ (\GComp \circ \AComp)^* \circ \GComp \ \tOr \ (z,z') \in \AllVars \times \InVars
\ \ \overset{\ref{thm:well_formed:a}}{\Rightarrow} \\
&(z,z') \notin \Prop. 
\end{align*}
Hence,  \((\Flows \cup \Env)^* \cap \Prop = \emptyset\). 
\end{proof}

\subsection{Properties of Composite Flows}

Let \(\InfFlow\) and \(\InfFlow'\) be interfaces.

\begin{proposition}\label{prop:composflow:alt}
The definition of composite flows can be rewritten as follows:
\begin{gather*}
\GComp_{\InfFlow, \InfFlow'} = \GComp \cup \GComp' \cup (\GComp \circ \GComp')^+ \cup
(\GComp' \circ \GComp)^+ 
\cup ((\GComp' \circ \GComp)^+ \circ \GComp') \cup
((\GComp \circ \GComp')^+ \circ \GComp).
\end{gather*}
\end{proposition}

\begin{proof} 
\begin{align*}
&\GComp_{\InfFlow, \InfFlow'} \overset{\text{def. } \ref{def:compos:flows}}{=} \\
&(\ID{\AllVars'} \cup \GComp) \circ (\GComp' \circ \GComp)^* \circ (\ID{\OutVars} \cup \GComp') = &\\
&\GComp' 
\cup (\GComp' \circ \GComp)^+
\cup ((\GComp' \circ \GComp)^+ \circ \GComp')\ \cup
 \GComp\cup (\GComp \circ (\GComp' \circ \GComp)^+)
\cup (\GComp \circ (\GComp' \circ \GComp)^* \circ \GComp') = &\\
&\GComp' \cup \GComp \cup (\GComp' \circ \GComp)^+ \cup
(\GComp \circ \GComp')^+ \ \cup 
 ((\GComp' \circ \GComp)^+ \circ \GComp') \cup
((\GComp \circ \GComp')^+ \circ \GComp). &\text{\qedhere}
\end{align*}
\end{proof}

From the previous proposition we can derive that 
composite flows are commutative.

\begin{proposition}\label{prop:composflow:comm}
\(\GComp_{\InfFlow, \InfFlow'} = \GComp_{\InfFlow', \InfFlow}.\)
\end{proposition}
\begin{proof}
Follows directly from Proposition \ref{prop:composflow:alt}.
\end{proof}

\begin{proposition}\label{prop:composflow:reflex}
\(\GComp_{\InfFlow,\InfFlow'}\) is a reflexive relation in \(\OutVars_{\InfFlow, \InfFlow'}\).
\end{proposition}
\begin{proof}
By definition of interface \(\GComp\) and 
\(\GComp'\) are reflexive relations.
Moreover, we can derive that \(\ID{\OutVars_{\InfFlow, \InfFlow'}} \subseteq \GComp \cup \GComp'\).
Then, by definition of composite flows, it follows that, 
\(\GComp \cup \GComp' \subseteq \GComp_{\InfFlow, \InfFlow'}\), and so
\(\ID{\OutVars_{\InfFlow, \InfFlow'}} \subseteq \GComp_{\InfFlow, \InfFlow'}\).
Hence \(\GComp_{\InfFlow,\InfFlow'}\) is a reflexive relation in \(\OutVars_{\InfFlow, \InfFlow'}\).
\end{proof}

In the lemma bellow we prove some properties of composite flows that are used in the proofs of important results later.
The first property tell us that, when the set of output ports is disjoint, 
then the only way to flow to a given output variable is through the flows allowed by that variable's interface. 
This property is very important to prove results related to the assumptions derived for the composition, which are defined later.
The second property tell us that considering more interfaces can only increase the possible flows.
Finally, the third property states that our operator always compute the same set of possible flows between multiple interfaces independently of the order in which each interface is considered.
This last property tell us that the composition's guarantee is associative.

\begin{lemma}
\label{lemma:flows}
Let \(\InfFlow\), \(\InfFlow'\) and \(\InfFlow''\)  be interfaces.
\begin{enumerate}[label=(\alph*)]
	\item \textbf{Suffix:} 
    If \((z,y) \in \GComp_{\InfFlow, \InfFlow'}\) and \(y \in \OutVars \setminus \OutVars'\), then
	\begin{itemize}
	\item \((z,y) \in \GComp \cup (\GComp_{\InfFlow, \InfFlow'} \circ \GComp)\) and
	\item  if \((z,y) \notin  \GComp\), then for all \((z,s) \in \GComp_{\InfFlow, \InfFlow'}\) 
	and \((s,y) \in \GComp\), we have \(s \in \OutVars'\).
	\end{itemize}
	\label{lemma:flows:suffix}
	\item \textbf{Monotonicity:} \(\GComp_{\InfFlow, \InfFlow'} \subseteq \GComp_{\InfFlow, \InfFlow' \IFCompos \InfFlow''}\).
	\label{lemma:flows:mono}

	\item \textbf{Associativity:} \(\GComp_{\InfFlow \IFCompos \InfFlow', \InfFlow''} = \GComp_{\InfFlow, \InfFlow' \IFCompos \InfFlow''}\).
	\label{lemma:flows:associative}
\end{enumerate}
\end{lemma}

\begin{proof}
Consider arbitrary interfaces \(\InfFlow\), \(\InfFlow'\) and \(\InfFlow''\).  
\begin{enumerate}[label=(\alph*)]
\item Assume that 
\((z,y) \in \GComp_{\InfFlow, \InfFlow'}\) with \(y \in \OutVars \setminus \OutVars'\).
\begin{align*}
(z,y) \in & \ \ \GComp_{\InfFlow, \InfFlow'} \ \ 
\underset{y \in \OutVars \setminus \OutVars'}{\overset{\text{prop. } \ref{prop:composflow:alt}}{\Leftrightarrow}} &\\
(\star) (z,y) \in &\ \ \GComp 
\cup \big((\GComp' \circ (\GComp \circ \GComp')^* \circ  \GComp)
\cup (\GComp \circ \GComp')^+ \circ  \GComp\big) 
 \overset{\text{prop. } \ref{prop:composflow:alt}}{\Rightarrow} &\\
(z,y) \in & \ \ \GComp \cup (\GComp_{\InfFlow, \InfFlow'} \circ \GComp). & 
\end{align*}

By \((\star)\), it follows that, if \((z,y)\notin \GComp\), then  \((z,y) = (z,s) \circ (s,y)\) with \((z,s) \in \GComp_{\InfFlow, \InfFlow'}\), \((s,y) \in \GComp\) and
\(s \in \OutVars'\).

\item

\begin{align*}
(z,y) \in & \ \ \GComp_{\InfFlow, \InfFlow'} \ \ 
\overset{\text{def. } \ref{def:compos:flows}}{\Leftrightarrow} \\
(z,y) \in &\ \ (\text{Id}_{\AllVars'} \cup \GComp) \circ (\GComp' \circ \GComp)^* \circ (\text{Id}_{\OutVars} \cup \GComp') \ \ 
\underset{\AllVars' \subseteq \AllVars_{\InfFlow', \InfFlow''}}{\overset{\GComp' \subseteq \GComp_{\InfFlow' \IFCompos \InfFlow''}}{\Rightarrow}} \\
(z,y) \in &\ \ (\text{Id}_{\AllVars_{\InfFlow', \InfFlow''}} \cup \GComp) \circ (\GComp_{\InfFlow' \IFCompos \InfFlow''} \circ \GComp)^* \circ (\text{Id}_{\OutVars} \cup \GComp_{\InfFlow' \IFCompos \InfFlow''}) \overset{\text{def. } \ref{def:compos:flows}}{\Leftrightarrow} \\
(z,y) \in & \ \ \GComp_{\InfFlow, \InfFlow' \IFCompos \InfFlow''}.  
\end{align*}

\item 
%
We start by proving that:
\(\GComp_{\InfFlow \IFCompos \InfFlow', \InfFlow''} \subseteq \GComp_{\InfFlow, \InfFlow' \IFCompos \InfFlow''}\).

Consider an arbitrary \((z,y) \in \GComp_{\InfFlow \IFCompos \InfFlow', \InfFlow''}\). We consider first the case that 
\(y \in \OutVars''\). Then, by definition of composite flows, we can derive that what we want to prove is equivalent to the following statement over (alternated) paths, for all \(n \in \mathbb{N}\), \(0 \leq i \leq n/2\) and \(0 \leq j \leq m/2\):
\begin{align*}
&\tIf (z,y) = \step_n \circ \ldots \circ \step_1,\ \text{with } \step_{2i} \in \GComp_{\InfFlow \IFCompos \InfFlow'} \tAnd \step_{2i+1} \in \GComp'',\\
&\text{ then there exists } \step'_m, \ldots, \step'_1 \tSt (z,y) = \step'_m \circ \ldots \circ \step'_1,
 \step'_{2j} \in \GComp,\ \step'_{2j+1} \in \GComp_{\InfFlow' \IFCompos \InfFlow''}, \tAnd\\
& \qquad \ \  \text{there exists } 0 \leq (k_j + r_j) \leq n \tSt \step'_{2j+1} = \step_{k_j+r_j} \circ \ldots \circ \step_{k_j + 0}.
\end{align*} 
We proceed by induction on \(n\).

\begin{description}
	\item [Base Case (\(n=1\)):]
	Consider arbitrary  \((z,y) \in \GComp'\), then, by definition of composite flows, \((z,y) \in  \GComp_{\InfFlow' \IFCompos \InfFlow''}\).
	So, \(m = 1\), \(k_1 = 1\) and  \(r_1 = 0\).
	
	\item [Inductive case:] Assume as induction hypothesis that our property holds for paths of size \(n\).
	Assume that there exists  \(\step_{n+1}, \ldots, \step_1\) such that for all \(0 \leq i \leq n/2\):
	\begin{align*}
	 (\star) & (z,y) = \step_{n+1}  \circ \ldots \circ \step_1,
	\text{ with } \step_{2i} \in \GComp_{\InfFlow \IFCompos \InfFlow'} \tAnd \step_{2i+1} \in \GComp''.
	\end{align*}
	By induction hypothesis we know that, \(0 \leq j \leq m_n/2\):
	\begin{align*}
	&\text{there exists } (s,y) = \step'_{m_{n}} \circ \ldots \circ \step'_1 \tSt (z,y) = \step_{n+1} \circ (s,y), \\
	&\qquad \step'_{2j} \in \GComp, \step'_{2j+1} \in \GComp_{\InfFlow' \IFCompos \InfFlow''} \tAnd\\
    &\qquad \text{there exists } 0 \leq (k_j + r_j) \leq n \tSt \step'_{2j+1} = \step_{k_j+r_j} \circ \ldots \circ \step_{k_j + 0}.
	\end{align*}

	\begin{description}
	\item [Case \(n+1\) is odd:]  Then, by \((\star)\), \(\step_{n+1} \in \GComp''\)
	and \(\step_{n} \in \GComp_{\InfFlow \IFCompos \InfFlow'}\).

	If \(m_n\) is even, then \(\step_m \in \GComp\). So, the property holds for \(n+1\) with \(m_{n+1} = m+1\)  and \(p_{m_{n+1}} = p_{n+1}\).
	
	Otherwise, if \(m_n\) is odd, then \(\step_{m_n} \in \GComp_{\InfFlow' \IFCompos \InfFlow''}\) with \(\step'_{m_n} = \step_{k+r} \circ \ldots \circ \step_{k + 0}\), for some \(k\) and \(r\).
	Additionally, we know that \(\step_{k+r} = \step_n\).

	\begin{description}
	\item [Case \(\step_n \in \GComp\):] It holds with \(m_{n+1} = m +1\) and 
			\(\step'_{m_{n+1}} = \step_{n+1}\).
	
	\item [Case \(\step_n \in \GComp'\):] It  holds with \(m_{n+1} = m\) and 
		\(\step'_{m_{n+1}} = \step_{n+1} \circ \step_{k+r} \circ \ldots \circ \step_{k + 0}\). 
	\end{description}
	
	\item [Case \(n+1\) is even:] Then, \(\step_{n+1} \in \GComp_{\InfFlow \IFCompos \InfFlow'}\). Follows from an analogous reasoning.
	\end{description}
\end{description}

	We can prove analogously that the property holds for arbitrary 
	\((z,y) \in \GComp_{\InfFlow \IFCompos \InfFlow', \InfFlow''}\) with
	\(y \in \OutVars \cup \OutVars'\). 
	
	We prove analogously that \(\GComp_{\InfFlow \IFCompos \InfFlow', \InfFlow''} \supseteq \GComp_{\InfFlow, \InfFlow' \IFCompos \InfFlow''}\). \qedhere
\end{enumerate}

\end{proof}

\subsection{Properties of Propagated Assumptions}

Let \(\InfFlow\), \(\InfFlow'\) and \(\InfFlow''\) be interfaces.

\begin{proposition}
\label{prop:propagate:irreflexive}
If \(\InfFlow \IFCompt \InfFlow'\), then 
\(\PAs_{\InfFlow,\InfFlow'}\) is an irreflexive relation.
\end{proposition}
\begin{proof}
Consider two compatible interfaces \(\InfFlow\) and \(\InfFlow'\).
Assume towards a contradiction that there exists \((z,z) \in  \PAs_{\InfFlow \rightarrow \InfFlow'}\). 
By definition of propagated assumptions:

\(\exists s \in X\cap Y' \tSt (z,s) \in \Assumption \tAnd (z,s) \in \GComp_{\InfFlow,\InfFlow'}.\)

\noindent Then, by \(s \in Y'\),  
\((z,s) \in \Assumption \cap (\AllVars_{\InfFlow, \InfFlow'} \times \OutVars_{\InfFlow, \InfFlow'})\).
So, by \(F \IFCompt F'\), it follows that \((z,s) \notin \GComp_{\InfFlow,\InfFlow'}\).

We can prove analogously that \((z,z) \notin  \PAs_{\InfFlow' \rightarrow \InfFlow}\). 
\end{proof}

\begin{lemma}\label{lemma:propAssumptions_mono}
Let \(\InfFlow_1\), \(\InfFlow_1'\) and \(\InfFlow_2\) be interfaces.
If  \(\InfFlow_1' \IFRef \InfFlow_1\), then
\(\PAs_{\InfFlow_1', \InfFlow_2} \subseteq \PAs_{\InfFlow_1, \InfFlow_2}\).
\end{lemma}
\begin{proof}
Assume that \(\InfFlow_1' \IFRef \InfFlow_1\).
Consider arbitrary \((z,z') \in  \PAs_{\InfFlow_1' \rightarrow \InfFlow_2}\).
Then, by definition of propagated assumptions,
\(\exists s \in \InVars_1' \cap \OutVars_2 \tSt (z,s) \in \Assumption_1' \tAnd (z',s) \in \GComp_{\InfFlow_1', \InfFlow_2}.\)

By \(\InfFlow_1' \IFRef \InfFlow_1\), we know that \(\Assumption_1' \subseteq \Assumption_1 \) and \(\GComp_1' \subseteq \GComp_1\).
Then, it follows by definition of composite guarantees that
\(\GComp_{\InfFlow_1', \InfFlow_2} \subseteq \GComp_{\InfFlow_1, \InfFlow_2}\) and so \((z',s) \in \GComp_{\InfFlow_1, \InfFlow_2}\).
By \((z,s) \in \Assumption'_1\) and \(\Assumption_1' \subseteq \Assumption_1 \), we have \((z,s) \in \Assumption_1\).
So, by definition of propagated assumptions, \((z,z') \in  \PAs_{\InfFlow_1, \InfFlow_2}\). So, \(\PAs_{\InfFlow_1' \rightarrow \InfFlow_2} \subseteq \PAs_{\InfFlow_1, \InfFlow_2}\).

We can prove analogously that \(\PAs_{\InfFlow_2 \rightarrow \InfFlow_1'} \subseteq \PAs_{\InfFlow_1, \InfFlow_2}\).
\end{proof}

\begin{lemma}
\label{lemma:composition}
Let \(\InfFlow\), \(\InfFlow'\) and \(\InfFlow''\) be interfaces, such that  \(\InfFlow \IFCompt \InfFlow'\) and \(\InfFlow \IFCompos \InfFlow' \IFCompt \InfFlow''\).
\begin{enumerate}[label=({\arabic*})]  
  \item If \((z,z') \in \PAs_{\InfFlow' \rightarrow \InfFlow''}\), then \((z,z') \in \PAs_{\InfFlow \IFCompos \InfFlow' \rightarrow \InfFlow''}\).
   \label{lemma:composition:propAGI}
  
  \item If \((z,z') \in \PAs_{\InfFlow'' \rightarrow \InfFlow'}\), then \((z,z') \in \PAs_{\InfFlow'' \rightarrow \InfFlow \IFCompos \InfFlow'}\).
  \label{lemma:composition:propAIG}
  
  \item If \((z,z') \in  \PAs_{\InfFlow, \InfFlow' \IFCompos \InfFlow''}\), then \((z,z') \in  \PAs_{\InfFlow \IFCompos \InfFlow', \InfFlow''} \cup \PAs_{\InfFlow, \InfFlow'}\).
  \label{lemma:composition:propagate}
\end{enumerate}
\end{lemma}

\begin{proof}
Consider arbitrary interfaces \(\InfFlow,\ \InfFlow' \tAnd \InfFlow''\).
Assume that:
\begin{enumerate*}[label=(a\arabic*)]
  \item \(\InfFlow \IFCompt \InfFlow'\); and \label{lemma:composition:a1}
  \item \(\InfFlow \IFCompos \InfFlow' \IFCompt \InfFlow''\). \label{lemma:composition:a2}
\end{enumerate*}

\begin{enumerate}[label=\emph{({\arabic*})}]  
\item Assume that \((z,z') \in \PAs_{\InfFlow' \rightarrow \InfFlow''}\). By definition of propagated assumptions~(def. \ref{def:prop:assumptions}), there exists
\(s \in \InVars' \cap \OutVars'' \tSt (z,s) \in \Assumption' \tAnd (z',s) \in \GComp_{\InfFlow', \InfFlow''}.\)

By \ref{lemma:composition:a2}, \(s'\notin \OutVars_{\InfFlow, \InfFlow'}\) and, by definition of composition (def. \ref{def:i:composition}), 
\((z,s) \in \Assumption_{\InfFlow \IFCompos \InfFlow'}\). By monotonicity of composite flows (lemma \ref{lemma:flows}\ref{lemma:flows:mono}), it follows that
\(\GComp_{\InfFlow',\InfFlow''} \subseteq \GComp_{\InfFlow \IFCompos \InfFlow',\InfFlow''}\). 
So, \((z',s) \in \GComp_{\InfFlow \IFCompos \InfFlow',\InfFlow''}\),
and by definition of propagated assumptions,   
\((z,z') \in \PAs_{\InfFlow \IFCompos \InfFlow' \rightarrow \InfFlow''}\).

\item  Assume that \((z,z') \in \PAs_{\InfFlow'' \rightarrow \InfFlow'}\). By definition of propagated assumptions (def. \ref{def:prop:assumptions}), there exists
\(s \in \InVars'' \cap \OutVars' \tSt (z,s) \in \Assumption'' \tAnd (z',s) \in \GComp_{\InfFlow', \InfFlow''}.\) 
By monotonicity of composite flows (Lemma \ref{lemma:flows}\ref{lemma:flows:mono}), it follows that
\(\GComp_{\InfFlow',\InfFlow''} \subseteq \GComp_{\InfFlow \IFCompos \InfFlow', \InfFlow''}\), so \((z',s) \in \GComp_{\InfFlow'', \InfFlow \IFCompos \InfFlow'}\).
Then, by \((z,s) \in \Assumption''\), it follows that \(\PAs_{\InfFlow'' \rightarrow \InfFlow \IFCompos \InfFlow'}\).

\item Assume that \((z,z') \in  \PAs_{\InfFlow, \InfFlow' \IFCompos \InfFlow''}\).
 We want to prove that \({(z,z') \in  \PAs_{\InfFlow \IFCompos \InfFlow', \InfFlow''} \cup  \PAs_{\InfFlow, \InfFlow'}}\).
\begin{description}
\item [Case \((z,z') \in  \PAs_{\InfFlow \rightarrow \InfFlow' \IFCompos \InfFlow''}\):]
By definition of propagated assumptions~(def. \ref{def:prop:assumptions}), there exists 
\(s \in \InVars \cap \OutVars_{\InfFlow'\IFCompos \InfFlow''} \tSt (z,s) \in \Assumption \tAnd (z',s) \in 
\GComp_{\InfFlow,\InfFlow' \IFCompos \InfFlow''}.\)
By associativity of composite flows (Lemma \ref{lemma:flows}\ref{lemma:flows:associative}),
\((z',s) \in  \GComp_{\InfFlow \IFCompos \InfFlow',\InfFlow''}\).
We proceed by cases on \(s \in \OutVars_{\InfFlow'\IFCompos \InfFlow''}\).

\begin{description}
\item [Case  \(s \in \OutVars'\):] 
By definition of composition,  \(s \in \OutVars_{\InfFlow \IFCompos \InfFlow'}\). 
By
suffix property of composite flows (Lemma \ref{lemma:flows}\ref{lemma:flows:suffix}) and \((z',s) \in  \GComp_{\InfFlow \IFCompos \InfFlow',\InfFlow''}\):
\((z',s) \in \GComp_{\InfFlow \IFCompos \InfFlow'} \cup (\GComp_{\InfFlow \IFCompos \InfFlow', \InfFlow''} \circ \GComp_{\InfFlow \IFCompos \InfFlow'}).\)

Consider the case that \((z',s) \in \GComp_{\InfFlow \IFCompos \InfFlow'}\).
Then, by \((z,s) \in \Assumption\), it follows 
\({(z,z') \in  \PAs_{\InfFlow, \InfFlow'}}\).

Otherwise, \((z',s) \in (\GComp_{\InfFlow \IFCompos \InfFlow', \InfFlow''} \circ \GComp_{\InfFlow \IFCompos \InfFlow'}).\) 
Consider arbitrary \((z',s') \in \GComp_{\InfFlow \IFCompos \InfFlow',\InfFlow''}\) 
and \((s',s) \in~\GComp_{\InfFlow \IFCompos \InfFlow'}\).
Then, by Lemma \ref{lemma:flows}\ref{lemma:flows:suffix}, \(s' \in \OutVars''\),
and by \ref{lemma:composition:a2}, \(s' \notin \OutVars_{\InfFlow, \InfFlow'}\).
Then, it must be the case that  \(s' \in \InVars_{\InfFlow, \InfFlow'}\).
By \((s',s) \in \GComp_{\InfFlow \IFCompos \InfFlow'}\) and 
\((z,s) \in \Assumption\), it follows that 
\((z,s') \in \PAs_{\InfFlow \rightarrow \InfFlow'}\). 
Then, by definition of composition and
\(s' \in \InVars_{\InfFlow \IFCompos \InfFlow'}\), 
 \((z,s') \in \Assumption_{\InfFlow \IFCompos \InfFlow'}\).
Again, by definition of propagated assumptions, \((z',s') \in \GComp_{\InfFlow \IFCompos \InfFlow', \InfFlow''}\) and \(s' \in \InVars_{\InfFlow \IFCompos \InfFlow'} \cap \OutVars''\), it follows that \((z,z') \in \PAs_{\InfFlow \IFCompos \InfFlow' \rightarrow \InfFlow''}\).

\item [Case  \(s \in \OutVars''\):] 
By \ref{lemma:composition:a2}, \(s \notin \OutVars_{\InfFlow \IFCompos \InfFlow'}\).
Then, by definition of composition, \({(z,s) \in \Assumption_{\InfFlow \IFCompos \InfFlow'}}\). 
So, by  \((z',s) \in  \GComp_{\InfFlow \IFCompos \InfFlow',\InfFlow''}\) and definition of
propagated assumptions, \((z,z') \in  \PAs_{\InfFlow \IFCompos \InfFlow' \rightarrow \InfFlow''}\).
\end{description}

\item [Case \((z,z') \in  \PAs_{\InfFlow' \IFCompos \InfFlow'' \rightarrow \InfFlow}\):] 
By definition of propagated assumptions ~(def. \ref{def:prop:assumptions}), there exists \(s \in \InVars_{\InfFlow' \IFCompos \InfFlow''} \cap \OutVars\) s.t.\
\((z,s) \in \Assumption_{\InfFlow' \IFCompos \InfFlow''} \tAnd (z',s) \in 
\GComp_{\InfFlow' \IFCompos \InfFlow'',\InfFlow}.\)
By associativity of composite flows (Lemma \ref{lemma:flows}\ref{lemma:flows:associative}) and \ref{lemma:composition:a1},
\((z',s) \in \GComp_{\InfFlow \IFCompos \InfFlow',\InfFlow''}\).
We proceed now by cases on \((z,s) \in \Assumption_{\InfFlow' \IFCompos \InfFlow''}\).

\begin{description}
\item [Case \((z,s) \in \Assumption'\):] By \(s \in \OutVars\) and definition of composition, 
\(s \in \OutVars_{\InfFlow, \InfFlow'}\). Then, by suffix property of composite flows (Lemma \ref{lemma:flows}\ref{lemma:flows:suffix}), 
\((z',s) \in \GComp_{\InfFlow \IFCompos \InfFlow'} \cup (\GComp_{\InfFlow \IFCompos \InfFlow', \InfFlow''} \circ \GComp_{\InfFlow \IFCompos \InfFlow'}).\)
The rest is analogous to the case \((z,z') \in  \PAs_{\InfFlow \rightarrow \InfFlow' \IFCompos \InfFlow''}\) with \(s\in \OutVars'\).

\item [Case \((z,s) \in \Assumption''\):]
By  \((z',s) \in \GComp_{\InfFlow \IFCompos \InfFlow', \InfFlow''}\), then 
\((z,z') \in \PAs_{\InfFlow'' \rightarrow \InfFlow \IFCompos \InfFlow'}\).

\item [Case \((z,s) \in \PAs_{\InfFlow', \InfFlow''}\):]
By previous lemmas \ref{lemma:composition}\ref{lemma:composition:propAGI} and \ref{lemma:composition}\ref{lemma:composition:propAIG}, 
\((z,s) \in  \PAs_{\InfFlow \IFCompos \InfFlow', \InfFlow''}\).
Then, by \((z',s) \in \GComp_{\InfFlow' \IFCompos \InfFlow'',\InfFlow}\) and 
definition of propagated assumptions, 
\((z,z') \in  \PAs_{\InfFlow \IFCompos \InfFlow', \InfFlow''}\).\qedhere
\end{description}
\end{description}
\end{enumerate}

\end{proof}

\subsection{Proof for Proposition~\ref{prop:commut}}
\emph{Let \(\InfFlow\) and \(\InfFlow'\) be interfaces. \(\InfFlow \IFCompt \InfFlow'\) iff \(\InfFlow' \IFCompt \InfFlow\), and 
\(\InfFlow \IFCompos \InfFlow' = \InfFlow' \IFCompos \InfFlow\).}
\begin{proof} By commutativity of union and intersection of sets:
\begin{itemize}
\item \(\Guarantee_{\InfFlow, \InfFlow'} =  \OutVars \cup \OutVars' = \OutVars' \cup \OutVars = \Guarantee_{\InfFlow', \InfFlow}\);
\item \(\InVars_{\InfFlow, \InfFlow'} = 
(\InVars \cup \InVars') \setminus \OutVars_{\InfFlow, \InfFlow'} = 
(\InVars' \cup \InVars) \setminus \OutVars_{\InfFlow', \InfFlow} =  \InVars_{\InfFlow', \InfFlow}\);
\item \(\PAs_{\InfFlow, \InfFlow'} = 
\PAs_{\InfFlow \rightarrow \InfFlow'} \cup \PAs_{\InfFlow' \rightarrow \InfFlow} = 
\PAs_{\InfFlow' \rightarrow \InfFlow} \cup \PAs_{\InfFlow \rightarrow \InfFlow'} = \PAs_{\InfFlow', \InfFlow}\);
\item 
\(\Assumption_{\InfFlow, \InfFlow'} = 
\Assumption \cup \Assumption' \cup \PAs_{\InfFlow, \InfFlow'} \cap (\AllVars_{\InfFlow, \InfFlow'} \times \InVars_{\InfFlow, \InfFlow'}) = \)

\(\Assumption' \cup \Assumption \cup \PAs_{\InfFlow', \InfFlow} \cap (\AllVars_{\InfFlow', \InfFlow} \times \InVars_{\InfFlow', \InfFlow}) = 
\Assumption_{\InfFlow', \InfFlow}\).

\item By Proposition \ref{prop:composflow:comm},  
\(\GComp_{\InfFlow, \InfFlow'} = \GComp_{\InfFlow', \InfFlow}\).

\item \(\Prop_{\InfFlow, \InfFlow'} = \Prop \cup \Prop' \cup \DProp{\Assumption_{\InfFlow, \InfFlow'}, \Guarantee_{\InfFlow, \InfFlow'}} = \Prop' \cup \Prop \cup \DProp{\Assumption_{\InfFlow', \InfFlow}, \Guarantee_{\InfFlow', \InfFlow}} = \Prop_{\InfFlow', \InfFlow}\).
\end{itemize}
Thus, \(\InfFlow \IFCompos \InfFlow' = \InfFlow' \IFCompos \InfFlow\).

\noindent Additionally, we can prove analogously that
\(\InfFlow \IFCompt \InfFlow'\) iff \(((\Assumption' \cup \Assumption) \cap (\AllVars_{\InfFlow', \InfFlow} \times \OutVars_{\InfFlow', \InfFlow}))
\subseteq \Guarantee_{\InfFlow',\InfFlow}\).
So, \(\InfFlow \IFCompt \InfFlow'\) iff \(\InfFlow \IFCompt \InfFlow'\).
\end{proof}

\subsection{Proof for Theorem \ref{thm:comp:wellformed}}

\emph{Let \(\InfFlow\) and \(\InfFlow'\) be well-formed interfaces.
If \(\InfFlow \IFCompt \InfFlow'\), then \(\InfFlow \IFCompos \InfFlow'\) is a well-formed interface.}

\begin{proof} 
Consider arbitrary well-formed interfaces \(\InfFlow\) and \(\InfFlow'\).
Assume that \(\InfFlow \IFCompt \InfFlow'\).

\begin{description}
\item [We start by proving that each relation in the tuple is a no-flow.]\hfill

By definition of no-flow relation, we need to prove that they are all irreflexive relations.
By definition of interface and Proposition \ref{prop:propagate:irreflexive}, \(\Assumption_{\InfFlow, \InfFlow'}\) is an irreflexive relation.
By Proposition \ref{prop:composflow:reflex},  \(\GComp_{\InfFlow,\InfFlow'}\) is a reflexive relation, and so \(\Guarantee_{\InfFlow,\InfFlow'}\) is an irreflexive relation.
Finally, by \(\DProp{\Assumption_{\InfFlow,\InfFlow'}, \Guarantee_{\InfFlow,\InfFlow'}} \subseteq  \Guarantee_{\InfFlow,\InfFlow'}\) and definition of interface, 
\(\Prop_{\InfFlow, \InfFlow'}\) is irreflexive, as well.

\item [We prove now that it satisfies the well-formed property:]\hfill

\begin{center}
\((\ID{\OutVars_{\InfFlow,\InfFlow'}} \cup \AComp_{\InfFlow,\InfFlow'}) \circ (\GComp_{\InfFlow,\InfFlow'} \circ \AComp_{\InfFlow,\InfFlow'})^* \circ \GComp_{\InfFlow,\InfFlow'} \cap \PropComp_{\InfFlow,\InfFlow'} = \emptyset.\)
\end{center}

Consider arbitrary \((z,z') \in (\ID{\OutVars_{\InfFlow,\InfFlow'}} \cup \AComp_{\InfFlow,\InfFlow'}) \circ (\GComp_{\InfFlow,\InfFlow'} \circ \AComp_{\InfFlow,\InfFlow'})^* \circ \GComp_{\InfFlow,\InfFlow'}\). 
Then,  \(z'\in \OutVars_{\InfFlow, \InfFlow'}\).
By definition of derived properties, \((z,z') \notin \DProp{\Assumption_{\InfFlow,\InfFlow'}, \Guarantee_{\InfFlow,\InfFlow'}}\).
So, we are missing to prove that \((z,z') \notin \Prop \cup \Prop'\).

Lets consider the case that \(z' \in \OutVars\).
Then, by \(\InfFlow \IFCompt \InfFlow'\),  \(z'\notin \OutVars'\). 
Thus, \((z,z') \notin \AllVars' \times \OutVars'\) and, by definition of interface,
\((z,z') \notin \Prop'\). 
Moreover, by Lemma \ref{lemma:flows_and_prop:it} and 
\(\InfFlow \IFCompt \InfFlow\), 
\((z,z') \in ((\ID{\AllVars} \cup \AComp) \circ (\GComp \circ \AComp)^* \circ \GComp)\). 
Hence, by \(\InfFlow\) being well-formed, \((z,z') \notin \Prop\), as well.

The case for \(z' \in \OutVars'\) is analogous. \qedhere
\end{description}

\end{proof}

In the lemmas below we prove that
any alternated path in \(\AComp_{\InfFlow,\InfFlow'}\) and \(\GComp_{\InfFlow, \InfFlow'}\)
from a variable of \(\InfFlow\) or \(\InfFlow'\) to an output variable of the same interface can be translated to a path 
using only the complement of that interface's assumptions and guarantees.

\begin{lemma}\label{lemma:flows_and_prop:it}
Let \(\InfFlow\) and \(\InfFlow'\) be interfaces s.t.\ \(\InfFlow \IFCompt \InfFlow'\),
and \((z,z') \in \AllVars \times \OutVars\). 

\noindent
If 
\((z,z') \in (\ID{\AllVars_{\InfFlow,\InfFlow'}} \cup \AComp_{\InfFlow,\InfFlow'}) \circ (\GComp_{\InfFlow,\InfFlow'} \circ \AComp_{\InfFlow,\InfFlow'})^* \circ \GComp_{\InfFlow,\InfFlow'}\), then \((z,z') \in (\ID{\AllVars} \cup \AComp) \circ (\GComp \circ \AComp)^* \circ \GComp.\)
\end{lemma}

\begin{proof}
Consider arbitrary interfaces \(\InfFlow\) and \(\InfFlow'\) and assume that \(\InfFlow \IFCompt \InfFlow'\).
Additionally, consider arbitrary  \((z,z') \in \AllVars \times \OutVars\).
We prove by induction that, for all \(n\in \mathbb{N}\):
\begin{align*}
\tIf &(z,z') \in (\ID{\AllVars_{\InfFlow,\InfFlow'}} \cup \AComp_{\InfFlow,\InfFlow'}) \circ (\GComp_{\InfFlow,\InfFlow'} \circ \AComp_{\InfFlow,\InfFlow'})^n \circ \GComp_{\InfFlow,\InfFlow'},\\
\tthen &(z,z') \in (\ID{\AllVars} \cup \AComp) \circ (\GComp \circ \AComp)^n \circ \GComp.
\end{align*}

\textbf{Base case \(n = 0\):} It holds by Lemma \ref{lemma:flows_and_prop:simple}, proved below.

\textbf{Induction step:} Assume by induction hypothesis (IH)
that the statement holds for \(n\).

Consider arbitrary:
\((z,z') \in (\ID{\AllVars_{\InfFlow,\InfFlow'}} \cup \AComp_{\InfFlow,\InfFlow'}) \circ (\GComp_{\InfFlow,\InfFlow'} \circ \AComp_{\InfFlow,\InfFlow'})^{n+1} \circ \GComp_{\InfFlow,\InfFlow'}\).
Then, by application of the induction hypothesis:
\((z,z') \in (\ID{\AllVars} \cup \AComp) \circ (\GComp \circ \AComp)^{n} \circ
\GComp \circ \AComp_{\InfFlow,\InfFlow'}
\circ \GComp_{\InfFlow,\InfFlow'}\).

Consider  arbitrary \((z,s) \in (\ID{\AllVars} \cup \AComp) \circ (\GComp \circ \AComp)^{n} \circ \GComp\),
\((s,s') \in  \AComp_{\InfFlow,\InfFlow'}\)  and \((s',z') \in \GComp_{\InfFlow,\InfFlow'}\).
Then, \((s,z') \in \AllVars \times \OutVars\) and, by Lemma \ref{lemma:flows_and_prop:simple},
\((s,z') \in \AComp \circ \GComp\).
Hence
\((z,z') \in (\ID{\AllVars} \cup \AComp) \circ (\GComp \circ \AComp)^{n+1} \circ \GComp\).
\end{proof}

\begin{lemma}\label{lemma:flows_and_prop:simple}
\emph{Let \(\InfFlow\) and \(\InfFlow'\) be interfaces s.t.\ \(\InfFlow \IFCompt \InfFlow'\). Let \((z,z') \in \AllVars \times \OutVars\).}
\begin{enumerate}[label=(\alph*)]
\item If \((z,z') \in \GComp_{\InfFlow,\InfFlow'} \),
then \((z,z') \in \GComp \cup (\AComp \circ \GComp)\).

\item If \((z,z') \in \AComp_{\InfFlow,\InfFlow'} \circ \GComp_{\InfFlow,\InfFlow'}\),
then \((z,z') \in \AComp \circ \GComp\).
\end{enumerate}
\end{lemma}

\begin{proof}
Consider arbitrary interfaces \(\InfFlow\) and \(\InfFlow'\).
Assume that \(\InfFlow \IFCompt \InfFlow'\).
Additionally, consider arbitrary  \((z,z') \in \AllVars \times \OutVars\).

\begin{enumerate}[label=(\alph*)]
\item Assume that \((z,z') \in \GComp_{\InfFlow,\InfFlow'}\).
By \(z' \in \OutVars\) and Proposition \ref{lemma:flows}\ref{lemma:flows:suffix},
\((z,z') \in \GComp \cup (\GComp_{\InfFlow,\InfFlow'} \circ \GComp)\).

The case that \((z,z') \in \GComp\) is trivial.

Consider now the case that \((z,z') \in (\GComp_{\InfFlow,\InfFlow'} \circ \GComp)\). Consider arbitrary \((z,s) \in \GComp_{\InfFlow,\InfFlow'}\)
and \((s,z') \in \GComp\).
We know by Proposition  \ref{lemma:flows}\ref{lemma:flows:suffix} that
\(s \in \OutVars'\). 
Then, by \(\InfFlow \IFCompt \InfFlow'\), \((z,s) \in \AllVars_{\InfFlow,\InfFlow'} \times \OutVars_{\InfFlow,\InfFlow'}\) and \((z,s) \in \GComp_{\InfFlow,\InfFlow'}\),
it follows that 
\((z,s) \notin ((\Assumption \cup \Assumption') \cap  \AllVars_{\InfFlow,\InfFlow'} \times \OutVars_{\InfFlow,\InfFlow'})\).
So, in particular, \((z,s) \notin \Assumption\).
Hence, \((z,z') \in \AComp \circ \GComp\).

\item  Assume that \((z,z') \in \AComp_{\InfFlow,\InfFlow'} \circ \GComp_{\InfFlow,\InfFlow'}\). By \(z' \in \OutVars\), we proceed by cases w.r.t.\ Proposition  \ref{lemma:flows}\ref{lemma:flows:suffix}.

If \((z,z') \in \AComp_{\InfFlow,\InfFlow'} \circ \GComp\), then,
there exists \((z,s) \in \AComp_{\InfFlow,\InfFlow'}\) and
\((s,z') \in \GComp\). As \((z,s) \in \AComp_{\InfFlow,\InfFlow'}\), then \(s\in \InVars_{\InfFlow, \InfFlow'}\), and by \((s,z') \in \GComp\), \(s\in \AllVars\).
Thus, by definition of variables between interfaces,  \(s\in \InVars\).
By definition of composite assumptions, \((z,s) \notin \Assumption_{\InfFlow, \InfFlow'}\), and as a consequence 
\((z,s) \notin \Assumption\). So, \((z,s) \in \AComp\).

If \((z,z') \in \AComp_{\InfFlow,\InfFlow'} \circ \GComp_{\InfFlow,\InfFlow'} \circ \GComp\), then there exists \((z,s) \in \AComp_{\InfFlow,\InfFlow'}\), \((s,s') \in \GComp_{\InfFlow,\InfFlow'}\)
and \((s',z') \in \GComp\).
By  Proposition  \ref{lemma:flows}\ref{lemma:flows:suffix}, we know that \(s' \in \OutVars'\). Additionally, by \(\InfFlow \IFCompt \InfFlow'\) and \(s' \in \AllVars\), then \(s' \in \InVars\).
Assume towards a contradiction that \((z,s') \in \Assumption\).
Then, by definition of propagated assumptions and \((s,s') \in \GComp_{\InfFlow,\InfFlow'}\),
we have \((z,s) \in \PAs_{\InfFlow,\InfFlow'}\).
This contradicts the fact that \((z,s) \in \AComp_{\InfFlow,\InfFlow'}\). 
Thus, \((z,s') \notin \Assumption\) and so \((z,z') \in \AComp \circ \GComp\). \qedhere
\end{enumerate}
\end{proof}

\subsection{Proof for Theorem \ref{thm:incremental}}

\emph{Let \(F,\ G\) and \(I\) be interfaces. If  \(F \IFCompt G \) and 
\(F \IFCompos G \IFCompt I \), then \(G \IFCompt I \) and 
\(F  \IFCompt G \IFCompos I \).}

\begin{proof}
Consider arbitrary interfaces \(F,\ G \tAnd I\).
Assume that:
\begin{enumerate*}[label=(a\arabic*),itemindent=1em]
  \item \(F \IFCompt G \); and \label{thm:incremental:a1}
  \item \(F \IFCompos G \IFCompt I \). \label{thm:incremental:a2}
\end{enumerate*}

\begin{description}
\item [We start by proving that \(G \IFCompt I\).]\hfill

From \ref{thm:incremental:a1}, \ref{thm:incremental:a2} and Lemma 
\ref{lemma:disjointY},
it follows that \(\OutVars_G \cap \OutVars_I = \emptyset\), i.e.\  \(G\) and \(I\) are
composable.
We are missing to prove that:
\(((\Assumption_G \cup \Assumption_I) \cap  (\AllVars_{G, I} \times \OutVars_{G, I})) \subseteq \Guarantee_{G,I}.\)
Let \((z,s) \in \InVars_{G, I} \times \OutVars_{G, I}\) and 
\((z,s) \in \Assumption_G \cup \Assumption_I\).
Note that by \ref{thm:incremental:a2}, \(s \notin \OutVars_{F}\).
We want to prove that \((z,s) \in \Guarantee_{G,I}\).
As  \(s \notin \OutVars_{F}\) and by definition of composition, 
if \((z,s) \in \Assumption_G\), then \((z,s) \in \Assumption_{F \IFCompos G}\).
So, it follows that \((z,s) \in \Assumption_{F \IFCompos G} \cup \Assumption_I \). Then, by \ref{thm:incremental:a2}, we know that \((z,s) \in \Guarantee_{F \IFCompos G,I}\).
By monotonicity of composite flows (Lemma \ref{lemma:flows}\ref{lemma:flows:mono}) \(\GComp_{G,I} \subseteq \GComp_{F \IFCompos G,I}\).
Then, it follows that \((z,s) \in \Guarantee_{G,I}\).

\item [We prove now that \( F \IFCompt G \IFCompos I\).]\hfill

From \ref{thm:incremental:a1}, \ref{thm:incremental:a2} and Lemma 
\ref{lemma:disjointY},
it follows that \(\OutVars_F \cap \OutVars_{G \IFCompos I} = \emptyset\), i.e.\  \(F\) and \(G \IFCompos I\) are
composable.
We are missing to prove that:
\(((\Assumption_F \cup \Assumption_{G \IFCompos I}) \cap (\InVars_{F, {G \IFCompos I}} \times \OutVars_{F, {G \IFCompos I}})) 
\subseteq \Guarantee_{F, G \IFCompos I}.\)

\noindent Let \((z,s) \in \InVars_{F, G \IFCompos I} \times \OutVars_{F, G \IFCompos I}\) and 
\((z,s) \in  \Assumption_F \cup \Assumption_{G \IFCompos I}\).
We prove 
\((z,s) \in \Guarantee_{F, G \IFCompos I} \) by cases.

\begin{description}
\item [Case \((z,s) \in \Assumption_F\) and \(s \in  \OutVars_G\):] 
Assume towards a contradiction that \((z,s) \notin \Guarantee_{F, G \IFCompos I}\).
So,  \((z,s) \in \GComp_{F, G \IFCompos I}\).
By associativity of composite flows (Lemma \ref{lemma:flows} \ref{lemma:flows:associative}), \((z,s) \in \GComp_{F \IFCompos G,I}\).
By definition of composition, then \(s \in  \OutVars_{F \IFCompos G}\).
Thus,  by \ref{thm:incremental:a1} and  \((z,s) \in \Assumption_F\), \((\star) (z,s) \in \Guarantee_{F,G}\).
Additionally, 
by suffix property of composite flows (Lemma \ref{lemma:flows} \ref{lemma:flows:suffix}), \((z,s) \in \GComp_{F \IFCompos G} \cup (\GComp_{F \IFCompos G,I} \circ \GComp_{F \IFCompos G})\).

If \((z,s) \in \GComp_{F \IFCompos G}\), then we have a contradiction with \((\star)\).

Otherwise, there exists \(s' \in \OutVars_{I}\) s.t.:
\((\star\star) (z,s') \in \GComp_{F \IFCompos G,I} \tAnd
(s',s) \in \GComp_{F \IFCompos G}.\)
By \((z,s) \in \Assumption_F\), \((s',s) \in \Guarantee_{F \IFCompos G}\) and definition of propagated assumptions, 
\((z,s') \in \PAs_{F,G}\). So, by \(s' \in \InVars_{F \IFCompos G}\) and definition of composition, \((z,s') \in \Assumption_{F \IFCompos G}\). As  \(s' \in \InVars_{F \IFCompos G} \cap \OutVars_{I}\) and by \ref{thm:incremental:a2}, \((z,s') \in \Guarantee_{F \IFCompos G,I}\).
This contradicts \((\star\star)\).

Hence \((z,s) \notin \GComp_{F, G \IFCompos I}\), i.e.\  \((z,s) \in \Guarantee_{F, G \IFCompos I}\).

\item [Case \((z,s) \in \Assumption_F\) and \(s \in  \OutVars_I\):] By definition of composition and \ref{thm:incremental:a2}, \((z,s) \in \Guarantee_{F \IFCompos G, I}\).
And by associativity of composite flows (Lemma \ref{lemma:flows}\ref{lemma:flows:associative}),
\((z,s) \in \Guarantee_{F,  G \IFCompos I}\).

\item [Case \((z,s) \in \Assumption_{G \IFCompos I}\):] \hfill

\begin{description}
\item [Case \((z,s) \in \Assumption_{I}\):] By \ref{thm:incremental:a2}, \((z,s) \in \Guarantee_{F \IFCompos G, I}\).
And by associativity of composite flows (Lemma \ref{lemma:flows}\ref{lemma:flows:associative}),
\((z,s) \in \Guarantee_{F,  G \IFCompos I}\).

\item [Case \((z,s) \in \Assumption_{G}\):] From an analogous reasoning to the case that \((z,s) \in \Assumption_F\),
it follows that \((z,s) \in \Guarantee_{F,  G \IFCompos I}\). \qedhere

\end{description}

%
%
\end{description}
\end{description}
\end{proof}

\subsection{Proof for Corollary \ref{thm:assoc}}

\emph{If \(F \IFCompt G\) and \(F \IFCompos G \IFCompt  I\), then}
\((F \IFCompos G) \IFCompos  I = F \IFCompos (G \IFCompos I).\)

\begin{proof}
Assume that \(\star\) \(F \IFCompt G\) and \(F \IFCompos G \IFCompt  I\).
Then, by theorem \ref{thm:incremental}, it follows that: 
\(\star\star\) \(G \IFCompt I\) and \(F \IFCompt G \IFCompos I\).
By definition of composition, it is easy to prove that: 
(\(\dagger\)) \(\InVars_{F \IFCompos G,I} = \InVars_{F, G  \IFCompos I}\),
\(\OutVars_{F \IFCompos G,I} = \OutVars_{F, G  \IFCompos I}\), and 
\(\AllVars_{F \IFCompos G,I} = \AllVars_{F, G  \IFCompos I}\).
And, by Lemma \ref{lemma:flows}\ref{lemma:flows:associative}, \(\Guarantee_{F \IFCompos G, I} = \Guarantee_{F, G \IFCompos I}\).

\begin{description}
\item[We prove now that \(\Assumption_{F \IFCompos G,I} = \Assumption_{F, G  \IFCompos I}\).]

\begin{align*}
&\Assumption_{F \IFCompos G,I} \quad 
\overset{\text{Def. \ref{def:i:composition}}}{=}\\
& (\Assumption_{F} \cup \Assumption_G \cup \PAs_{F,G} \cup \Assumption_I \cup \PAs_{F \IFCompos G, I})\  \cap (\AllVars_{F \IFCompos G,I} \times \InVars_{F \IFCompos G,I})
\underset{\text{Lemma } \ref{lemma:composition}\ref{lemma:composition:propagate}}{\overset{(\star), (\star\star)}{=}}\\
&(\Assumption_{F} \cup \Assumption_G \cup \PAs_{F,G} \cup \Assumption_I \cup \PAs_{F, G \IFCompos I}  \cup \PAs_{G,I})\ \cap (\AllVars_{F \IFCompos G,I} \times \InVars_{F \IFCompos G,I})
\underset{\text{Lemma } \ref{lemma:composition} \ref{lemma:composition:propAGI}-\ref{lemma:composition:propAIG}}{\overset{(\star), (\star\star)}{=}}\\
& (\Assumption_{F} \cup \Assumption_G \cup \Assumption_I \cup \PAs_{F, G \IFCompos I}  \cup \PAs_{G,I})\ \cap (\AllVars_{F \IFCompos G,I} \times \InVars_{F \IFCompos G,I}) 
\overset{(\dagger)}{=} \\
&(\Assumption_{F} \cup \Assumption_G \cup \Assumption_I \cup \PAs_{F, G \IFCompos I}  \cup \PAs_{G,I}) \ \cap (\AllVars_{F, G  \IFCompos  I} \times \InVars_{F, G  \IFCompos  I}) 
\qquad \ \ \overset{\text{Def. \ref{def:i:composition}}}{=} \\
&(\Assumption_F \cup \Assumption_{G \IFCompos I} \cup  \PAs_{F, G \IFCompos I}) \cap (\AllVars_{F, G  \IFCompos I} \times \InVars_{F, G  \IFCompos I})
\qquad \ \ \  \overset{\text{Def. \ref{def:i:composition}}}{=} \\
& \Assumption_{F, G  \IFCompos I}.
\end{align*}

\item[Finally, we prove that \(\Prop_{F \IFCompos G, I} = \Prop_{F, G \IFCompos I}\).]
Note that by our previous results, it follows that:
\(\DProp{\Assumption_{F \IFCompos G,I}, \Guarantee_{F \IFCompos G,I}} = \DProp{\Assumption_{F,G \IFCompos I}, \Guarantee_{F,G \IFCompos I}}\).
\begin{align*}
& \Prop_{F \IFCompos G, I}
\overset{\text{Def. \ref{def:i:composition}}}{=}&\\
& \Prop_{F \IFCompos G} \cup \Guarantee_{I} \cup \DProp{\Assumption_{F \IFCompos G, I}, \Prop_{F \IFCompos G, I}} \overset{\text{Def. \ref{def:i:composition}}}{=}&\\
&\Prop_F \cup \Prop_{G} \cup \DProp{\Assumption_{F \IFCompos G}, \Prop_{F \IFCompos G}}  \cup \Prop_I \cup \DProp{\Assumption_{F \IFCompos G,I}, \Prop_{F \IFCompos G,I}}  
\underset{\text{Lemma \ref{lemma:mono:g}}}{\overset{(\star),(\star\star)}{=}}&\\
&\Prop_F \cup \Prop_{G} \cup \Guarantee_I \cup \DProp{\Assumption_{F \IFCompos G,I}, \Prop_{F \IFCompos G,I}} \ \  \overset{}{=}&\\
&\Prop_F \cup \Prop_{G} \cup \Prop_I \cup \DProp{\Assumption_{F, G \IFCompos I}, \Prop_{F,G \IFCompos I}}
\underset{\text{Lemma \ref{lemma:mono:g}}}{\overset{(\star),(\star\star)}{=}}&\\
&\Prop_F \cup \Prop_{G} \cup \Prop_I \cup \DProp{\Assumption_{F, G \IFCompos I}, \Prop_{F,G \IFCompos I}} \cup \DProp{\Assumption_{G \IFCompos I}, \Prop_{G \IFCompos I}}
\overset{\text{Def. \ref{def:i:composition}}}{=}&\\
&\Prop_{F, G \IFCompos I}. & \text{\qedhere}
\end{align*}

\end{description}
\end{proof}

\subsection{Proof for Theorem \ref{thm:independent_imp}}

\emph{Let \(\InfFlow_1'\), \(\InfFlow_1\) and \(\InfFlow_2\) be well-formed interfaces. 
If \(\InfFlow_1' \IFRef \InfFlow_1\) 
and \(\InfFlow_1 \IFCompt \InfFlow_2\), 
then \(\InfFlow_1' \IFCompt \InfFlow_2\) and \(\InfFlow_1' \IFCompos \InfFlow_2 \IFRef \InfFlow_1 \IFCompos \InfFlow_2\).
}

\begin{proof}
Assume that \(\InfFlow_1' \IFRef \InfFlow_1\) and \(\InfFlow_1 \IFCompt \InfFlow_2\).
Then, by \(\InfFlow_1' \IFRef \InfFlow_1\):
\begin{enumerate*}[label=(r\arabic*)]
\item \(\Assumption_1' \subseteq \Assumption_1 \); \label{thm:independent:refA}
\item \(\Guarantee_1 \subseteq \Guarantee_1'\); \label{thm:independent:refG}
\item \(\Prop_1  \subseteq \Prop_1'\).  \label{thm:independent:refP}
\end{enumerate*}

\begin{description}
\item [We start by proving that \(\InfFlow_1' \IFCompt \InfFlow_2\).]\hfill

By \ref{thm:independent:refA},
\((\Assumption_1' \cup \Assumption_2) \subseteq (\Assumption_1 \cup \Assumption_2)\).
By definition of composite flows and \ref{thm:independent:refG}, \((\dagger) \Guarantee_{\InfFlow_1, \InfFlow_2} \subseteq \Guarantee_{\InfFlow_1', \InfFlow_2}\).
Then,  by \(\InfFlow_1 \IFCompt \InfFlow_2\), we have:
\((\Assumption_1' \cup \Assumption_2) \cap (\InVars_{\InfFlow_1, \InfFlow_2} \times \OutVars_{\InfFlow_1, \InfFlow_2})\subseteq \Guarantee_{\InfFlow_1', \InfFlow_2}.\)
So, \(\InfFlow_1' \IFCompt \InfFlow_2\).

\item [We prove now that \(\InfFlow_1' \IFCompos \InfFlow_2 \IFRef \InfFlow_1 \IFCompos \InfFlow_2\).]\hfill

By our results above for the assumptions and guarantees composite, we 
are only missing 
to prove that \(\Prop_{\InfFlow_1 \IFCompos \InfFlow_2} \subseteq \Prop_{\InfFlow_1' \IFCompos \InfFlow_2}\).

Consider arbitrary \((z,y) \in \Prop_{\InfFlow_1 \IFCompos \InfFlow_2}\). 
If \((z,y) \in \Prop_1 \cup \Prop_2\), then, by \ref{thm:independent:refP}, if follows that \((z,y) \in \Prop_1' \cup \Prop_2\).

If \((z,y) \in \DProp{\Assumption_{\InfFlow_1 \IFCompos \InfFlow_2}, \Guarantee_{\InfFlow_1 \IFCompos \InfFlow_2}}\), then, by definition of derived properties:
\begin{enumerate}[label = (\roman*)]
\item  \((z,y) \in \Guarantee_{\InfFlow_1 \IFCompos \InfFlow_2}\), and
\item   \((z,y) \notin (\ID{\AllVars_{\InfFlow_1,\InfFlow_2}} \cup \AComp_{\InfFlow_1,\InfFlow_2}) 
\circ (\GComp_{\InfFlow_1,\InfFlow_2} \circ \AComp_{\InfFlow_1,\InfFlow_2})^* \circ \GComp_{\InfFlow_1,\InfFlow_2}\). \label{thm:independent:iG2}
\end{enumerate}
By \(\Guarantee_{\InfFlow_1, \InfFlow_2} \subseteq \Guarantee_{\InfFlow_1', \InfFlow_2}\), then \((z,y) \in \Guarantee_{\InfFlow_1',\InfFlow_2}\), as well.
We prove by induction that for all \(n \in \mathbb{N}\) and \((z,y) \in \Guarantee_{\InfFlow_1,\InfFlow_2}\):
\begin{gather*}
\tIf (z,y) \notin (\ID{\AllVars_{\InfFlow_1,\InfFlow_2}} \cup \AComp_{\InfFlow_1,\InfFlow_2}) 
\circ (\GComp_{\InfFlow_1,\InfFlow_2} \circ \AComp_{\InfFlow_1,\InfFlow_2})^n \circ \GComp_{\InfFlow_1,\InfFlow_2}\\
\tthen (z,y) \notin (\ID{\AllVars_{\InfFlow_1',\InfFlow_2}} \cup \AComp_{\InfFlow_1',\InfFlow_2}) 
\circ (\GComp_{\InfFlow_1',\InfFlow_2} \circ \AComp_{\InfFlow_1',\InfFlow_2})^n \circ \GComp_{\InfFlow_1',\InfFlow_2}.
\end{gather*}

\noindent Consider arbitrary \((z,y) \in \Guarantee_{\InfFlow_1,\InfFlow_2}\).

\textbf{Base case \(n=0\): }
Note that \((z,y) \notin \GComp_{\InfFlow_1,\InfFlow_2}\).
Assume that
\((z,y) \notin (\AComp_{\InfFlow_1,\InfFlow_2}\circ\GComp_{\InfFlow_1,\InfFlow_2})\).
Then, for all \((z,s) \in \AComp_{\InfFlow_1,\InfFlow_2}\) there exists
\((s,y) \notin \GComp_{\InfFlow_1,\InfFlow_2}\).
By \(\InfFlow_1' \IFRef \InfFlow_1\) and Lemma \ref{lemma:propAssumptions_mono},
we have \( (\dagger \dagger) \Assumption_{\InfFlow_1' \IFCompos \InfFlow_2} \subseteq \Assumption_{\InfFlow_1 \IFCompos \InfFlow_2}\).
So, it follows from \((\dagger)\) and \((\dagger \dagger)\), that for all 
\((z,s) \in \AComp_{\InfFlow_1',\InfFlow_2}\) there exists 
\((s,y) \in \Guarantee_{\InfFlow_1',\InfFlow_2}.\)
Thus, \((z,y) \notin \AComp_{\InfFlow_1',\InfFlow_2}\circ\GComp_{\InfFlow_1',\InfFlow_2}\), as well.

\textbf{Induction step:} We assume as induction hypothesis
that the statement holds for \(n\).
Consider arbitrary:
\((z,y) \notin (\ID{\AllVars_{\InfFlow_1,\InfFlow_2}} \cup \AComp_{\InfFlow_1,\InfFlow_2}) 
\circ (\GComp_{\InfFlow_1,\InfFlow_2} \circ \AComp_{\InfFlow_1,\InfFlow_2})^{n+1} \circ \GComp_{\InfFlow_1,\InfFlow_2}.\)

By induction hypothesis,
\begin{align*}
&(z,y) \notin (\ID{\AllVars_{\InfFlow_1',\InfFlow_2}} \cup \AComp_{\InfFlow_1',\InfFlow_2}) 
\circ (\GComp_{\InfFlow_1',\InfFlow_2} \circ \AComp_{\InfFlow_1',\InfFlow_2})^{n} \circ \GComp_{\InfFlow_1',\InfFlow_2} \circ 
\AComp_{\InfFlow_1,\InfFlow_2} \circ \GComp_{\InfFlow_1,\InfFlow_2}.
\end{align*}
Then, for all 
\((z,s) \in (\ID{\AllVars_{\InfFlow_1',\InfFlow_2}} \cup \AComp_{\InfFlow_1',\InfFlow_2}) 
\circ (\GComp_{\InfFlow_1',\InfFlow_2} \circ \AComp_{\InfFlow_1',\InfFlow_2})^{n} \circ \GComp_{\InfFlow_1',\InfFlow_2}\) there exists\linebreak \(
(s,y) \notin \AComp_{\InfFlow_1,\InfFlow_2} \circ \GComp_{\InfFlow_1,\InfFlow_2}\).
By the same reasoning applied to the base case, it follows that
for all 
\((z,s) \in (\ID{\AllVars_{\InfFlow_1',\InfFlow_2}} \cup \AComp_{\InfFlow_1',\InfFlow_2}) 
\circ (\GComp_{\InfFlow_1',\InfFlow_2} \circ \AComp_{\InfFlow_1',\InfFlow_2})^{n} \circ \GComp_{\InfFlow_1',\InfFlow_2}\) there exists 
\((s,y) \notin \AComp_{\InfFlow_1',\InfFlow_2} \circ \GComp_{\InfFlow_1',\InfFlow_2}\).
Thus, 
\((z,y) \notin (\ID{\AllVars_{\InfFlow_1',\InfFlow_2}} \cup \AComp_{\InfFlow_1',\InfFlow_2}) 
\circ (\GComp_{\InfFlow_1',\InfFlow_2} \circ \AComp_{\InfFlow_1',\InfFlow_2})^{n+1} \circ \GComp_{\InfFlow_1',\InfFlow_2}\).

Hence by definition of derived properties, \(\Prop_{\InfFlow_1 \IFCompos \InfFlow_2} \subseteq \Prop_{\InfFlow_1' \IFCompos \InfFlow_2}\). And, by definition of refinement, \(\InfFlow_1' \IFCompos \InfFlow_2 \IFRef \InfFlow_1 \IFCompos \InfFlow_2\). \qedhere
\end{description}
\end{proof}

\subsection{Proof for Proposition \ref{prop:impl}}

\emph{Let \(\InfFlow\) and \(\InfFlow'\) be interfaces.
Let \(\InfFlowC = (\InVars, \OutVars, \Flows)\) and 
\(\InfFlowC' = (\InVars', \OutVars', \Flows')\) be components:
If \(\InfFlowC \IFImpl \InfFlow\) and \(\InfFlowC' \IFImpl \InfFlow'\), then 
\(\InfFlowC \IFCompos \InfFlowC' \IFImpl \InfFlow \IFCompos \InfFlow'\).}

\begin{proof} Consider arbitrary interfaces \(\InfFlow\) and \(\InfFlow'\); and components
 \(\InfFlowC = (\InVars, \OutVars, \Flows)\) and
\(\InfFlowC' = (\InVars', \OutVars', \Flows')\).

Assume that \begin{enumerate*}[label=(\roman*)]
\item \(\InfFlowC \IFImpl \InfFlow\) and  \label{thm:impl:a1}
\item \(\InfFlowC' \IFImpl \InfFlow'\). \label{thm:impl:a2}
\end{enumerate*}
By definition of implements, \(\Flows \subseteq \GComp\) and  \(\Flows' \subseteq \GComp'\).
By definition of component, both \(\Flows\) and \(\Flows'\) flow relations, i.e. they are reflexive and transitively closed.
We want to prove that \((\Flows \cup \Flows')^* \subseteq \GComp_{\InfFlow \IFCompos \InfFlow'}\).
\begin{align*}
&(\Flows \cup \Flows')^*  
\overset{\text{denesting rule}}{=} 
\Flows^* \circ (\Flows' \circ \Flows^*)^* 
\overset{\text{flows are closed for *}}{=}&\\
&(\ID{\AllVars'} \cup \Flows) \circ (\Flows' \circ \Flows)^* \circ (\ID{\OutVars} \cup \Flows') 
\underset{\ref{thm:impl:a2}}{\overset{\ref{thm:impl:a1}}{\subseteq}}&\\
&(\ID{\AllVars'} \cup \GComp) \circ (\GComp' \circ \GComp)^* \circ (\ID{\OutVars} \cup \GComp') = 
\GComp_{\InfFlow \IFCompos \InfFlow'}.
&\text{\qedhere}
\end{align*}
\end{proof}

\subsection{Environments and Composition}

\begin{definition}
\label{def:compos:c:restrict}
For any two components \(\InfFlowC\) and
\(\InfFlowC'\) their \emph{composition w.r.t.\ a set of variables} \(\mathcal{V}\) is defined as
 \(\InfFlowC \IFCompos_{\mathcal{V}} \InfFlowC' =  (\InVars_{\InfFlow, \InfFlow'}, \OutVars_{\InfFlow, \InfFlow'}, (\Flows \cup \Flows')^* \cap (\AllVars_{\InfFlow, \InfFlow'} \times \mathcal{V}))\).
\end{definition}

\begin{definition}
\label{def:c:inverse}
The component with inversed roles w.r.t.\ component \(\InfFlowC = (\InVars, \OutVars, \Flows)\) is defined as \(\InfFlowC^{-1} = (\OutVars, \InVars, \Flows)\).
\end{definition}

\begin{proposition}
Let \(\InfFlow\) and \(\InfFlow'\) be compatible interfaces with \(\InVars \cap \InVars' = \emptyset\).
Let \(\InfFlowC_{\Env} = (\OutVars, \InVars, \Env)\) and \(\InfFlowC'_{\Env} = (\OutVars', \InVars', \Env')\) be components.
If \( \InfFlowC \IFCompos_{\InVars_{\InfFlowC, \InfFlowC'}} \InfFlowC' \IFImpl \InfFlow \IFCompos \InfFlow'\)
and 
\((\InfFlowC \IFCompos_{\OutVars_{\InfFlowC, \InfFlowC'}} \InfFlowC')^{-1} \IFImpl \InfFlow \IFCompos \InfFlow'\), then
\(\InfFlowC_{\Env} \IFImpl \InfFlow\) and \(\InfFlowC_{\Env}' \IFImpl \InfFlow'\).
\end{proposition}

\begin{proof} Assume that \(\InfFlow \IFCompt \InfFlow\) s.t.\ \(\InVars \cap \InVars = \emptyset\) and:
\begin{enumerate}[label=(a\arabic*)]
\item \( \InfFlowC \IFCompos_{\InVars_{\InfFlowC, \InfFlowC'}} \InfFlowC' \IFImpl \InfFlow \IFCompos \InfFlow'\)
\label{prop:impl:a1}
and 
\item \((\InfFlowC \IFCompos_{\OutVars_{\InfFlowC, \InfFlowC'}} \InfFlowC')^{-1} \IFImpl \InfFlow \IFCompos \InfFlow'\).
\label{prop:impl:a2}
\end{enumerate}

Consider arbitrary \((z,s) \in \Env \times \Env'\).
We want to prove that if \((z,s) \in \AllVars \times \InVars\) or
\((z,s) \in \AllVars' \times \InVars'\), then
\((z,s) \notin \Assumption\) or \((z,s) \notin \Assumption'\), respectively.
We proceed by cases.
\begin{description}
\item [Case \(s \in  \InVars_{\InfFlow, \InfFlow'}\):]
Follows directly from \ref{prop:impl:a1} and definition of composite assumptions.

\item [Case \(s \in  \OutVars\):] 
Then, by definition of interface, \((z,s) \notin \AllVars \times \InVars\).

Assume that \((z,s) \in \AllVars' \times \InVars'\). 
By  \ref{prop:impl:a2}, \(\InVars \cap \InVars = \emptyset\) and \(s' \in \OutVars_{\InfFlow, \InfFlow'}\), 
\((z,s) \in \GComp_{\InfFlow, \InfFlow'}\).
Then,
by definition of composition and \(\InfFlow \IFCompt \InfFlow'\), it follows that
\((z,s) \notin \Assumption'\).
\item [Case \(s \in  \OutVars'\):] Analogous.
\end{description}
\end{proof}

We can remove the restriction that input variables of the interfaces being composed  need to be disjoint by not allowing reflexive pairs to be considered for components composition. 
Then, composition would be defined as: 
\(\InfFlowC \IFCompos \InfFlowC' =  (\InVars_{\InfFlow, \InfFlow'}, \OutVars_{\InfFlow, \InfFlow'}, ((\Flows \setminus \ID{\OutVars}) \cup (\Flows' \setminus \ID{\OutVars'}))^*)\).

\subsection{Proof for Theorem \ref{thm:sharedref}}

\emph{Let \(\InfFlow\), \(\InfFlow'\) and \(\InfFlow''\) be well-formed interfaces.
\(\InfFlow \IFSharedRef \InfFlow'\) is a well-formed interface, and
if \(\InfFlow'' \IFRef \InfFlow\)
and \(\InfFlow'' \IFRef \InfFlow'\), then 
\(\InfFlow'' \IFRef \InfFlow \IFSharedRef \InfFlow'\).}

\begin{proof}
Consider arbitrary well-formed interfaces \(\InfFlow\), \(\InfFlow'\) and \(\InfFlow''\).
\begin{description}
\item [We prove that \(\InfFlow \IFSharedRef \InfFlow'\) is a well-formed interface.] 
Note that, by definition of shared refinement \((\star)\) \(\GComp_{\InfFlow \IFSharedRef \InfFlow'} \subseteq \GComp \cup \GComp' \) and 
\(\AComp \cup \AComp' \subseteq \AComp_{\InfFlow \IFSharedRef \InfFlow'}\)
Consider arbitrary \((z,y) \in \Prop_{\InfFlow \IFSharedRef \InfFlow'}\).
\begin{description}
\item [Case \((z,y) \in \Prop\):]
Assume towards a contradiction that:
\vspace{-0.2cm}
\begin{center}
\((z,y) \in ((\ID{\AllVars} \cup \AComp_{\InfFlow \IFSharedRef \InfFlow'}) \circ (\GComp_{\InfFlow \IFSharedRef \InfFlow'} \circ \AComp_{\InfFlow \IFSharedRef \InfFlow'})^* \circ \GComp_{\InfFlow \IFSharedRef \InfFlow'})\).
\end{center}

We prove next the following statement about alternated paths, 
for all \(n \in \mathbb{N}\):
\begin{align*}
\tIf &(z,y) \in ((\ID{\AllVars} \cup \AComp_{\InfFlow \IFSharedRef \InfFlow'}) \circ (\GComp_{\InfFlow \IFSharedRef \InfFlow'} \circ \AComp_{\InfFlow \IFSharedRef \InfFlow'})^n \circ \GComp_{\InfFlow \IFSharedRef \InfFlow'}),\\
\tthen & (z,y) \in ((\ID{\AllVars} \cup \AComp) \circ (\GComp \circ \AComp)^n \circ \GComp).
\end{align*}

\textbf{Base case \((n=0)\):}
Assume that 
\((z,y) \in \GComp_{\InfFlow \IFSharedRef \InfFlow'} \cup (\AComp_{\InfFlow \IFSharedRef \InfFlow'} \circ \GComp_{\InfFlow \IFSharedRef \InfFlow'})\).

If \((z,y) \in \GComp_{\InfFlow \IFSharedRef \InfFlow'}\), then by \((\star)\), \((z,y) \in \GComp\).

Consider arbitrary \((z,s) \in \AComp_{\InfFlow \IFSharedRef \InfFlow'}\) and 
\((s,y) \in \GComp_{\InfFlow \IFSharedRef \InfFlow'}\).

If \((z,s) \in \AComp\), then by \((\star)\), \((z,y) \in \AComp \circ \GComp\).

If \((z,s) \in \Assumption\) then, by definition of shared refinement, 
\((z,s) \notin \Assumption'\). So, by \((z,y) \in \Prop\) and 
definition of propagated guarantees, \((s,y) \in \hat{\Guarantee}_{\InfFlow \rightarrow \InfFlow'}\).
Thus, by definition of shared refinement, it cannot be the case that 
\((s,y) \in \GComp_{\InfFlow \IFSharedRef \InfFlow'}\), and so 
\((z,y) \notin \AComp_{\InfFlow \IFSharedRef \InfFlow'} \circ \GComp_{\InfFlow \IFSharedRef \InfFlow'}\).

\textbf{Induction step:} Assume as induction hypothesis that it holds for \(n\).
Consider arbitrary \((z,y) \in ((\ID{\AllVars} \cup \AComp_{\InfFlow \IFSharedRef \InfFlow'}) \circ (\GComp_{\InfFlow \IFSharedRef \InfFlow'} \circ \AComp_{\InfFlow \IFSharedRef \InfFlow'})^{n+1} \circ \GComp_{\InfFlow \IFSharedRef \InfFlow'})\).

By induction hypothesis: 
\((z,y) \in ((\ID{\AllVars} \cup \AComp) \circ (\GComp \circ \AComp)^{n} \circ \GComp \circ \AComp_{\InfFlow \IFSharedRef \InfFlow'} \circ \GComp_{\InfFlow \IFSharedRef \InfFlow'})\).
Then, we can prove analogouly to the base case that for all
\((z,s) \in ((\ID{\AllVars} \cup \AComp) \circ (\GComp \circ \AComp)^{n} \circ \GComp\) and \((s,y) \in \AComp_{\InfFlow \IFSharedRef \InfFlow'} \circ \GComp_{\InfFlow \IFSharedRef \InfFlow'}\), we have \((s,y) \in \AComp \circ \GComp\).\\

Hence \((z,y) \in ((\ID{\AllVars} \cup \AComp) \circ (\GComp \circ \AComp)^* \circ \GComp)\).
By \(\InfFlow\) being well-formed, this contradicts our initial assumption that 
\((z,y) \in \Prop\).

\item [Case \((z,y) \in \Prop'\):] Analogous.
\end{description}

\item [We prove that \(\InfFlow'' \IFRef \InfFlow \IFSharedRef \InfFlow'\).]

Consider arbitrary \(\InfFlow''\) s.t.\ \(\InfFlow'' \IFRef \InfFlow\) and
\(\InfFlow'' \IFRef \InfFlow'\). Then,
by definition of refinement:
\begin{itemize}
\item \(\Assumption'' \subseteq \Assumption\) and  \(\Assumption'' \subseteq \Assumption'\);
\item \(\Prop \subseteq \Prop''\) and  \(\Prop' \subseteq \Prop''\).
\end{itemize}

Then, \(\Assumption'' \subseteq \Assumption \cap \Assumption'\) and 
\(\Prop \cup \Prop' \subseteq \Prop''\).

We are missing to prove that 
\(\Guarantee \cup \Guarantee' \cup \hat{\Guarantee}_{\InfFlow, \InfFlow'} \subseteq \Guarantee''\).
Assume towards a contradiction that 
there exists
\((z,y) \in \Guarantee \cup \Guarantee' \cup \hat{\Guarantee}_{\InfFlow, \InfFlow'}\) s.t.\ 
\((z,y) \notin \Guarantee''\).

If \((z,y) \in \Guarantee \cup \Guarantee'\)
then, by \(\InfFlow'' \IFRef \InfFlow\) and
\(\InfFlow'' \IFRef \InfFlow'\), 
\((z,y) \in \Guarantee''\). This is a contradiction.

Consider the case that \((z,y) \in \hat{\Guarantee}_{\InfFlow \rightarrow \InfFlow'}\). Then, by definition of propagated guarantees, there exists
\((z',z) \in \Assumption\), \((z',z) \notin \Assumption'\) s.t.\
\((z',y) \in \Prop\). Then, by \(\InfFlow'' \IFRef \InfFlow\),
\((z',y) \in \Prop''\) and \((z',z) \notin \Assumption''\).
So, by \((z',z) \in \AComp''\) and \((z,y) \in \GComp''\),
\((z',y) \in \AComp'' \circ \GComp''\).
This contradicts our assumption that \(\InfFlow''\) is well-formed, because \((z',y) \in \Prop\).
We can prove analogously that it cannot be the case that 
there exists
\((z,y) \in \hat{\Guarantee}_{\InfFlow' \rightarrow \InfFlow}\) s.t.\ 
\((z,y) \notin \Guarantee''\).

Hence \(\InfFlow'' \IFRef \InfFlow \IFSharedRef \InfFlow'\). \qedhere
\end{description}
\end{proof}

\section{Stateful Interfaces: Proofs}
\label{sec:app:stateful}

In what follows, 
\(\IFFul = (\InVars,\OutVars,\States, \InitState, \Transition, \AssumptionFul,\GuaranteeFul, \PropFul)\)
and 
\(\IFFul' = (\InVars',\OutVars',\States', \InitState', \Transition', \AssumptionFul',\GuaranteeFul', \PropFul')\)
are stateful interfaces, and
\(\IFCFul = (\InVars,\OutVars,\States_{\IFCFul}, \InitState_{\IFCFul}, \Transition_{\IFCFul}, \FlowsFul)\)
and 
\(\IFCFul_{\Env} = (\OutVars, \InVars, \States_{\Env}, \InitState_{\Env}, \Transition_{\Env}, \EnvFul)\) are stateful components.

\subsection{Proof for Proposition \ref{prop:well_formed:full}}

\emph{Let \(\IFFul\) be a well-formed interface, and 
\(\IFCFul \IFImpl \IFFul\) and \(\IFCFul_{\Env} \IFEnv \IFFul\). 
For all 
 \(\Simul \subseteq \States_{\IFCFul} \times \States\) 
and
\(\Simul_{\Env} \subseteq \States \times  \States_{\Env}\) 
that witness them, respectively, then:
\begin{enumerate}[label=(\roman*)]
\item \((\FlowsFul(\InitState_\IFCFul) \cup \EnvFul(\InitState_\Env))^* \cap \PropFul(\InitState) = \emptyset\); and 
\item for all \(\std \in \States\) that are reachable from \(\InitState\), 
if \((\std_\IFCFul, \std) \in \Simul\) and \((\std, \std_\Env) \in \Simul_{\Env}\), 
then \((\FlowsFul(\std_\IFCFul) \cup \EnvFul(\std_\Env))^* \cap \PropFul(\std) = \emptyset.\)
\end{enumerate}
}

\begin{proof}
Consider arbitrary well-formed interface 
\(\IFFul\), and components \(\IFCFul\)
and \(\IFCFul_{\Env}.\)
Assume that:
\begin{enumerate}[label=(a\arabic*)]
\item \(\IFCFul \IFImpl \IFFul\) and \(\Simul \subseteq \States_{\IFCFul} \times \States\) witnesses it; and
\label{prop:wellformed:ful:1}
\item \(\IFCFul_{\Env} \IFEnv \IFFul\) and 
\(\Simul_{\Env} \subseteq \States \times \States_{\Env}\)
witnesses it.
\label{prop:wellformed:ful:2}
\end{enumerate}

\begin{enumerate}[label=(\roman*)]
\item By \ref{prop:wellformed:ful:1} and \ref{prop:wellformed:ful:2},
\(\IFCFul(\InitState_\IFCFul) \IFImpl \IFFul(\InitState)\)
and \(\IFCFul_{\Env}(\InitState_\Env) \IFImpl \IFFul(\InitState)\).
Then,  by Theorem \ref{thm:comp:wellformed} for 
stateless interfaces and \(\IFFul\) being well-formed, it follows that 
\((\FlowsFul(\InitState_\IFCFul) \cup \EnvFul(\InitState_\Env))^* \cap \PropFul(\InitState) = \emptyset\).

\item Consider arbitrary \(\std \in \States\) that is reachable from \(\InitState\).
Additionally, consider arbitrary  \(\std_\IFCFul\) and \(\std_\Env\) s.t.\
\((\std_\IFCFul, \std) \in \Simul  \tAnd (\std, \std_\Env) \in \Simul_{\Env}\).

By \ref{prop:wellformed:ful:1}, \(\IFCFul(\std_\IFCFul) \IFImpl \IFFul(q)\)
and, \ref{prop:wellformed:ful:2}, \(\IFCFul_{\Env}(\std_\Env) \IFImpl \IFFul(q)\).
By \(\IFFul\) being well-formed and by \(\std\) being acessible from the initial state \(\InitState\), then \(\IFFul(q)\) is a well-formed (stateless) interface.
Hence, by Theorem \ref{thm:comp:wellformed} for 
stateless interfaces, it follows that 
\((\FlowsFul(\std_\IFCFul) \cup \EnvFul(\std_\Env))^* \cap \PropFul(\std) = \emptyset\). \qedhere
\end{enumerate}
\end{proof}

\subsection{Proof for Proposition \ref{prop:comp:wellformed:ful}}

\emph{For all well-formed interfaces \(\IFFul_1\) and \(\IFFul_2\):
If \(\IFFul_1 \IFCompt \IFFul_2\), then \(\IFFul_1 \IFCompos \IFFul_2\) is a well-formed interface.}

\begin{proof}
Assume that \(\IFFul_1 \IFCompt \IFFul_2\).
Then, \((\star)\ \IFFul_1(\InitState_1) \IFCompt \IFFul_2(\InitState_2)\).
Moreover, we know by definition of composition that for all
\((\std_1, \std_2) \in \States_{\IFFul, \IFFul'}\) s.t.\ \((\std_1, \std_2) \neq (\InitState_1, \InitState_2)\) we have \(\IFFul_1(\std_1) \IFCompt \IFFul_2(\std_2)\).
So, it follows that for all states \((\std_1',\std_2')\) accessible by
 \((\InitState_1, \InitState_2)\) in \(\IFFul_1 \IFCompos \IFFul_2\),
we have \(\IFFul_1(\std_1') \IFCompt \IFFul_2(\std_2')\).
So, by theorem \ref{thm:comp:wellformed} for stateless interfaces,
\((\star\star)\) \(\IFFul_1(\std_1') \IFCompos \IFFul_2(\std_2')\) is a well-formed stateless interface.
Then, by \((\star)\) and \((\star\star)\),
it follows that \(\IFFul_1 \IFCompos \IFFul_2\) is well-formed.
\end{proof}

\subsection{Proof for Proposition \ref{prop:incremental:ful}}

\emph{For all interfaces $\mathbb{F}$, $\mathbb{G}$ and $\mathbb{I}$, if  \(\mathbb{F} \IFCompt \mathbb{G}\) and 
\((\mathbb{F} \IFCompos \mathbb{G}) \IFCompt \mathbb{I}\), 
then \(\mathbb{G} \IFCompt \mathbb{I}\) and 
\(\mathbb{F}  \IFCompt (\mathbb{G} \IFCompos \mathbb{I})\).}

\begin{proof}
Consider arbitrary interfaces $\mathbb{F}$, $\mathbb{G}$ and $\mathbb{I}$.
By definition of compatibility, we want to prove that:
\begin{align*}
&\tIf \mathbb{F}(\InitState_\mathbb{F}) \IFCompt \mathbb{G}(\InitState_\mathbb{G}) \tAnd 
(\mathbb{F}(\InitState_\mathbb{F}) \IFCompos  \mathbb{G}(\InitState_\mathbb{G})) \IFCompt  \mathbb{I}(\InitState_\mathbb{I})\\
&\tthen \mathbb{G}(\InitState_\mathbb{G}) \IFCompt  \mathbb{I}(\InitState_\mathbb{I})
\tAnd \mathbb{F}(\InitState_\mathbb{F}) \IFCompt (\mathbb{G}(\InitState_\mathbb{G}) \IFCompos  \mathbb{I}(\InitState_\mathbb{I})).
\end{align*}
This follows directly from theorem~\ref{thm:incremental}.
\end{proof}

\subsection{Proof for Proposition~\ref{prop:impl:ful}}

\emph{If \(\IFCFul \IFImpl \IFFul\) and \(\mathbbm{g} \IFImpl \mathbb{G}\), 
then \(\IFCFul \IFCompos \mathbbm{g} \IFImpl \IFFul \IFCompos \mathbb{G}\).}

\begin{proof}
Assume that:
\begin{enumerate*}[label=(\alph*)]
\item \(\IFCFul \IFImpl \IFFul\); and \label{thm:impl:ful:a1}
\item \(\mathbbm{g} \IFImpl \mathbb{G}\). \label{thm:impl:ful:a2}
\end{enumerate*}
Then, there exists \(\Simul_\IFCFul\) and \(\Simul_\mathbbm{g}\)
that witnesses \ref{thm:impl:ful:a1} and \ref{thm:impl:ful:a2}, respectively.

Consider the relation:
\(\Simul = \{((\std_\IFCFul, \std_\mathbbm{g}), (\std_\IFFul, \std_\mathbb{G}))
\ |\ \std_\IFFul \in \Simul_\IFCFul(\std_\IFCFul) \tAnd \std_\mathbb{G} \in \Simul_{\mathbbm{g}}(\std_\mathbbm{g})\}.\)

Clearly, by  \ref{thm:impl:ful:a1} and \ref{thm:impl:ful:a2}, 
\(((\InitState_\IFCFul, \InitState_\mathbbm{g}), (\InitState_\IFFul, \InitState_\mathbb{G})) \in \Simul\).
Then,
\(\IFCFul(\InitState_\IFCFul) \IFImpl \IFFul(\InitState_\IFFul)\)
and 
\(\mathbbm{g}(\InitState_\mathbbm{g}) \IFImpl \mathbb{G}(\InitState_\mathbb{G})\).
So, by Proposition~\ref{prop:impl} for stateless interfaces, it follows that 
\(\IFCFul(\InitState_\IFCFul) \CFCompos \mathbbm{g}(\InitState_\mathbbm{g}) \IFImpl 
\IFFul(\InitState_\IFFul) \IFCompos \mathbb{G}(\InitState_\mathbb{G})\).

Consider arbitrary \(((\std_\IFCFul, \std_\mathbbm{g}), (\std_\IFFul, \std_\mathbb{G})) \in \Simul\). Then, by  \ref{thm:impl:ful:a1} and \ref{thm:impl:ful:a2}, 
there exists \((\std_\IFCFul', \std_\IFFul') \in \Simul_\IFCFul\) s.t.\
\(\IFCFul(\std_\IFCFul') \IFImpl \IFFul(\std_\IFFul')\):
and there exists
\((\std_\mathbb{G}', \std_\mathbbm{g}') \in \Simul_{\mathbbm{g}}\) s.t.\
\(\IFCFul(\std_\mathbbm{g}') \IFImpl \IFFul(\std_\mathbb{G}')\).
Thus, by definition of~H, 
\(((\std_\IFCFul', \std_\mathbbm{g}'), (\std_\IFFul',\std_\mathbb{G}')) \in H\). And, by 
by Proposition~\ref{prop:impl},
\(\IFCFul(\std_\IFCFul') \CFCompos \mathbbm{g}(\std_\mathbbm{g}') \IFImpl 
\IFFul(\std_\IFFul') \IFCompos \mathbb{G}(\std_\mathbb{G}')\).

Hence, \(\Simul\) is a simulation relation for 
\(\IFCFul \IFCompos \mathbbm{g} \IFImpl \IFFul \IFCompos \mathbb{G}\).
\end{proof}

\subsection{Proof for Proposition \ref{thm:refinement:full}}

\emph{Let \(\IFFul_1 \IFRef \IFFul_2\).}
\begin{enumerate*}[label=(\alph*)]
\item \emph{If \(\IFCFul \IFImpl \IFFul_1\), then \(\IFCFul \IFImpl \IFFul_2\).}
\item \emph{If \(\IFCFul_{\Env} \IFEnv \IFFul_2\), then \(\IFCFul_{\Env} \IFEnv \IFFul_1\).}
\end{enumerate*}

\begin{proof}
Assume that \(\IFFul_1 \IFRef \IFFul_2\).
Then, there exists a simulation relation \(\Simul_{\IFRef} \subseteq \States_1 \times \States_2\) that witnesses it.

\begin{enumerate}[label=(\alph*)]
\item 
Assume that \(\IFCFul \IFImpl \IFFul_1\). Then, there exists a simulation relation \(\Simul_{\IFImpl} \subseteq \States_\IFCFul \times \States_1\)
 that witnesses it.
Consider the relation \(\Simul =  \Simul_{\IFImpl} \circ \Simul_{\IFRef}\).

By Definitions  \ref{def:ref:full} and \ref{def:implements:full}, 
\((\InitState_\IFCFul, \InitState_1) \in \Simul_{\IFImpl}\) and
\((\InitState_1, \InitState_2) \in \Simul_{\IFRef}\). 
So, \((\InitState_\IFCFul, \InitState_2) \in \Simul\)
Additionally, 
\(\IFFul_1(\InitState_1) \IFRef \IFFul_2(\InitState_2)\) and 
\(\IFCFul(\InitState_\IFCFul) \IFImpl \IFFul_1(\InitState_1)\).
Then, by Proposition \ref{thm:refinement},
\(\IFCFul(\InitState_\IFCFul) \IFImpl \IFFul_2(\InitState_2)\).

Consider arbitrary \((\std_\IFCFul, \std_2) \in \Simul\).
By construction of \(\Simul\) there exists 
\((\std_\IFCFul, \std_1) \in \Simul_{\IFImpl}\)
and 
\((\std_1, \std_2) \in \Simul_{\IFRef}\).
We want to prove that:
\[\tIf \std_\IFCFul' \in \Transition_\IFCFul(\std_\IFCFul) \tThen
\text{there exists } \std_{2}' \in \Transition_{2}(\std_{2})
\tSt (\std_\IFCFul', \std_{2}') \in \Simul \tAnd \IFCFul(\std_\IFCFul') \IFImpl \IFFul_2(\std_2').\]

Assume that \(\std_\IFCFul' \in \Transition_\IFCFul(\std_\IFCFul)\).
By \((\std_\IFCFul, \std_{1}) \in \Simul_{\IFImpl}\),
then there exists a state  
\(\std_{1}' \in \Transition_{1}(\std_{1})\)
s.t.\ \((\std_\IFCFul', \std_{1}') \in \Simul_{\IFImpl}\).
So, \(\FlowsFul(\std_\IFCFul') \subseteq \Compl{\GuaranteeFul}_{\IFFul_1}(\std_{1}')\).
Additionally, by \((\std_1, \std_2) \in \Simul_{\IFRef}\)
and \(\std_{1}' \in \Transition_{1}(\std_{1})\), 
there exists \(\std_{2}' \in \Transition_{2}(\std_{2})\) s.t.\
\(\GuaranteeFul_{\IFFul_2}(\std_{2}') \subseteq \GuaranteeFul_{\IFFul_1}(\std_{1}')\).
Thus, \((\std_\IFCFul', \std_{2}') \in \Simul\) and \(\FlowsFul(\std_\IFCFul') \subseteq \Compl{\GuaranteeFul_1}(\std_1') \subseteq \Compl{\GuaranteeFul_2}(\std_2')\).
So, by definition of implements for stateless interfaces, \(\IFCFul(\std_\IFCFul') \IFImpl \IFFul_2(\std_2')\).

Hence \(\Simul\) is a witness for \(\IFCFul \IFImpl \IFFul\).

\item Assume that \(\IFCFul_{\Env} \IFImpl \IFFul_2\). Then, there exists a simulation relation \(\Simul_{\IFImpl} \subseteq \States_2 \times \States_{\Env}\)
 that witnesses it.
Consider the relation \(\Simul = \Simul_{\IFRef} \circ  \Simul_{\IFImpl}\).
We can prove analogously to the previous case that \(H\) witnesses 
\(\IFCFul_{\Env} \IFEnv \IFFul_1\).
 \qedhere
\end{enumerate}
\end{proof}

\subsection{Proof for Theorem \ref{thm:independent_imp:ful}}

\emph{For all well-formed interfaces \(\IFFul_1'\), \(\IFFul_1\) and \(\IFFul_2\), 
if \(\IFFul_1' \IFRef \IFFul_1\) and \(\IFFul_1 \IFCompt \IFFul_2\), 
then \(\IFFul_1' \IFCompt \IFFul_2\) and \(\IFFul_1' \IFCompos \IFFul_2 \IFRef \IFFul_1 \IFCompos \IFFul_2\).}

\begin{proof}
Assume that:
\begin{enumerate*}[label=(\alph*)]
\item \(\IFFul_1' \IFRef \IFFul_1\); and \label{thm:independent_imp:ful:a1}
\item \(\IFFul_1 \IFCompt \IFFul_2\). \label{thm:independent_imp:ful:a2}
\end{enumerate*}
\begin{description}
\item [\(\IFFul_1' \IFCompt \IFFul_2\).]
It follows from \ref{thm:independent_imp:ful:a1} and theorem \ref{thm:independent_imp} for stateless interfaces.

\item [We prove now that \(\IFFul_1' \IFCompos \IFFul_2 \IFRef \IFFul_1 \IFCompos \IFFul_2\).]

From \ref{thm:independent_imp:ful:a1}, there exists a relation 
\(\Simul_\IFRef \subseteq \States_1' \times \States_1\) that witnesses the refinement.
Consider the following simulation relation:
\[\Simul = \{((\std_{\IFFul_1'}, \std_{\IFFul_2}),(\std_{\IFFul_1}, \std_{\IFFul_2})) \ |\  (\std_{\IFFul_1'},\std_{\IFFul_1}) \in \Simul_\IFRef  \tAnd \IFFul_1(\std_{\IFFul_1}) \IFCompt  \IFFul_2(\std_{\IFFul_2}) \}.\]

By \ref{thm:independent_imp:ful:a1} and \ref{thm:independent_imp:ful:a2}, 
\(((\InitState_{\IFFul_1'}, \InitState_{\IFFul_2}),(\InitState_{\IFFul_1}, \InitState_{\IFFul_2})) \in \Simul\).
Additionally, \(\IFFul_1'(\InitState_{\IFFul_1'}) \IFRef \IFFul_1(\InitState_{\IFFul_1})\).
Then, by Theorem \ref{thm:independent_imp} for stateless interface, 
\(\IFFul_1'(\InitState_{\IFFul_1'}) \IFCompos \IFFul_2(\InitState_{\IFFul_2}) \IFRef \IFFul_1(\InitState_{\IFFul_1}) \IFCompos \IFFul_2(\InitState_{\IFFul_2})\).

Consider arbitrary \(((\std_{\IFFul_1'}, \std_{\IFFul_2}),(\std_{\IFFul_1}, \std_{\IFFul_2})) \in \Simul\).
Consider arbitrary \(O \in \OutStep_{\IFFul_1' \IFCompos \IFFul_2}((\std_{\IFFul_1'}, \std_{\IFFul_2}))\).
Then, there exists \(\Guarantee'\) and \(\Prop'\) s.t.\
\(O = \OutStep_{\IFFul_1' \IFCompos \IFFul_2}((\std_{\IFFul_1'}, \std_{\IFFul_2}), \Guarantee', \Prop')\).
By \(\Simul_\IFRef\) witnessing \ref{thm:independent_imp:ful:a1} and Definition \ref{def:moves:in:out},
there exists 
\(\Guarantee \subseteq \Guarantee'\) and \(\Prop \subseteq \Prop'\)
s.t.\
\(O' = \OutStep_{\IFFul_1 \IFCompos \IFFul_2}((\std_{\IFFul_1}, \std_{\IFFul_2}), \Guarantee, \Prop)\).

Consider arbitrary \(I \in \InStep_{\IFFul_1 \IFCompos \IFFul_2}((\std_{\IFFul_1}, \std_{\IFFul_2}))\).
Then, by \(\Simul_\IFRef\) witnessing \ref{thm:independent_imp:ful:a1} and Definition \ref{def:moves:in:out}, there exists \(\Assumption\) s.t.\
\(I' = \InStep_{\IFFul_1 \IFCompos \IFFul_2}((\std_{\IFFul_1}, \std_{\IFFul_2}), \Assumption)\).
By \(\Simul_\IFRef\) witnessing \ref{thm:independent_imp:ful:a1},
there exists 
\(\Assumption' \subseteq \Assumption\) 
s.t.\
\(I = \InStep_{\IFFul_1' \IFCompos \IFFul_2}((\std_{\IFFul_1'}, \std_{\IFFul_2}), \Assumption')\).

Consider arbitrary \(((\std'_{\IFFul_1'}, \std'_{\IFFul_2}),(\std'_{\IFFul_1}, \std'_{\IFFul_2})) \in (O \cap I) \times (O' \cap I')\).
Then, by  \ref{thm:independent_imp:ful:a1} and \(H\) definition,
\(\IFFul_1'(\std'_{\IFFul_1'}) \IFRef \IFFul_1(\std'_{\IFFul_1})\) and 
\(\IFFul_1(\std'_{\IFFul_1}) \IFCompt  \IFFul_2(\std'_{\IFFul_2})\)
So, by Theorem \ref{thm:independent_imp} for stateless interfaces, 
\(\IFFul_1'(\std_1') \IFCompos \IFFul_2(\std_2)  \IFRef \IFFul_1(\std_1) \IFCompos \IFFul_2(\std_2)\).

Hence \(\Simul\) is a witness relation for 
\(\IFFul_1' \IFCompos \IFFul_2 \IFRef \IFFul_1 \IFCompos \IFFul_2\). \qedhere
\end{description}

\end{proof}

\section{Models for Theorem \ref{thm:semantics:comparison}}
\label{sec:app:semantics}

Consider the set of traces depicted in the table \ref{tab:ex:expressiv:model:1}.
Clearly for both the initial time and any time greater than 1 we have that \(x\) is independent of both \(y\) and \(z\). \footnote{Note that we are considering interleavings of possible outcomes at a given point in time. If there is only one possible outcome, then two variables are vacuously independent.}
At time 1 we have that:
\begin{itemize} 	
\item \(x\) is not independent of \(y\), because we are missing a trace that interleaves
the trace \(\pi_1\) and \(\pi_2\), i.e.\ a trace with \(x\) evaluated to 1 and \(y\) to 0 at time 1;
\item \(x\) is not independent of \(z\), because we are missing a trace that interleaves
the trace \(\pi_1\) and \(\pi_3\), i.e.\ a trace with \(x\) evaluated to 1 and \(z\) to 1 at time 1.
\end{itemize}
Then, the set of traces in the table \ref{tab:ex:expressiv:model:1} are not a model for the structure-aware semantics.
However, this is a model for the unstructured interpretation.
This formula allows \(t\) to be a function of both \(\pi\) and \(\pi'\),
unlike the structure-aware in which  \(t\) can only depend on \(\pi\).
Note that the interleaving of \(x\) and \(y\) for \(\pi_1\)
and \(\pi_3\) at time 1 is the trace \(\pi_1\).
Then, \(f_t(\pi_1, \pi_2) = 1\) while \(f_t(\pi_1, \pi_3) = 2\).

We verified both models using Z3 prover. We present now our encoding.

We defined data types to identify variables and traces.
\begin{lstlisting}[frame=single]
(declare-datatypes () ((Var X Y Z)))
(declare-datatypes () ((Traces TR1 TR2 TR3)))
\end{lstlisting}

Below we show the encoding for structure-aware semantics.
The other semantics are defined by changing the position of 
\verb|exists ((t Int))|.
\begin{lstlisting}[frame=single]
; Structure-aware Semantics
(assert (
  forall ((pi1 Traces)) (
  exists ((t Int)) (
  forall ((pi2 Traces)) (
  exists ((pi3 Traces)) (
  exists ((x Bool)) (
  exists ((z Bool)) (
      and 
        (> t 0) 
        (
          and
            (= (setTraces pi1 t X) x) 
            (= (setTraces pi3 t X) x)
            (= (setTraces pi2 t Z) z)
            (= (setTraces pi3 t Z) z)
        )
        (
          forall ((tP Int)) (
          exists ((xP Bool)) (
          exists ((y Bool)) (
          ite (and (> tP -1) (< tP t)) (
            and 
              (= (setTraces pi1 tP X) xP) 
              (= (setTraces pi3 tP X) xP) 
              (= (setTraces pi2 tP Y) y) 
              (= (setTraces pi3 tP Y) y) 
          ) true )
          ))
        )
        (
          forall ((tG Int)) (
          exists ((xG Bool)) (
          exists ((zG Bool)) (
          ite (> tG t) (
          and
            (= (setTraces pi1 tG X) xG) 
            (= (setTraces pi3 tG X) xG)
            (= (setTraces pi2 tG Z) zG)
            (= (setTraces pi3 tG Z) zG)
           ) true ))))))))))))
\end{lstlisting}

The encoding for the model \(\setTraces_u\).
\begin{lstlisting}[frame=single]
; Model Tu
(define-fun setTraces ((x!0 Traces) (x!1 Int) (x!2 Var)) Bool (
  ite (= x!0 TR1) ( 
    ; TR1 at time 0
    ite (= x!1 0) (ite (= x!2 X) false 
    	(ite (= x!2 Y) false (ite (= x!2 Z) false true))) (
    ; TR1 at time 1
    ite (= x!1 1) (ite (= x!2 X) true 
    	(ite (= x!2 Y) true (ite (= x!2 Z) false true))) (   
    ; TR1 at time greater than 1
    ite (= x!2 X) false (ite (= x!2 Y) false 
    	(ite (= x!2 Z) false true)))
  )) (
    ite (= x!0 TR2) ( 
    ; TR2 at time 0
    ite (= x!1 0) (ite (= x!2 X) false
    	(ite (= x!2 Y) false (ite (= x!2 Z)false true))) (
    ; TR2 at time 1
    ite (= x!1 1) (ite (= x!2 X) false 
    	(ite (= x!2 Y) false (ite (= x!2 Z) false true))) (   
    ; TR2 at greater than 1
    ite (= x!2 X) false (ite (= x!2 Y) false
    	(ite (= x!2 Z) false true))) 
  )) (
    ite (= x!0 TR3) ( 
    ; TR3 at time 0
    ite (= x!1 0) (ite (= x!2 X) false 
    	(ite (= x!2 Y) false(ite (= x!2 Z) false true))) (
    ; TR3 at time 1
    ite (= x!1 1) (ite (= x!2 X) false 
    	(ite (= x!2 Y) true (ite (= x!2 Z) true true))) (   
    ; TR3 at greater than 1
    ite (= x!2 X) false (ite (= x!2 Y) false
    	(ite (= x!2 Z) false true)))
  )) true))))
\end{lstlisting}

The encoding for the model \(\setTraces_a\).
\begin{lstlisting}[frame=single]
; Model Ta
(define-fun setTraces ((x!0 Traces) (x!1 Int) (x!2 Var)) Bool (
  ite (= x!0 TR1) ( 
    ; TR1 at time 0
    ite (= x!1 0) (ite (= x!2 X) false 
    	(ite (= x!2 Y) false (ite (= x!2 Z) false true))) (
    ; TR1 at time 1
    ite (= x!1 1) (ite (= x!2 X) false 
    	(ite (= x!2 Y) false (ite (= x!2 Z) true true))) (   
    ; TR1 at time greater than 1
    ite (= x!2 X) false (ite (= x!2 Y) false 
    	(ite (= x!2 Z) false true)))
  )) (
    ite (= x!0 TR2) ( 
    ; TR2 at time 0
    ite (= x!1 0) (ite (= x!2 X) false 
    	(ite (= x!2 Y) false (ite (= x!2 Z)false true))) (
    ; TR2 at time 1
    ite (= x!1 1) (ite (= x!2 X) true 
    	(ite (= x!2 Y) true (ite (= x!2 Z) false true))) (   
    ; TR2 at greater than 1
    ite (= x!2 X) false (ite (= x!2 Y) false 
    	(ite (= x!2 Z) false true))) 
  )) (
    ite (= x!0 TR3) ( 
    ; TR3 at time 0
    ite (= x!1 0) (ite (= x!2 X) false 
    	(ite (= x!2 Y) false(ite (= x!2 Z) false true))) (
    ; TR3 at time 1
    ite (= x!1 1) (ite (= x!2 X) true 
    	(ite (= x!2 Y) false (ite (= x!2 Z) false true))) (   
    ; TR3 at greater than 1
    ite (= x!2 X) false (ite (= x!2 Y) false 
    	(ite (= x!2 Z) false true)))
  )) (
    ite (= x!0 TR4) (ite (= x!2 X) false 
    	(ite (= x!2 Y) false (ite (= x!2 Z) false true))) true
 )))))
\end{lstlisting}

\end{document}